# A Correlative Four-Dimensional Study of Phase Separation at the Subnanoscale to Nanoscale of a Ni-Al Alloy

Elizaveta Y. Plotnikov,[1] Zugang Mao,[1] Sung-Il Baik,[1,6] Mehmet Yildirim,[1,2] Yongsheng Li,[1,3] Daniel Cecchetti,[1] Ronald D. Noebe,[4] Georges Martin[1,5], and David N. Seidman[1,6,*]

[1]Northwestern University, Department of Materials Science and Engineering, 2220 Campus Drive, Evanston, IL 60208-3108, USA

[2]Selçuk Üniversitesi Mühendislik Fakültesi, Metalurji ve Malzeme Müh. Bölümü, B-101 42075- Selçuklu/KONYA, Turkey

[3]Nanjing University of Science and Technology, School of Materials Science and Engineering, 200 Xiaolingwei, Nanjing, 210094, China

[4]NASA Glenn Research Center, Materials and Structures Division, 21000 Brookpark Rd., Cleveland, OH 44135-3191, USA

[5] Directeur de Recherches Émérite, CEA Saclay, 9119, FRANCE

[6]Northwestern University Center for Atom-Probe Tomography, 2220 Campus Drive, Evanston, IL 60208-3108, USA

**Abstract**

The temporal evolution of ordered γ'(L1$_2$)-precipitates precipitating in a disordered γ(f.c.c.) matrix is studied in extensive detail for a Ni-12.5 Al at.% alloy aged at 823 K (550 ºC), for times ranging from 0.08 to 4096 h. Three-dimensional atom-probe tomography (3-D APT) results are compared to monovacancy-mediated lattice-kinetic Monte Carlo (LKMC$_1$) simulations on a rigid lattice, which include monovacancy-solute binding energies through 4$^{th}$ nearest-neighbor distances, for the same mean composition and aging temperature. The temporal evolution of the measured values of the mean radius, <R(t)>, number density, aluminum supersaturations, and volume fraction of the γ'(L1$_2$)-precipitates are compared to the predictions of a modified version of the Lifshitz-Slyozov-Wagner coarsening model due to Calderon, Voorhees et al. The resulting experimental rate constants are used to calculate the Gibbs interfacial free-energy between the γ(f.c.c.)- and γ'(L1$_2$)-phases, which enter the model, using data from two thermodynamic databases, and its value is compared to all extant values dating from 1966. The diffusion coefficient for coarsening is calculated utilizing the same rate constants and compared to all archival diffusivities, *not determined from coarsening experiments, and is demonstrated to be the inter-diffusivity, $\tilde{D}$, of Ni and Al.* The monovacancy-mediated LKMC$_1$ simulation results are in good agreement with our 3-D APT data. It is demonstrated that the compositional interfacial width, for the {100} interface, between the γ(f.c.c.)- and γ'(L1$_2$)-phases, decreases continuously with increasing aging time and <R(t)>, both for the 3-D APT results and monovacancy-mediated LKMC$_1$ simulations, in disagreement with an ansatz intrinsic to the so-called trans-interface diffusion-controlled coarsening model, which assumes the exact opposite trend for binary alloys.

**Keywords:** Nickel-based superalloys, Atom-probe tomography, Temporal evolution, Monovacancy-mediated lattice kinetic Monte Carlo, Lifshitz-Slyozov-Wagner (LSW) model

*Corresponding author: d-seidman@northwestern.edu, Tel: +1-847-491-4391, Fax: +1-847-467-2269

**Nomenclature**

| | |
|---|---|
| $a_0$ | lattice parameter of Ni$_3$Al(L1$_2$) |
| $\langle a \rangle$ | average of the lattice parameters of the γ(f.c.c.)- and γ'(L1$_2$)-phases |
| $B^{\gamma'}$ | bulk modulus of the γ'(L1$_2$)-phase |
| $b_{rand}$ | random number between 0 and 1 |
| $C_{EQ}^{V}$ | equilibrium monovacancy concentration in a pure metal |
| $C_i^{\gamma}$ | concentration of element $i$ in the γ(f.c.c.)-phase |
| $C_{Al}^{\gamma,eq}(R)$ | the equilibrium concentration of Al in the γ(f.c.c)-phase at the γ/γ' interface of a γ'(L1$_2$)-precipitate |
| $C_i^{\gamma,eq}(\infty)$ | equilibrium concentration of element $i$ in the γ(f.c.c.)- phase |
| $\langle C_i^{\gamma,ff}(t) \rangle$ | $ff$ (far-field) concentration of element $i$ in the γ(f.c.c.)- phase |
| $\left( C_i^{\gamma',eq}(\infty) - C_i^{\gamma,eq}(\infty) \right)$ | the difference between the equilibrium concentrations of the γ'(L1$_2$)- and γ(f.c.c.)-phases |
| $C_{LKMC}^{V}$ | monovacancy concentration in LKMC$_1$ simulations |
| $C_0$ | initial concentration of the precipitating solute element |
| $D_{experiment}^{coarsening}$ | diffusivity determined from 3-D APT coarsening experiments |
| $D_i^{region}$ | diffusivity of element $i$ ($i$ = Ni or Al) determined in a given region from monovacancy-mediated LKMC$_1$ simulations |
| $D_i^{necks}$ | diffusivity of element $i$ ($i$ = Ni or Al) determined in the necks connecting two γ'(L1$_2$)-precipitates from monovacancy-mediated LKMC$_1$ simulations |
| $D_i^{supercell}$ | diffusivity of element $i$ ($i$ = Ni or Al) determined in the supercell from monovacancy-mediated LKMC$_1$ simulations |
| $D_i^{\gamma-matrix}$ | diffusivity of element $i$ ($i$ = Ni or Al) determined in the γ(f.c.c.)-phase from monovacancy-mediated LKMC$_1$ simulations |
| $D_i^{\gamma' \, precipitates}$ | diffusivity of element $i$ ($i$ = Ni or Al) determined in the γ'(L1$_2$)-phase from monovacancy-mediated LKMC$_1$ simulations |
| $\tilde{D}$ | interdiffusivity of Ni and Al |
| $D_0$ | diffusivity pre-exponential factor |
| $E_{sp-p,q}^{i}$ | energy of atom $i$ at the saddle point ($sp$) between sites $p$ and $q$ |
| $F(\phi_{\gamma'})$ | correction factor applied to the calculation of $K$ to account for a nonzero volume fraction, $\phi_{\gamma'}(t)$ |
| $f(t)$ | fraction of γ'(L1$_2$)-precipitates interconnected by necks |
| $f_i$ | modeled linear fit |
| $G_m^{\gamma}{''}$ | curvature of the Gibbs molar free energy of mixing in the γ(f.c.c.)-phase |
| $g(R/\langle R(t) \rangle, t)$ | normalized and dimensionless quantity used for the ordinate axis of PSDs |
| $J_{APT}^{st}$ | stationary nucleation current as measured by APT experiments |
| $J^{st}_{CNT}$ | stationary nucleation current as calculated by CNT |
| $J^{st}_{experiment}$ | stationary nucleation current measured experimentally using 3-D APT |
| $J^{st}_{LKMC}$ | stationary nucleation current as calculated utilizing monovacancy-mediated LKMC$_1$ simulations |
| $K$ | rate constant for $\langle R(t) \rangle$ according to the LSW and CVMK models |
| $K_i^{\gamma'/\gamma}(t)$ | partitioning coefficient of element $i$ between the γ'- and γ-phases |
| $K_n$ | rate constant for $N_v(t)$ in the LSW and CVMK models |
| $k_B$ | Boltzmann's constant |
| $L$ | ratio of the elastic energy contribution to the interfacial Gibbs free energy contribution to the overall morphology of γ'(L1$_2$)-precipitates |
| $L^3$ | volume of monovacancy-mediated LKMC$_1$ simulation box |
| $l^\gamma$ | capillary length in the γ(f.c.c.)- phase |
| $N$ | total number of possible nucleation sites per unit volume |
| $N_{ppt}$ | effective number of γ'(L1$_2$)-precipitates per analyzed 3-D APT volume |
| $N_v(t)$ | number density per unit volume of γ'(L1$_2$)-precipitates |
| $n$ | total number of atoms enclosed within an isoconcentration surface |
| $n_j$ | number of jumps made by a monovacancy in monovacancy-mediated LKMC$_1$ simulation by either an Al or NI atom |
| $n_{tot}$ | total number of data points in a distribution |
| $p$ | temporal exponent for $\langle R(t) \rangle$ according to the LSW and CVMK models |
| $Q$ | activation energy for solute diffusion in a thermally activated process |
| $q$ | temporal exponent for $N_v(t)$ according to the LSW and CVMK models |
| $R$ | γ'(L1$_2$)-precipitate radius |
| $R^*$ | critical radius for nucleation |
| $\langle R(t) \rangle$ | time-dependent mean γ'(L1$_2$)-precipitate radius |
| $\langle R(t_0) \rangle$ | mean γ'(L1$_2$)-precipitate radius at the onset of stationary coarsening, which is not in general $t = 0$ |
| $r$ | temporal exponent for $\Delta C_i^\gamma(t)$ according to the LSW and CVMK models |
| $SS_{res}$ | sum of square residuals |
| $SS_{tot}$ | total sum of squares |
| $T$ | temperature in Kelvin |
| $t$ | aging time |
| $t_{LKMC}$ | monovacancy-mediated LKMC$_1$ simulation time |
| $t_0$ | time at which stationary coarsening commences in an alloy |
| $V_a^\gamma$ | atomic volume of the γ(f.c.c.)- phase |
| $V_a^{\gamma'}$ | atomic volume of the γ'(L1$_2$)- phase |
| $V_m^{\gamma'}$ | molar volume of a solute atom in the γ'(L1$_2$)-precipitate phase |
| $W_i^*$ | critical net reversible work required for the formation of a spherical nucleus containing $i$ atoms |
| $W_{p,q}^{i,v}$ | exchange frequency between an atom of type $i$ on a site $p$ and a monovacancy, $v$, on a NN site $q$ |
| $W_R$ | reversible work for the formation of a spherical nucleus |
| $W_R^*$ | critical net reversible work required for the formation of a critical spheroidal nucleus |
| $y_i$ | measured quantity |
| $\langle y \rangle$ | mean of all $y_i$ values |
| $Z$ | Zeldovich factor |
| $\alpha$ | monovacancy jump distance in monovacancy-mediated LKMC$_1$ simulations |
| $\beta^*$ | kinetic coefficient describing the rate of condensation of single atoms on the critical nuclei |
| $\Delta C_i^{\gamma}(t)$ | supersaturation of element $i$ in the γ(f.c.c.)-phase |
| $\Delta C_i^{\gamma'}(t)$ | supersaturation of element $i$ in the γ'(L1$_2$)-phase |
| $\Delta F_{ch}$ | chemical energy component of the total Helmholtz free energy of an alloy |
| $\Delta F_{el}$ | elastic energy component of the total Helmholtz free energy of an alloy |
| $\Delta H$ | change in enthalpy |
| $\Delta H_V^F$ | enthalpy of formation of a monovacancy |
| $\Delta S$ | entropy change |
| $\Delta S_V^F$ | entropy of formation of a monovacancy |
| $\delta(t)$ | compositional interfacial widths between the γ(f.c.c.)- and γ'(L1$_2$)-phases |
| $\varepsilon$ | lattice parameter misfit between the γ(f.c.c.)- and γ'(L1$_2$)-phases |
| $\varepsilon_{i-j}^k$ | atom-atom interaction energies |
| $\varepsilon_{v-i}^k$ | monovacancy-solute *ghost interactions* between an atom and a monovacancy in the $k^{th}$ NN shell |
| $\zeta$ | detection efficiency of the two-dimensional microchannel plate (MCP) |
| $\eta_{correlation}^{region,Ni-Al}$ | correlation factor for diffusion in a dilute Ni-Al system |
| $\kappa$ | rate constant for $\Delta C_i^{\gamma}(t)$ according to the LSW and CVMK models |
| $\langle \lambda_{edge-edge}(t) \rangle$ | average minimum edge-to-edge distance between γ'(L1$_2$)-precipitates |
| $\mu^{\gamma}$ | shear modulus of the γ(f.c.c.)-phase |
| $v^i$ | attempt frequency for the exchange between a Ni or Al atom and a monovacancy |
| $\Xi$ | correlation length: the interaction distance over which monovacancy-solute interactions occur |
| $\xi^2$ | coefficient of determination |
| $\rho$ | atomic number density of the γ'(L1$_2$)-phase |
| $\sum_{\text{broken bonds}} \left( n_{ij}^k \varepsilon_{i-j}^k + \varepsilon_{v-i}^k \right)$ | sum of the atomic interactions over all the bonds that are affected by having atom $i$ and the monovacancy $v$ exchange between sites $p$ and $q$, respectively |
| $\sigma$ | one standard deviation from the mean |
| $\sigma^{\gamma/\gamma'}$ | interfacial Gibbs free energy between γ(f.c.c.)- and γ'(L1$_2$)-phases |
| $\tau_j$ | time for Monte Carlo step $j$ |
| $\phi_{\gamma'}(t)$ | volume fraction of the γ'(L1$_2$)-precipitate phase |
| $\phi_{\gamma'}^{eq}$ | equilibrium volume fraction of the γ'(L1$_2$)-precipitate phase |
| 3-D APT | 3-D atom-probe tomography |
| AV | Akaiwa-Voorhees precipitate size distribution |
| BW | Brailsford-Wynblatt precipitate size distribution |
| CVMK | Calderon-Voorhees-Murray-Kostorz mean-field diffusion-limited coarsening model |
| CNT | classical nucleation theory |
| DFT | density functional theory |
| $ff$ | far-field |
| GCMC | Grand Canonical Monte Carlo |
| ICP-AES | inductively coupled plasma atomic emission spectroscopy |
| LDA | local-density approximation |
| LKMC | lattice kinetic Monte Carlo |
| LKMC$_1$ | LKMC parameters that take into account the monovacancy-solute binding energies out to 4$^{th}$ NN distances |
| LKMC$_2$ | LKMC parameters that artificially truncate the monovacancy-solute binding beyond the 1$^{st}$ NN, keeping all atomic interactions the same as for LKMC$_1$ |
| LSW | Lifshitz-Slyozov-Wagner diffusion-limited mean-field coarsening model |
| MCP | microchannel plate |
| NN | nearest-neighbor |
| PSD | precipitate size distribution |
| RTA | residence time algorithm |
| TEM | transmission electron microscopy |
| TIDC | trans-interface diffusion-controlled coarsening model |
| VASP | Vienna *ab initio* simulation package |

1. **Introduction**

Commercial Ni-based superalloys are used for single-crystal turbine blades in aircraft engines (military and commercial) and land-based natural-gas fired turbine engines because of their high strength, coarsening, creep, oxidation and corrosion resistance, and toughness at elevated temperatures [1-3]. These alloys consist of a disordered Ni-rich matrix (γ(f.c.c.)-phase) and coherent ordered Ni$_3$Al (L1$_2$ structure) precipitates (γ'(L1$_2$)-phase). Precipitation strengthening results from the nucleation, growth, and coarsening of the γ'(L1$_2$)-precipitate phase, though continued coarsening during usage at elevated temperatures will eventually lead to a loss in peak strength. Aluminum and Ti are commonly added to aid in the formation of the L1$_2$-phase precipitates. Refractory elements with small diffusivities, such as Mo, W, Nb, Ta, Zr, Ru, Re, and Hf, are added to decelerate the coarsening kinetics [1, 2]. The primary thermodynamic driving force for coarsening is the minimization of the total interfacial area per unit volume of the γ'(L1$_2$)-precipitates, which minimizes the total Gibbs free energy because it is the product of the total interfacial area and the interfacial Gibbs free energy, $\sigma^{\gamma/\gamma'}$, between the γ(f.c.c.)- and γ'(L1$_2$)-phases [3-6]. The coarsening rate in a commercial Ni-based superalloy is mainly controlled by: (1) the diffusivities of the alloying elements, which are strongly temperature dependent; (2) $\sigma^{\gamma/\gamma'}$, whose value decreases continuously with increasing temperature; and (3) the elastic stress field interactions among γ'(L1$_2$)-precipitates [7].

Our prior research has focused primarily on analyzing the nucleation, growth, and coarsening kinetics in ternary Ni-Al-Cr alloys aged at 873 K (600 °C) and quaternary Ni-Al-Cr-W, Ni-Al-Cr-Re, Ni-Al-Cr-Ta, and Ni-Al-Cr-Ru alloys aged mainly at 1073 K (800 °C) [8-31]. The present research concentrates on comparing the temporal evolution and kinetic pathways for phase separation in a binary Ni-12.5 Al at.% alloy aged at 823 K (550 °C). This extensively studied binary Ni-Al alloy should, in principle, be simpler than concentrated multicomponent alloys, but it has some surprising subtleties and complexities that are resolved in this article in great detail and compared with prior results.

*1.1. A short history of the interfacial Gibbs free energy of the Ni(f.c.c.)/Ni$_3$Al(L1$_2$) heterophase interface, $\sigma^{\gamma/\gamma'}$*

The value of $\sigma^{\gamma/\gamma'}$ can be extracted from experimental coarsening data by determining the rate constants of the mean radius of γ'(L1$_2$)-precipitates, $\langle R(t) \rangle$, and the supersaturation of solute in the γ(f.c.c.)-matrix, $\Delta C_{Al}^{\gamma}(t)$, utilizing one of several extant diffusion-limited mean-field coarsening models. Ardell and Nicholson used the Lifshitz-Slyozov [32] and Wagner [33] (LSW) coarsening models and the then available archival diffusivities to estimate $\sigma^{\gamma/\gamma'}$ experimentally, for a Ni-13.5Al at.% alloy aged at 898 K (625 °C), 1023 K (750 °C), and 1048 K (775 °C), to be ~30 mJ m$^{-2}$ [34]. Additionally, Ardell used a technique developed by Ben Israel and Fine [35] to measure solute depletion in the γ(f.c.c.)-matrix employing the change in the Curie temperature and the rate constant associated with $\langle R(t) \rangle$ for Ni-13.2 Al and Ni-12.9 Al at.% alloys aged at 898 (625 °C) and 988 K (715 °C), respectively, which yielded ~14 mJ m$^{-2}$ for $\sigma^{\gamma/\gamma'}$ [36]. Chellman and Ardell used the same methodology to calculate $\sigma^{\gamma/\gamma'}$ for four Ni-Al alloys (Ni-14.1 Al, Ni-15.9 Al, Ni-17.1 Al, and Ni-19.3 Al at.%) aged at 1073 K (800 °C) and obtained ~8 mJ m$^{-2}$ [37]. Hirata and Kirkwood used the LSW model and archival diffusivities to calculate $\sigma^{\gamma/\gamma'}$ for a Ni-12.3 Al at.% alloy, aged at 943 K (670 °C), 953 K (680 °C), 963 K (690 °C), and 968 K (695 °C), to be 17.4, 16.6, 19.8, and 24.3 mJ m$^{-2}$, respectively [38]. Marsh and Chen measured the rate constants for $\langle R(t) \rangle$ and the volume fraction of the γ'(L1$_2$)-phase, $\phi_{\gamma'}(t)$, from coarsening data of a Ni-12.5 Al at.% alloy aged at 823 K (550 °C), 873 K (600 °C), 923 K (700 °C), and 973 K (700 °C) to obtain $\sigma^{\gamma/\gamma'}$ values of 16.9, 21.7, 16.6, and 10.3 mJ m$^{-2}$, respectively [39]. Calderon, Voorhees, Murray, and Kostorz (CVMK) determined the $\langle R(t) \rangle$ rate constant from the coarsening data for a Ni-12.8 Al at.% alloy aged at 848 K (575 °C) and 863 K (590 °C) using a modified LSW model that takes into account the thermodynamics of the γ(f.c.c.)-phase for the very first time [40]; they obtained a $\sigma^{\gamma/\gamma'}$ value of ~40-80 mJ m$^{-2}$ based on an early approximate thermodynamic database for the γ(f.c.c.)-matrix-phase. These large differences in $\sigma^{\gamma/\gamma'}$ were attributed to the measurements being performed at different temperatures, because $\sigma^{\gamma/\gamma'}$ is temperature dependent [40]. Next, Ardell used a non-ideal solution value for the curvature of the molar Gibbs free energy of mixing of the γ(f.c.c.)-phase, $G_m^{\gamma}{''}$, from Calderon et al.'s research to reanalyze his prior results [36] and obtained an estimate of ~8 mJ m$^{-2}$ for $\sigma^{\gamma/\gamma'}$ at 898 K (625 °C) to 973 K (700 °C) ([41]. More recently, Ardell employed his trans-interface diffusion-controlled (TIDC) coarsening model to reanalyze his earlier results and obtained a value of ~20 mJ m$^{-2}$ for $\sigma^{\gamma/\gamma'}$ at 898 K (625 °C) and 988 K (715 °) [42]. A summary of all the previously determined experimental $\sigma^{\gamma/\gamma'}$ values for different binary Ni-Al alloys aged at different temperatures is listed in Table 1 and also displayed graphically in Fig. 1.

Surprisingly, there exist substantially fewer theoretical results for values of $\sigma^{\gamma/\gamma'}$ than experimental ones. Price and Cooper used density functional theory (DFT) to calculate $\sigma^{\gamma/\gamma'}$ and compared the effect of either including or ignoring spin-polarization on $\sigma^{\gamma/\gamma'}$ for the (100) interface [43]. They found that $\sigma^{\gamma/\gamma'}$ ranges from 63 mJ m$^{-2}$ when spin-polarization is included to 25 mJ m$^{-2}$ when it is ignored. Mishin used an embedded-atom method potential, which he developed, to calculate values of $\sigma^{\gamma/\gamma'}$ for interfaces parallel to the {100}, {110}, and {111} planes as a function of supercell size [44], which yielded $\sigma^{\gamma/\gamma'}$ values of 46, 28, and 12 mJ m$^{-2}$ for the {100}, {110}, and {111} planes, respectively, at 700 K. Mao et al. used first-principles calculations, including ferromagnetism and vibrational entropy, and Monte Carlo simulations to study the γ(f.c.c.)/γ'(L1$_2$) interface parallel to the {100}, {110}, and {111} planes as a function of temperature [45]. They found a linear decrease in $\sigma^{\gamma/\gamma'}$ from 0 to 1100 K (827 °C) for each {hkl} interface. Woodward et al. also used first-principles calculations to study the γ(f.c.c.)/γ'(L1$_2$) interface parallel to the {100}, {110}, and {111} planes as a function of temperature [46]. They observed a non-linear decrease in $\sigma^{\gamma/\gamma'}$ from 0 to 1300 K (1027 ºC), which varied with {hkl} and asymptotically approached ~10 mJ m$^{-2}$ above 600 K (323 ºC) for each {hkl} interface. Next, Mishin utilized the capillary fluctuation methodology to calculate $\sigma^{\gamma/\gamma'}$ for the {100} planes as a function of temperature and obtained a value of 14 mJ m$^{-2}$ at 800 K (527 ºC) [47]. A summary of all the theoretical/computationally determined $\sigma^{\gamma/\gamma'}$ values for binary Ni-Al alloys is presented in Table 2. Aluminum has a non-zero size effect in Ni, which differs from that of Ni$_3$Al, and hence there exists a dependence of the precipitate energy on precipitate radius, which is non-trivial even in the absence of precipitate-precipitate interactions: Section 4.2.2.1.

*1.2.  The Mean-Field Coarsening Models*

*1.2.1.  Lifshitz-Slyozov-Wagner (LSW) mean-field model for diffusion-limited coarsening*

The LSW model for a binary alloy assumes: (i) no elastic interactions among γ'(L1$_2$)-precipitates, thereby limiting $\phi_{\gamma'}(t)$ to zero; (ii) γ'(L1$_2$)-precipitates have a spherical morphology; (iii) coarsening occurs in a stress-free matrix; (iv) the γ'(L1$_2$)-precipitate diffusion fields do not overlap; (v) dilute solid-solution theory obtains in the γ(f.c.c.)-matrix; (vi) the linearized version of the Gibbs–Thomson equation obtains; (vii) coarsening is diffusion- controlled and occurs by the classical evaporation-condensation ("the big eat the small") mechanism; and (viii) γ'(L1$_2$)-precipitates coarsen with a fixed chemical composition, which is 100% solute. These assumptions are highly restrictive and extremely difficult to obtain experimentally. And yet surprisingly little experimental evidence (in systems where $\phi_{\gamma'}(t)$ is greater than zero) exists to support the prediction of the LSW temporal dependency of $\langle R(t) \rangle$. The experimental γ'(L1$_2$)-precipitate size distributions (PSDs) are consistently broader, flatter, and more symmetric than what the mean-field LSW model predicts [48-53]. Hence, the LSW PSD is agreed to be incorrect, as is probably the *modified* LSW PSD [54]. A more realistic approach, which is dependent on $\phi_{\gamma'}(t)$, was developed by Brailsford and Wynblatt (BW) [55]. Akaiwa and Voorhees (AV) used a numerical solution to the multi-particle diffusion problem to model PSDs during the later stages of coarsening, when $\phi_{\gamma'}(t)$ is close to its equilibrium value [56]. Additionally, Marsh and Glicksman (MG) developed a model that treats γ'(L1$_2$)-precipitates as an equivalent distribution of field cells, each of which is analyzed using a diffusion field [57]. The LSW, BW, AV, and MG models all assume that γ'(L1$_2$)-precipitates have a spherical morphology.

The LSW model for a binary alloy, subject to the above assumptions, yields the following relationships for: (1) $\langle R(t) \rangle$; (2) the number density per unit volume of γ'(L1$_2$)-precipitates, $N_v(t)$; and (3) the supersaturation, $\Delta C_i^{\gamma}(t)$:

$$\langle R(t) \rangle^p - \langle R(t_0) \rangle^p = K(t - t_0) \tag{1}$$

$$N_v(t)^q = K_n(t - t_0) \tag{2}$$

$$\Delta C_i^\gamma(t) = \langle C_i^{\gamma,ff}(t) \rangle - C_i^{\gamma,eq}(\infty) = \kappa(t - t_0)^{-1/r} \tag{3}$$

where $\langle R(t_0) \rangle$ is the mean γ'(L1$_2$)-precipitate radius at the onset of quasi-stationary coarsening at time $t_0$; $\Delta C_i^\gamma(t)$ is the difference between the far-field (*ff*) matrix concentration of component *i*, $\langle C_i^{\gamma,ff}(t) \rangle$, and the equilibrium concentration of component *i*, $C_i^{\gamma,eq}(\infty)$; and where $K$, $K_n$, and $\kappa$ are the associated rate constants. In the diffusion-limited LSW model the temporal exponents $p$, $q$, and $r$ are 3, -1, and 3, respectively. In our analyses we take $p$, $q$, and $r$ to be *unknown constants*, which are determined from 3-D atom-probe tomography (APT) data and monovacancy mediated lattice-kinetic Monte Carlo (LKMC$_1$) simulations, utilizing a nonlinear multivariate regression methodology for analyzing the data [58], Appendix A. If $p$, $q$, and $r$ are not equal to 3, -1, and 3, then the values of $K$, $K_n$, and $\kappa$ are dimensionally incorrect, which is not a generally appreciated fact.

*1.2.2. Calderon-Voorhees-Murray-Kostorz (CVMK) model for diffusion-limited mean-field coarsening*

CVMK developed a diffusion-limited mean-field coarsening model for binary alloys that: (1) incorporates a nonzero value of $\phi_{\gamma'}(t)$; (2) permits a nonzero solubility of the solute species in the γ(f.c.c.)-matrix; and (3) incorporates non-ideal solution thermodynamics for the γ(f.c.c)-matrix [40]. The CVMK model utilizes a capillary length, $l^\gamma$, for the γ(f.c.c.)-phase:

$$l^\gamma = \frac{2V_m^{\gamma'} \sigma^{\gamma/\gamma'}}{\left(C_i^{\gamma',eq}(\infty) - C_i^{\gamma,eq}(\infty)\right) G_m^{\gamma ''}} ; \tag{4}$$

where $V_m^{\gamma'}$ is the molar volume of the solute in the γ'(L1$_2$)-precipitate phase; and $\left(C_i^{\gamma',eq}(\infty) - C_i^{\gamma,eq}(\infty)\right)$ is the difference between the equilibrium concentrations of element *i* in the γ'(L1$_2$)- and γ(f.c.c.)-phases, respectively. This approach permitted CVMK to rewrite $K$ and $\kappa^\gamma$ (the value of $\kappa$ is measured in the γ(f.c.c.)-phase) utilizing Eqs. (1) and (3) in terms of $l^\gamma$ and the diffusivity, $D_{experiment}^{coarsening}$, obtained from coarsening experiments:

$$K = \frac{4D_{experiment}^{coarsening} l^{\gamma} F(\phi_{\gamma'})}{9\left(C_i^{\gamma',eq}(\infty) - C_i^{\gamma,eq}(\infty)\right)} \quad (5)$$

$$\kappa^{\gamma} = \left(\frac{4D_{experiment}^{coarsening}}{9\left(l^{\gamma}\right)^2 \left(C_i^{\gamma',eq}(\infty) - C_i^{\gamma,eq}(\infty)\right)}\right)^{-1/3} \quad (6)$$

where $F(\phi_{\gamma'})$ is a correction factor that includes the effect of a nonzero value of $\phi_{\gamma'}(t)$, which is calculated numerically [40, 59]. The CVMK model does not provide a corresponding correction factor for $\kappa^{\gamma}$.

Following the CVMK coarsening model, the γ'(L1$_2$)-precipitates nucleate initially from the γ(f.c.c)-matrix as spheroids, which subsequently grow and coarsen temporally. The γ'(L1$_2$)-precipitates in a Ni-12.5 Al at.% alloy aged at 823 K (550 °C) follow four distinct regimes of phase separation (or phase decomposition or precipitation): (I) quasi-stationary γ'(L1$_2$)-precipitate nucleation; (II) concomitant γ'(L1$_2$)-precipitate nucleation and growth; (III) concurrent growth and coarsening via the coagulation and coalescence mechanism as opposed to the classic evaporation-condensation (the big eat the small) mechanism; and (IV) quasi-stationary coarsening of γ'(L1$_2$)-precipitates. During the nucleation and growth stages of the aging process, coagulation and coalescence of γ'(L1$_2$)-precipitates (a phenomenon often ignored in mean-field coarsening models [60]) plays a significant role in the coarsening of γ'(L1$_2$)-precipitates in Ni-Al and Ni-Al-Cr alloys. Coagulation and coalescence occurs when an L1$_2$ partially-ordered neck forms between two adjacent γ'(L1$_2$)-precipitates. Mao et al. performed a correlative 3-D APT study and monovacancy-mediated LKMC$_1$ simulations to determine the role of the diffusion mechanism on precipitation in three different Ni-Al-Cr alloys aged at 873 K (600 °C), for times ranging from 0.17 to 4096 h [61, 62]. Their studies demonstrate that coagulation and coalescence is a consequence of overlapping nonequilibrium concentration profiles surrounding the γ'(L1$_2$)-precipitates, *which yield initially nonequilibrium diffuse interfaces: that is, non-atomically sharp. These diffuse interfaces are energetically unfavorable, leading to the coagulation and coalescence of two adjacent γ'(L1$_2$)-precipitates, and the complete disappearance of the necks with increasing aging times*. At later aging times, coagulation and coalescence is less prevalent because $N_v(t)$ is continuously decreasing and concomitantly the edge-to-edge distance between γ'(L1$_2$)-precipitates is continuously increasing, and coarsening is then dominated by the classical evaporation-condensation mechanism (the big eat the small), which also occurs during nucleation, regime I (Sections 3.2 and 4.2), when $N_v(t)$ is rapidly increasing and $\langle R(t) \rangle$ is approximately constant. At long aging times $N_v(t)$ decreases to a value at which the diffusion fields of the γ'(L1$_2$)-precipitates are no longer overlapping. Also observed in the Ni-Al-Cr alloys is the beginning of a transition, at aging times >1000 h, of the γ'(L1$_2$)-precipitates from a spheroidal-to-a-cuboidal morphology, and the alignment of the γ'(L1$_2$)-precipitates along a <100>-type direction to minimize elastic strain energy [34, 63-65], which is a phenomenon commonly called rafting [66]. The small elastic strain energy in a Ni-12.5 Al at.% alloy is caused by a lattice parameter misfit at 823 K (550 ºC) between the γ(f.c.c.)-matrix (lattice parameter = 3.58 Å) and the γ'(L1$_2$)-precipitates (lattice parameter = 3.59 Å), yielding a fractional value of 0.003 ± 0.001 for the lattice parameter misfit [67].

*1.3. A new methodology for calculating the interfacial Gibbs free energy of the Ni(f.c.c.)/Ni3Al(L12) interface, $\sigma^{\gamma/\gamma'}$*

Ardell [41] utilized the CVMK mean-field coarsening model [40], to develop a new approach for calculating $\sigma^{\gamma/\gamma'}$ from two of the three rate constants for a binary alloy by demonstrating that $l^{\gamma}$ can be rewritten in terms of $K$ and $\kappa^{\gamma}$, by combining Eqs. (5) and (6):

$$l^{\gamma} = K^{1/3} \kappa^{\gamma} F(\phi_{\gamma'})^{-1/3} . \qquad (7)$$

Combining Eqs. (4) and (7) yields an expression for $\sigma^{\gamma/\gamma'}$ that is independent of $D_{experiment}^{coarsening}$:

$$\sigma^{\gamma/\gamma'} = \frac{l^{\gamma}\left(C_i^{\gamma',eq}(\infty) - C_i^{\gamma,eq}(\infty)\right)G_m^{\gamma\,"}}{2V_m^{\gamma'}} = \frac{K^{1/3}\kappa^{\gamma}F(\phi_{\gamma'})^{-1/3}\left(C_i^{\gamma',eq}(\infty) - C_i^{\gamma,eq}(\infty)\right)G_m^{\gamma\,"}}{2V_m^{\gamma'}} \qquad (8)$$

This approach yields more accurate values for $\sigma^{\gamma/\gamma'}$ than those obtained using an archival value of $D$, which was the main approach employed prior to Ardell's model, and it permits $\sigma^{\gamma/\gamma'}$ to be calculated *independent of* $D_{experiment}^{coarsening}$. Once $l^{\gamma}$ is determined, $D_{experiment}^{coarsening}$ is then calculated by combining Eqs. (5) and (6) to yield:

$$D_{experiment}^{coarsening} = \frac{9K^{2/3}\left(C_i^{\gamma',eq}(\infty) - C_i^{\gamma,eq}(\infty)\right)}{4\kappa^{\gamma}F(\phi_{\gamma'})^{2/3}} . \qquad (9)$$

All the values of $\sigma^{\gamma/\gamma'}$ we calculate employ Eq. (8), which takes into account the effect of $\phi_{\gamma'}(t)$ on $K$, *are independent of* $D_{experiment}^{coarsening}$, which is calculated using Eq. (9). We demonstrate unequivocally, Section 4.8, that $D_{experiment}^{coarsening}$ is equal to the inter-diffusivity of Ni and Al, $\tilde{D}$. It is suggested strongly that all experimental values of $\sigma^{\gamma/\gamma'}$ determined, which don't use Ardell's approach (1995), not be taken seriously because they all utilize a value of a diffusivity from the archival literature, rather than determining both values independently of one another.

*1.4. Overview of our current research*

In our research, a Ni-12.5 Al at.% alloy aged at 823 K (550 ºC) is studied utilizing Vickers microhardness measurements, 3-D APT, monovacancy-mediated LKMC$_1$ simulations, and some transmission electron microscopy (TEM). Our research focuses strongly on analyzing: (1) Vickers microhardness of the alloy; (2) γ'(L1$_2$)-precipitate morphology; (3) γ'(L1$_2$)-precipitate volume fraction, $\phi_{\gamma'}(t)$; (4) γ'(L1$_2$)-precipitate number density, $N_v(t)$; (5) mean radius of γ'(L1$_2$)-precipitates, $\langle R(t) \rangle$; (6) Al concentration of the γ(f.c.c.)-phase, $C_{Al}^{\gamma}(t)$; (7) Al concentration of the γ'(L1$_2$)-phase, $C_{Al}^{\gamma'}(t)$; (8) supersaturation of Al in the γ(f.c.c.)-phase, $\Delta C_{Al}^{\gamma}(t)$; (9) supersaturation of Al in the γ'(L1$_2$)-phase, $\Delta C_{Al}^{\gamma'}(t)$; (10) partitioning coefficient of Al, $K_{Al}^{\gamma'/\gamma}(t)$, between the two phases; (11) partitioning coefficient of Ni, $K_{Ni}^{\gamma'/\gamma}(t)$, between the two-phases; (12) γ'(L1$_2$)-precipitate PSDs; (13) fraction of γ'(L1$_2$)-precipitates interconnected by necks, $f(t)$; (14) average *minimum edge-to-edge distance* between γ'(L1$_2$)-precipitates, $\langle \lambda_{edge-edge}(t) \rangle$; and (15) interfacial compositional widths between the γ(f.c.c.)- and γ'(L1$_2$)-phases, $\delta(t)$. Monovacancy-mediated LKMC$_1$ simulations of these 15 quantities are compared to all the experimental 3-D APT results.

This study constitutes the most complete and detailed investigation of the temporal evolution of the nanostructure and chemical compositions of a first-order phase transformation in any metallic two-phase alloy made to date. While there exists considerable data quantifying phase separation (or phase decomposition or precipitation) in Ni-Al alloys for an Al concentration that is in the [γ(f.c.c.) plus γ'(L1$_2$)] phase field, most research has focused on either the early regimes (nucleation and growth) or the later regimes (growth and coarsening), but not the four regimes. Additionally, while a prior study also focused on measurements of $\langle R(t) \rangle$ and $C_{Al}^{\gamma}(t)$, these measurements were taken from different experiments [36].

3-D APT allows us to derive all the quantities listed above (with the exception of the Vickers microhardness) from only 3-D APT experiments for each aging time, implying that our data is completely self-consistent. We further demonstrate that the monovacancy-mediated LKMC$_1$ simulations (Section 2.5) are in excellent agreement with the experimental 3-D APT results, which provide extremely valuable physical insights into the origin of $\delta(t)$. The latter has been a subject of recent great interest.

**2. Methodologies**

*2.1. Processing of specimens*

High-purity Ni and Al were vacuum induction melted under a small partial pressure of argon and chill cast in a 19 mm diameter copper mold to form a polycrystalline master ingot with a target composition of Ni-12.5 Al at. %. The composition of the ingot was determined to be Ni-12.5 ± 0.1Al at. %, utilizing inductively-coupled plasma atomic-emission spectroscopy (ICP-AES). Samples from the master ingot underwent a three-stage heat treatment. The first stage was homogenization in the γ(f.c.c.)-phase field in vacuum at 1573 K (1300 ºC) for 20 h. The second stage was a vacancy anneal in the γ(f.c.c.)-phase field in vacuum/argon at 1223 K (950 ºC) for 3 h, followed by a direct drop-quench into water. The latter temperature was chosen to be above the γ(f.c.c.)/γ'($L1_2$) solvus temperature, which was calculated to be 1180 K (907 ºC), utilizing Saunders's thermodynamic database [68] in Thermo-Calc [69]. The final stage was an aging anneal in the [γ(f.c.c.) plus γ'($L1_2$)-phase field] at 823 K (550 ºC) under flowing Ar for times ranging from 0.08 to 4096 h, followed by an ice-brine water-quench. Fig. 2 displays a partial Ni-Al phase diagram as determined experimentally by Ma and Ardell, Ardell and Nicholson, simulations using two different databases in Thermo-Calc, and our grand canonical Monte Carlo (GCMC) simulations [22, 34, 68-71]. The average composition of the alloy, Ni-12.5 Al at.%, is indicated by a vertical dashed line. The compositional trajectories of the γ(f.c.c.)- and γ'($L1_2$)-phases are indicated by two horizontal heavy black arrows, Sections 3.3 and 4.3.

## 2.2.   Micro-indentation measurements

Vickers microhardness measurements were conducted on specimens aged from 0.08 to 2607 h, utilizing a Struers Duramin-5 microhardness tester with an applied load of 300 g for 5 s at ambient temperature across a flat area of a bulk sample. The samples were prepared by hot mounting at 180 ºC, followed by grinding using 2400 grit SiC paper, and then polished to a 1 μm finish using an alumina solution. At least 16 separate indentation measurements were averaged for each aging time.

## 2.3.   Three-dimensional atom-probe tomography

This research was performed utilizing a pulsed-laser 3-D APT (a LEAP 4000X Si tomograph [72-82]), employing a picosecond ultraviolet (UV) laser (wavelength = 355 nm), at a target detection rate of 0.05 ions pulse$^{-1}$, a specimen temperature of 40.0 ± 0.3 K, a pulse repetition rate of 200 kHz, and an ambient gauge pressure of <6.7 x $10^{-8}$ Pa. A UV laser energy of 5 pJ pulse$^{-1}$ was determined to be the optimum value for this Ni-12.5 Al at.% alloy [83]. These experimental conditions were optimized to provide the highest possible compositional accuracy. 3-D APT data were analyzed utilizing the program IVAS 3.6 (Cameca Instruments, Madison, Wisconsin). The γ(f.c.c.)-matrix/γ'($L1_2$)-precipitate heterophase interfaces were delineated with Al isoconcentration surfaces utilizing the inflection-point technique [13], and compositional information was obtained using the proximity histogram methodology [84, 85]. The overall composition of each dataset was found to vary slightly from those given by ICP-AES testing, ±0.5 at.%, especially at longer aging times. Additionally, preferential evaporation of Ni was observed. To account for these variations, the overall composition of each data set was normalized to the ICP-AES composition measurement based on the measured value of $\phi_{\gamma'}(t)$, Appendix B. The standard errors for all quantities are calculated based on counting statistics and represent two standard deviations from the mean value [86]. A solutionized specimen ($t=0$) was analyzed by 3-D APT to establish whether or not γ'(L1$_2$)-precipitates were present in the initial quenched-in state, and none were detected: thus, $\phi_{\gamma'}(t=0)=0$. Additional details of the procedures are given elsewhere [18, 23, 30].

*2.4.    Transmission electron microscopy*

Conventional TEM specimens were prepared from standard 3 mm diam. discs. The 500 μm thick discs were ground mechanically to a thickness of ~100 μm. These discs were electropolished utilizing a Struers Tenupol-5 double-jet electropolisher operating at 21 Vdc at a temperature of 253 K (-20 °C), using liquid nitrogen as a coolant. The electrolyte consisted of 8 vol.% perchloric acid and 14 vol.% 2-butoxyethanol in 78 vol.% methanol. Conventional TEM investigations were performed employing a Hitachi HT-7700 instrument operating at 120 kV, utilizing a double-tilt sample holder. The ordered γ'(L1$_2$)-precipitates were imaged employing a centered dark-field condition utilizing a superlattice reflection of the ordered L1$_2$-structure of the γ'(L1$_2$)-phase.

*2.5.    Monovacancy-mediated lattice-kinetic Monte Carlo simulations*

We utilized monovacancy-mediated LKMC$_1$ simulations [87, 88] to study the temporal evolution of phase separation at 823 K. Monovacancy-mediated LKMC$_1$ simulations are distinctly different from Monte Carlo simulations, utilizing the Metropolis algorithm, because the former method employs a residence-time algorithm (RTA) [89], which takes into account the physics of the atomic jump frequency via a monovacancy. Indeed, the underlying physical mechanism is thermally-activated diffusion via a monovacancy mechanism; specifically, a monovacancy exchanges places with one of its 12 first-nearest-neighbors (1$^{st}$ NN). It is well-established that diffusion in f.c.c. elements and their alloys occur by a monovacancy mechanism. The primitive atomic positions are an array of rhombohedral cells of the f.c.c. lattice and the volume of the simulation box is $L^3$, where $L$ is 128 sites in the nucleation regime and 256 sites in the coarsening regime. One monovacancy is introduced into the simulation volume and therefore each lattice site is occupied by either one Ni or one Al atom or by the monovacancy. The exchange frequency, $W_{p,q}^{i,v}$, between an atom of type $i$ on a site $p$ and the vacancy, $v$, on a first NN site $q$ is given by:

15 | P a g e

$$W_{p,q}^{i,v} = v^i \exp\left(-\frac{E_{sp-p,q}^i - \sum_{\text{broken bonds}} \left(n_{ij}^k \varepsilon_{i-j}^k + \varepsilon_{v-i}^k\right)}{k_B T}\right) \qquad (10)$$

where $v^i$ is the attempt frequency for the exchange; $E_{sp-p,q}^i$ is the energy of atom $i$ at the saddle point ($sp$) between sites $p$ and $q$; where $\sum_{\text{broken bonds}} \left(n_{ij}^k \varepsilon_{i-j}^k + \varepsilon_{v-i}^k\right)$ is the sum of the atomic interactions over all the bonds affected by having atom $i$ and the monovacancy, $v$, exchange between sites $p$ and $q$, respectively [90]; $k_B$ is Boltzmann's constant; and $T$ is the temperature in Kelvin. The thermodynamics of the Ni-Al alloy are embodied in the values of $\varepsilon_{i-j}^k$, the atom-atom interaction energies, and $\varepsilon_{v-i}^k$, *the monovacancy-solute ghost interactions* between an atom and a monovacancy in the $k^{th}$ NN shell. Ghost interactions take into account the effect of the perturbation of the electronic structure, in the vicinity of the monovacancy, on the total energy. *All the atom-atom interaction energies in the alloy and the monovacancy-solute binding energies are estimated utilizing first-principles calculations* performed employing the plane-wave pseudo-potential total-energy method within the local-density approximation (LDA) [91], as implemented in the Vienna *ab initio* simulation package (VASP) [61, 92-97]. The kinetic database requires additional parameters: values for $v^i$ and the $E_{sp-p,q}^i$ terms were taken from the literature [11, 98, 99]. Table 3 summarizes the values of $\varepsilon_{i-j}^k$ for the 1st to 4th NNs utilized for the Ni-Ni, Al-Al, and Ni-Al interactions, as well as the $E_{sp-p,q}^i$ and $v^i$ values for Ni and Al that were used in the LKMC$_1$ simulations for the present work. These parameters are further elucidated in [71]. The notation LKMC$_1$ is for the case when the interactions extend to fourth NNs.

Monovacancy-mediated LKMC$_1$ computer time was converted to real physical time for comparison with the APT experimental results using the jump frequency of each first NN atom surrounding a monovacancy. Each transition is performed with an associated physical time using a residence-time algorithm, RTA [89]. The time for each MC step is:

$$\tau_j = \left(\sum_i W_{p,q}^{i,v}\right)^{-1}. \qquad (11)$$

The direction in which a monovacancy jumps is chosen with a certain probability for each MC step, while satisfying the following condition:

$$\sum_{i=1}^{j-1} W_{p,q}^{i,v} \leq \frac{b_{rand}}{\tau} \leq \sum_{i=1}^{j} W_{p,q}^{i,v} \qquad (12)$$

where $b_{rand}$ is a random number between 0 and 1.

The monovacancy concentrations, $C^V_{LKMC}$, are 4.77×10$^{-7}$ at.fr. for 128$^3$ cells and 5.96×10$^{-8}$ at.fr. for 256$^3$ cells, respectively. These values are significantly greater than the equilibrium monovacancy concentration in the Ni-Al alloy at 823 K (550 °C) because the diffusional properties and the precipitation kinetic pathways depend on the jump frequencies, the monovacancy-mediated LKMC$_1$ time is rescaled by a factor given by the ratio of the monovacancy concentration in the simulated system to its value ideally in the bulk Ni-Al system. The equilibrium monovacancy concentration in a pure metal, $C^V_{EQ}$, is given by [100]:

$$C^V_{EQ} = \exp\left(\frac{\Delta S^F_V}{k_B}\right) \exp\left(-\frac{\Delta H^F_V}{k_B T}\right) ; \quad (13)$$

where $\Delta S^F_V$ and $\Delta H^F_V$ are the monovacancy formation entropy and formation enthalpy, respectively. For pure Ni, $\Delta S^F_V$ =1.19 x 10$^{-4}$ eV K$^{-1}$ (1.38 $k_B$) [101], $\Delta H^F_V$ =1.72 eV [102], and therefore $C^V_{EQ}(823\,K)$ = 1.17 x 10$^{-10}$ at.fr., which is a factor 4,077 times smaller than $C^V_{LKMC}$. Accordingly, the recorded time in monovacancy-mediated LKMC$_1$ simulations is rescaled utilizing the equilibrium vacancy concentration as follows:

$$t = t_{LKMC} C^V_{LKMC} \exp\left(\frac{\Delta H^F_V}{k_B T}\right) \quad (14)$$

where $t_{LKMC}$ is the simulation "time." Pareige et al. [11] fitted their LKMC simulations at an aging time of 4 h to their 3-D APT results. The new value of $\Delta H^F_V$ for the higher solute concentrations in a Ni-Al-Cr alloy is 1.93 eV. A similar procedure was performed to find $\Delta H^F_V$ for a Ni-Al alloy, with compositions ranging from Ni-12.5 to Ni-13.4 Al at.%, and a range of values between 1.85 to 1.95 eV was obtained. For the Ni-12.5 Al at.% alloy a $\Delta H^F_V$ value of 1.85 eV is employed at 823 K (550 °C).

The atomic configurations in the alloy as a function of time and the resulting Ni and Al concentration profiles are obtained using LKMC parameter set 1 (LKMC$_1$) [61, 62]. Parameter set LKMC$_1$ takes into account the monovacancy-solute binding energies out to 4$^{th}$ NN distances, whereas LKMC$_2$ truncates artificially the monovacancy-solute binding beyond the 1$^{st}$ NN distance, keeping all atomic interactions the same. *Hence, LKMC$_1$ and LKMC$_2$ simulate alloys with the same thermodynamics, but they have distinctly different correlations between successive vacancy jumps, which are shown to have profound effects on the temporal evolution of phase-separation (or phase decomposition or precipitation) and $\delta(t)$* [45, 61].

## 3. Results

### 3.1. Temporal evolution of the Vickers microhardness values

The temporal evolution of the Vickers microhardness values for Ni-12.5 Al at.% aged for 0.08 to 2607 h at 823 K (550 °C) is displayed in Fig. 3. The value of the Vickers microhardness remains fairly constant for <4 h of aging, where the alloy is in the nucleation and nucleation plus growth regimes, regimes I and II (Sections 3.2 and 4.2). After the initiation of regime III (growth and coarsening) the Vickers microhardness proceeds to increase continuously as the alloy enters the quasi-stationary coarsening regime, IV, through 1024 h. This indicates that the increase in $\phi_{\gamma'}(t)$ is responsible for the increase of the Vickers microhardness, Section 3.2. The Vickers microhardness achieves a maximum value (peak hardness) of 247 ± 11 at 1024 h and then decreases to 233 ± 5 at 2607 h, which is commonly referred to as over aging.

### 3.2. Temporal evolution of the gamma-prime (L1$_2$ structure) precipitate volume fraction, $\phi_{\gamma'}(t)$, number density, $N_v(t)$, and mean radius, $\langle R(t) \rangle$

The temporal evolution of the γ'(L1$_2$)-precipitate morphology, presented in 50 x 50 x 100 nm$^3$ 3-D APT reconstructions, of the alloy aged for 0.25, 1, 256, 1024, 2607, and 4096 h is displayed in Fig. 4a-f. Spheroidal γ'(L1$_2$)-precipitates are detected commencing at the earliest aging time [$\langle R(t=0.08\,h) \rangle$ = 0.79 ± 0.20 nm], which grow and coarsen temporally to $\langle R(t=4096\,h) \rangle$ = 14.59 ± 1.62 nm, where they have a cuboidal morphology, faceted on {100} planes. The corresponding increase in $\langle R(t) \rangle$ is a factor of 18.47. The γ'(L1$_2$)-precipitates at 256 h [$\langle R(t=256\,h) \rangle$ = 5.65 ± 0.22 nm, Fig. 4c] are still spheroidal in morphology, while γ'(L1$_2$)-precipitates at 1024 h [$\langle R(t=1024\,h) \rangle$ = 9.43 ± 0.65 nm, Fig. 4d], are undergoing the spheroidal-to-cuboidal morphological transition to minimize elastic strain energy, which is commonly observed in Ni-Al alloys [30, 34, 63, 64, 103].

Some of the γ'(L1$_2$)-precipitates at 1024 h, Fig. 4d, are spheroidal, while others are already cuboids once the radii of the precipitates become larger. At 2607 h [$\langle R(t=2607\,h) \rangle$ =10.72 ± 0.87 nm, Fig. 4e] *all the visible* γ'(L1$_2$)-precipitates are cuboidal, and at 4096 h, the γ'(L1$_2$)-precipitates have a cuboidal morphology and are aligned along a <100>-type direction, Fig. 4f, which is commonly called rafting [66]. Fig. 4g is a dark-field TEM micrograph for an aging time of 1024 h, which can be directly compared to the 3-D APT results in Fig. 4d. The γ'(L1$_2$)-precipitates, displayed in the white against the black γ(f.c.c.)-matrix background, are aligned along a <100>-type direction, which is less obvious in Fig. 4d because 3-D APT reconstructions have a smaller field-of-view than TEM micrographs. The better spatial resolution of the 3-D APT reconstructions demonstrates, however, that all the γ'(L1$_2$)-precipitates are not yet cuboids at 1024 h once the radii of the precipitates become larger, which Fig. 4g does not indicate very well. The value of $\langle R(t=1024h) \rangle$ obtained by TEM is 8.05 ± 0.16 nm, and the 3-D APT value at 1024 h is 9.43 ± 0.65 nm, which is in approximate agreement.

Fig. 5 displays the temporal evolution of $\phi_{\gamma'}(t)$, $N_v(t)$, and $\langle R(t) \rangle$ of the γ'(L1$_2$)-precipitates as a function of aging time. The 3-D APT results, through 4096 h, are compared to results from the monovacancy-mediated LKMC$_1$ simulations through 800 h. The horizontal dashed line in Fig. 5a indicates that the equilibrium volume fraction, $\phi_{\gamma'}^{eq}$ = 13.48%, calculated using Dupin et al.'s phase diagram [70], is consistent with the $\phi_{\gamma'}(t=4096h)$ value measured by 3-D APT, 12.59 ± 1.40%. The temporal exponent of $N_v(t)$, $q$, is continuously evolving as the alloy proceeds from the growth and coarsening regime, III, into the quasi-stationary coarsening regime, IV, and it is anticipated to approach -1 at long aging times, Eq. (2), as described by the N(umerical)-model [104]. The temporal exponent $q$ is calculated from the 3-D APT $N_v(t)$ data for the five longest aging times (64, 256, 1024, 2607, and 4096 h), yielding $q$ = -0.75 ± 0.03. Fig. 5b indicates the four regimes of phase separation: (I) quasi-stationary γ'(L1$_2$)-precipitate nucleation; (II) concomitant γ'(L1$_2$)-precipitate nucleation and growth; (III) concurrent growth and coarsening of γ'(L1$_2$)-precipitates via the coagulation and coalescence mechanism; and (IV) quasi-stationary coarsening of γ'(L1$_2$)-precipitates. $N_v(t)$ is continuously increasing in regimes I and II, and continuously decreasing in regions III and IV, which implies that nucleation of γ'(L1$_2$)-precipitates has ceased by the end of regime II.

The $\langle R(t) \rangle$ values in Fig. 5c are determined using the *spherical volume equivalent radius method* [105, 106], whereby the volume equivalent radius, $R$, of each γ'(L1$_2$)-precipitate is calculated using the total number of atoms in a γ'(L1$_2$)-precipitate given by the *so-called cluster analysis algorithm in IVAS 3.6* [107, 108]:

$$R = \left( \frac{3n}{4\pi\rho\zeta} \right)^{1/3} \tag{15}$$

where $n$ is the total number of atoms enclosed within an isoconcentration surface; $\rho$ is the atomic number density of the γ'(L1$_2$)-phase, 86.22 atoms nm$^{-3}$; and $\zeta$ is the detection efficiency of the 2-D microchannel plate (MCP), 50%, in the LEAP4000X Si tomograph. The quantity $R$ assumes a spherical morphology for all γ'(L1$_2$)-precipitates; it can be used to compare the dimensions of spheroidal γ'(L1$_2$)-precipitates at aging times of <100 h with cuboidal γ'(L1$_2$)-precipitates that appear at aging times >100 h, including the transition from spheroids-to-cuboids [105, 106]. The 3-D APT data points within regime IV in Fig. 5c (64, 256, 1024, 2607, and 4096 h of aging) are fitted to Eq. (1) using a nonlinear multivariate regression analysis [58], which yields a temporal exponent of $1/p = 0.34 \pm 0.02$ for $\langle R(t) \rangle$, and a rate constant, $K = 2.09 \pm 0.10 \times 10^{-31}$ m$^3$ s$^{-1}$. This experimental value of $1/p$ is in excellent agreement with the LSW value of 1/3. Once again we utilized a nonlinear multivariate regression analysis [58] as opposed to assuming a temporal exponent of 1/3 and then plotting $\langle R(t) \rangle^3$ versus time and calculating the coefficients of determination for this exponent. Appendix A explicates the reasons for not using the latter approach, which linearizes Eq. (1) and yields coefficients of determination that are similar for values of 1/p ranging from two to four, demonstrating that it is an insensitive methodology and therefore should not be employed.

The numerical values of the 3-D APT results are listed in Table 4. The monovacancy-mediated LKMC$_1$ simulation results displayed in Fig. 5 are in generally good agreement with the 3-D APT results, but slightly underestimate $\phi_{\gamma'}(t)$ between 4 and 256 h of aging.

### 3.3. Temporal evolution of the compositions of the gamma(f.c.c.)- and gamma-prime(L1$_2$ structure)-phases

$C_{Al}^{\gamma}(t)$ and $C_{Al}^{\gamma'}(t)$ are displayed in Fig. 6a-b, respectively, from the 3-D APT derived proximity histograms [84, 85] and the monovacancy-mediated LKMC$_1$ simulations. The numerical values from the 3-D APT results are listed in Table 5. The γ(f.c.c.)-phase experiences an initial state, where $C_{Al}^{\gamma}(t)$ remains approximately constant for aging times <1 h. At aging times >1 h, $C_{Al}^{\gamma}(t)$ and $C_{Al}^{\gamma'}(t)$ decrease continuously while the thermodynamic driving force (Al supersaturation) is concomitantly decreasing. The monovacancy-mediated LKMC$_1$ simulation results are in very good agreement with the experimental data for $C_{Al}^{\gamma'}(t)$; they, however, slightly underestimate $C_{Al}^{\gamma}(t)$. The equilibrium concentrations of Al in the γ(f.c.c.)- and γ'(L1$_2$)-phases, 11.14 ± 0.32 and 23.14 ± 0.47 at.%, respectively, are calculated by fitting the 3-D APT data, Fig. 6 to Eq. (3), employing a nonlinear multivariate regression analysis [58]. Employing our GCMC simulations [22] to calculate the equilibrium concentrations of Al in the γ(f.c.c.)- and γ'(L1$_2$)-phases yields values of 10.54 ± 0.15 and 23.24 ± 0.21 at.%, respectively. The data in Fig. 6a-b are also plotted on the partial Ni-Al phase diagram, Fig. 2, where the temporal compositional trajectories of both phases are indicated by horizontal bold black arrows going toward the left as the degree of phase separation increases with the Al supersaturation concomitantly decreasing.

In the [γ(f.c.c.) plus γ'(L1$_2$)] phase-field, Fig. 2, the temporal compositional trajectory of the γ(f.c.c.)-phase evolves toward the solvus curve separating the γ(f.c.c.)-phase-field from the [γ(f.c.c.) plus γ'(L1$_2$) phase-field], and at 4096 h it terminates on the overlapping solvus curves determined by Ma and Ardell [71] and Dupin et al. [70]. The initial composition of the γ'(L1$_2$)- phase ($C_{Al}^{\gamma'} = 33.97 \pm 8.49$ at.%) is very far from the [γ(f.c.c.) plus γ'(L1$_2$)]/ γ'(L1$_2$) solvus curves at 0.08 h of aging, which is initially in the indicated [γ'(L1$_2$) plus NiAl(B2)] phase-field. At 4096 h, the composition of the γ'(L1$_2$)-phase approaches the solvus curve due to Saunders [68], Section 4.3.

The 3-D APT and monovacancy-mediated LKMC$_1$ simulation results for the temporal evolutions of $\Delta C_{Al}^{\gamma}(t)$ and $\Delta C_{Al}^{\gamma'}(t)$ are displayed in Fig. 6c-d, respectively, and their temporal exponents are also calculated using a nonlinear multivariate regression analysis technique [58]. The monovacancy-mediated LKMC$_1$ simulation results agree with the 3-D APT results for $\Delta C_{Al}^{\gamma}(t > 0.5\,h)$, Fig. 6c. In contrast, for aging times <0.5 h, the monovacancy-mediated LKMC$_1$ results exhibit a decrease in $\Delta C_{Al}^{\gamma}(t < 0.5\,h)$, while the 3-D APT results remain approximately constant, within experimental error, indicating that 3-D APT does not, for the LEAP 4000X Si, have sufficient detection efficiency to measure a decrease in $\Delta C_{Al}^{\gamma}(t)$ for very small γ'(L1$_2$)-precipitates.

The 3-D APT results for $\Delta C_{Al}^{\gamma'}(t)$ decrease slightly more rapidly than the monovacancy-mediated LKMC$_1$ results, Fig. 6d. Eq. (3) is utilized to analyze the 3-D APT data, again employing a nonlinear multivariate regression analysis technique [58]. The 3-D APT temporal exponents for $\Delta C_{Al}^{\gamma}(t)$ and $\Delta C_{Al}^{\gamma'}(t)$ are both $1/r = 0.33 \pm 0.03$, which are in excellent agreement with the LSW value of -1/3. The resulting values of $\kappa^{\gamma}$ and $\kappa^{\gamma'}$ from the 3-D APT data for the γ(f.c.c.)-matrix and γ'(L1$_2$)-precipitates are 0.25 ± 0.01 and 0.68 ± 0.09 s$^{1/3}$, respectively. The corresponding $\kappa^{\gamma}$ and $\kappa^{\gamma'}$ values from the monovacancy-mediated LKMC$_1$ results are 0.18± 0.01 and 0.47± 0.03 s$^{1/3}$, respectively, which are in approximate agreement.

The partitioning coefficients $K_{Al}^{\gamma'/\gamma}(t)$ and $K_{Ni}^{\gamma'/\gamma}(t)$ are displayed in Fig. 7, where $K_i^{\gamma'/\gamma}(t)$ is defined as the ratio $C_i^{\gamma'}(t)$ to $C_i^{\gamma}(t)$, where the concentrations are in atomic fraction (at. fr.). The horizontal dashed line corresponds to $K_i^{\gamma'/\gamma}(t) = 1$, which indicates a complete absence of preferential partitioning behavior. Fig. 7 demonstrates that for aging at 823 K (550 °C) Al partitions to the γ'(L1$_2$)-phase and Ni to the γ(f.c.c.)-phase. Both $K_i^{\gamma'/\gamma}(t)$ ratios are slightly time dependent for aging times <0.5 h, after which $K_{Al}^{\gamma'/\gamma}(t)$ approaches 2.1 and $K_{Ni}^{\gamma'/\gamma}(t)$ approaches 0.9.

3.4. *Temporal evolution of the γ'(L1$_2$)-precipitate size distributions (PSDs)*

The PSDs for aging times ranging from 0.08 to 2607 h are displayed in Fig. 8a, and the PSD for 4096 h is displayed in Fig. 8b. The total number of γ'(L1$_2$)-precipitates for each aging time utilized to construct the PSDs is smaller than the *effective number of γ'(L1$_2$)-precipitates, $N_{ppt}$*, Table 4, because only γ'(L1$_2$)-precipitates that are fully enclosed within the 3-D APT reconstruction volume, which varies from 4.2 x $10^5$ to 9.8 x $10^6$ nm$^3$ per data set, are utilized to generate the PSDs [30]. *The monovacancy-mediated LKMC$_1$ results are not used to generate PSDs because the analyzed volume is too small to yield satisfactory statistics.* The PSDs are plotted as a function of the normalized quantity ($R/\langle R(t)\rangle$) on the abscissa axis, so that PSDs for different aging times are directly comparable, even though $\langle R(t)\rangle$ is different for each aging time, Fig. 5c. A bin size of 0.2 nm is employed for $R/\langle R(t)\rangle$. For the ordinate axis, the number of γ'(L1$_2$)-precipitates in each bin is divided by two quantities: (1) the total number of enclosed γ'(L1$_2$)-precipitates (indicated in Fig. 8 for each aging time), and (2) the bin size (0.2 nm). The resulting normalized and dimensionless quantity, $g(R/\langle R(t)\rangle,t)$ [21], is plotted on the ordinate axis. In this way the area under each histogram is normalized to unity, and the PSDs for different aging times are directly comparable. The PSDs for 256, 1024, 2607, and 4096 h are compared to four model PSDs: the LSW PSD, a modified LSW model PSD [54], the BW PSD [55], and the AV PSD [56].

The LSW PSD has the following analytical form:

$$\frac{4}{9}\left(\frac{R}{\langle R(t)\rangle}\right)^2 \left(\frac{3}{3+R/\langle R(t)\rangle}\right)^{7/3} \left(\frac{1.5}{1.5-R/\langle R(t)\rangle}\right)^{11/3} \exp\left(\frac{R/\langle R(t)\rangle}{R/\langle R(t)\rangle-1.5}\right), \quad 0 < \frac{R}{\langle R(t)\rangle} < 1.5$$

$$0, \quad \frac{R}{\langle R(t)\rangle} \geq 1.5$$

(16)

Unlike the LSW PSD, which is independent of $\phi_{\gamma'}(t)$, the modified LSW, BW, and AV PSDs are calculated using $\phi_{\gamma'}^{eq}$ = 13.48% [70]. The modified LSW PSD comes from a numerical solution of this model [54]. The BW PSD comes from an analytical solution of the BW model [55]. The AV PSD comes from a simulation, which uses their numerical model to study a system of 80,000 γ'(L1$_2$)-precipitates, whose radii initially follow an LSW distribution, which coarsen until 100 remain, at which point the system has achieved a stationary-condition [56].

### 3.5. Temporal evolution of the fraction of γ'(L1$_2$)-precipitates interconnected by necks, f(t), and the minimum edge-to-edge distances, $\langle \lambda_{edge-edge}(t)\rangle$, between neighboring γ'(L1$_2$)-precipitates

Fig. 9a displays $f(t)$ and Fig. 9b displays $\langle \lambda_{edge-edge}(t)\rangle$, for the 3-D APT experiments and monovacancy-mediated LKMC$_1$ simulations; the 3-D APT derived values are included in

Table 4. An increase in $f(t)$ corresponds to a decrease in $\langle \lambda_{edge-edge}(t) \rangle$ and vice versa, as γ'(L1$_2$)-precipitates need to be close to one another for coalescence and coagulation to occur [61]. The quantity $\langle \lambda_{edge-edge}(t) \rangle$ can be measured directly from the 3-D APT data set using the approach developed by Karnesky et al. [109]. Alternatively, Nembach's analytical equation for calculating $\langle \lambda_{edge-edge}(t) \rangle$ between γ'(L1$_2$)-precipitates in a regular array is given by [110]:

$$\langle \lambda_{edge-edge}(t) \rangle = \sqrt{\frac{2\pi \langle R(t)^3 \rangle}{3\phi_{\gamma'}(t)\langle R(t) \rangle}} - 2\langle R(t) \rangle . \quad (17)$$

The first-term on the right-hand side of Eq. (17) is the mean center-to-center distance between γ'(L1$_2$)-precipitates and subtracting $2\langle R(t) \rangle$ from it yields $\langle \lambda_{edge-edge}(t) \rangle$; note that $\langle R(t)^3 \rangle$ is not equal to $\langle R(t) \rangle^3$. The two $R(t)$ related terms in the numerator and denominator of Eq. (17) do not cancel as $\langle R(t)^3 \rangle$ is greater than $\langle R(t) \rangle^3$ for all aging times due to the asymmetry of the PSDs, Fig. 8. Based on Nembach's approach [110], $\langle \lambda_{edge-edge}(t) \rangle$ is large for the shortest aging time; for 0.08 h, $\langle \lambda_{edge-edge}(t) \rangle$ = 272.71 ± 15.19 nm, as $N_v(t)$ is initially small ($N_v(t)$ = 7.88 ± 1.97 x 10$^{21}$ m$^{-3}$). The quantity $\langle \lambda_{edge-edge}(t) \rangle$ then decreases with increasing aging time and finally it increases because $N_v(t)$ decreases at long aging times. The value of $\langle \lambda_{edge-edge}(t) \rangle$ is approximately a constant between 1 and 64 h [10 < $\langle \lambda_{edge-edge}(t) \rangle$ < 15 nm], and eventually it increases to a final value of 37.38 ± 4.15 nm at 4096 h.

    Karnesky et al.'s approach calculates distances between γ'(L1$_2$)-precipitates based on the X-Y-Z location of each γ'(L1$_2$)-precipitate's center and its $R$ value. It yields smaller values of $\langle \lambda_{edge-edge}(t) \rangle$ than does Nembach's model.

    3-D APT results display a maximum in $f(t)$ at 1 h (39.68 ± 2.81%), while the monovacancy-mediated LKMC$_1$ simulation results display a maximum in $f(t)$ at 2 h (34.47%). The 3-D APT results for $f(t)$ at 2 h (39.53 ± 2.80%) are, however, almost identical to the 3-D APT results for $f(t)$ at 1 h, with a difference of only 0.4%, and hence the monovacancy-mediated LKMC$_1$ simulations agree with 3-D APT results for this quantity.

    In summary, the monovacancy-mediated LKMC$_1$ results are in good agreement with the 3-D APT data for $\langle \lambda_{edge-edge}(t) \rangle$ calculated using Nembach's approach for aging times ≥ 0.25 h,

and slightly overestimate the 3-D APT data for $\langle \lambda_{edge-edge}(t) \rangle$ calculated using Karnesky et al.'s approach, Fig. 9b. Monovacancy-mediated LKMC$_1$ results for $f(t)$ underestimate somewhat the experimental 3-D APT values at aging times greater than about 0.25 h. *In LKMC$_1$, due to the size of the simulation box, the periodic boundary conditions introduce artificially certain long-range ordering of the γ'(L1$_2$)-precipitates, which can account for this underestimation.*

### 3.6. Temporal evolution of the interfacial compositional width, $\delta(t)$, between the gamma (f.c.c.)- and gamma-prime(L1$_2$ structure)-phases

Fig. 10a displays the concentration profiles for Ni and Al on both sides of the γ(f.c.c.)/γ'(L1$_2$) interface for 1, 4, 256, and 4096 h as measured by 3-D APT. A positive distance is defined as into the γ'(L1$_2$)-precipitates, while a negative distance is into the γ(f.c.c.)-matrix. At 1 h the γ(f.c.c.)/γ'(L1$_2$) (100) interfacial width is initially somewhat diffuse [$\delta(t)$=1.71±0.08 nm], with large values of $\Delta C_{Al}^{\gamma}$ and $\Delta C_{Al}^{\gamma'}$, Fig. 6c-d. An atomically sharp interface is given by a step function, which is an inadequate description of the interfaces we observe. With increasing aging time, the compositions of each phase are quasi-stationary and the γ(f.c.c.)/γ'(L1$_2$) (100) interface width becomes somewhat sharper [$\delta(t=4096\,h)$=1.12±0.12 nm], but they are never equal to zero. Moreover, while the concentration profiles for each phase change dramatically from 1 to 4 h and from 4 to 256 h, the changes from 256 to 4096 h are small. This is also demonstrated by the asymptotic approaches of $C_{Al}^{\gamma}$ and $C_{Al}^{\gamma'}$ toward their equilibrium concentrations, Fig. 6a-b. Fig. 10b displays the concentration profiles for Ni and Al for 0.25, 4, 16, and 256 h as determined from monovacancy-mediated LKMC$_1$ simulations. Similarly, to the 3-D APT concentration profiles, the γ(f.c.c.)/γ'(L1$_2$) interface is initially somewhat diffuse, but it becomes sharper with increasing aging time as demonstrated by the time dependence of $\delta(t)$ for the (100) interface, but it is not equal to zero.

Fig. 11a displays $\delta(t)$ for the {100} interface, obtained from the 3-D APT data, and the monovacancy-mediated LKMC$_1$ simulations as a function of aging time. The values for $\delta(t)$ for the {100} interfaces are measured by fitting the concentration profiles across the γ(f.c.c.)/γ'(L1$_2$) interface at each aging time to a spline curve, with the plateaus of the concentration profiles matching the *far-field (ff)* concentrations determined for the γ(f.c.c.)- and γ'(L1$_2$)-phases [111]. A spline fit produces a piecewise-defined function, where the distance between each data point in a data set is fit to a cubic interpolation [112]. The horizontal distance between the 10$^{th}$ and 90$^{th}$ percentiles of the concentration difference between the γ(f.c.c.)- and γ'(L1$_2$)-phases is defined to be $\delta(t)$ [111], which is the definition employed commonly in phase-field modeling. The spatial resolution of each concentration profile is 0.1 nm, which is less than the lattice parameter of Ni$_3$Al, 0.359 nm [67]. Note that *all* of the γ'(L1$_2$)-precipitates have a morphology with {100} facets starting at 2607 h, Fig. 4e, with the value of $\delta(t)$ decreasing with increasing aging time, as $t^{-0.08\pm0.01}$, Fig. 11a. Fig. 11b combines the $\langle R(t) \rangle$ data from Fig. 5c with the $\delta(t)$ data from Fig. 11a to display $\delta(t)$ as a function of $\langle R(t) \rangle$ for each aging time. The value of $\delta(t)$ is decreasing continuously as $\langle R(t) \rangle$ is increasing, varying as $\langle R(t) \rangle^{-0.47\pm0.03}$. The value of $\delta(t)$ is asymptotically approaching a constant value at the longest aging time, 4096 h, which is definitely not equal to zero. We measured $\delta(t)$ at long aging times (>1000 h) for the {100}-type interface planes because the γ'(L1$_2$)-precipitates are *mainly* cuboidal at 1024 h and are additionally aligning and rafting along <100>-type directions: TEM micrograph, Fig. 4g. The significance of the results presented in Fig. 11 concerning the time dependency of $\delta(t)$ and the dependency of $\delta(t)$ on $\langle R(t) \rangle$ are discussed in Section 4.6.

*3.7. Calculation of four different diffusivities based on monovacancy-mediated lattice-kinetic Monte Carlo simulations*

*3.7.1. The monovacancy's trajectories in four distinct regions*

Fig 12, taken from LKMC$_1$ simulations (Section 2.5 and Table 3), displays three stages of precipitate evolution (right-hand column) together with successive locations of the monovacancy at work in LKMC$_1$, during this process (left-hand column). In the left-hand column, the γ(f.c.c.)-matrix appears as the yellow background, and successive positions of the monovacancy are indicated by red-dots. In the right-hand column, Ni and Al atoms are only displayed in the γ'(L1$_2$)-precipitate, respectively in green and red.

In Fig.12, The two first rows (a, b and c, d) deal with the coagulation and coalescence of two γ'(L1$_2$)-precipitates at, respectively, 1 and 4 hours, at 823 K (550 ºC) in a 128 x 128 x 128 lattice sites super-cell. The third row (Fig. 12 e, f) displays the growth of a single precipitate at 400 h, at the same temperature, in a 256 x 256 x 256 lattice sites super-cell.

The partial ordering of Ni and Al atoms within the γ'(L1$_2$)-precipitates and the neck region are clearly evident (Figs. 12b and 12d). At 1 and 4 h the γ'(L1$_2$)-precipitates do not yet exhibit {100}-type facets, while these facets are clearly evident at 400 h, consistent with the 3-D APT reconstructions, Fig. 4. Additionally, the interfacial regions between the γ(f.c.c.)-matrix and γ'(L1$_2$)-precipitates (Figs. 12b, 12d, and 12f) are all qualitatively diffuse: less so at 1 and 4 h than at 400 h. Table 6 presents the normalized times, $t/t_{matrix}$, the monovacancy spends in the four distinct regions: (1) the γ(f.c.c.)-matrix; (2) the γ'(L1$_2$)-precipitates; (3) the partially ordered neck connecting two γ'(L1$_2$)-precipitates; and (4) the super-cell for 1 and 4 h, Fig. 12, normalized to the volume of the cell. The monovacancy is 3.12 to 3.25 times more likely to be found inside the γ'(L1$_2$)-precipitates than in the γ(f.c.c.)-matrix, and 3.58 to 3.85 times more likely to be found inside the neck region connecting the two γ'(L1$_2$)-precipitates than in the γ(f.c.c.)-matrix at 1 and 4 h. Note that the normalized times decrease with increasing aging time, Table 6.

Figs. 12b, 12d, and 12f display the positions of the Ni (green) and Al (red) atoms for γ'(L1$_2$)-precipitates at 1, 4 and 400 h, respectively. Figs. 12b, 12d and 12f complement and supplement Figs. 12a, 12c and 12e, with the former showing the positions of the Ni and Al atoms, and the latter displaying the positions of the single monovacancy as a function of time. The partial ordering of the Ni and Al atoms within the γ'(L1$_2$)-precipitates and the neck region are clearly evident: Figs. 12b and 12d. At 1 and 4 h the γ'(L1$_2$)-precipitates do not yet exhibit {100}-type facets, while these facets are clearly evident at 400 h, consistent with the 3-D APT reconstructions in Fig. 4. Additionally, the interfacial regions between the γ(f.c.c.)-matrix and γ'(L1$_2$)-precipitates, Figs. 12b, 12d, and 12f, are all qualitatively diffuse: less so at 1 and 4 h than at 400 h.

### 3.7.2. Calculations of the four different diffusivities, $D_i^{region}$

Table 6 presents four different calculated diffusivities, $D_i^{region}$, based on the following standard equation [90, 100], Eq. 18, where $D_i^{region}$ includes the correlation factor for a monovacancy diffusion mechanism. The correlation factor for random walk in an f.c.c. crystal structure is 0.78145 for a tracer diffusion experiment [113]. The binary alloy studied herein has strong correlations between diffusional fluxes, which are less than 0.78145. Bocquet utilized a novel jump frequency model, taking into account the full range of solute–vacancy interactions up to third NN sites, and determined in the dilute Ni-Cr system, 0.41304 [114]. Ramunni studied diffusion coefficients of solute atoms in the Ni–Al and Al–U systems employing the classical molecular statics technique (CMST), where migrating solute atoms interchange with 1$^{st}$ NN vacancies in the temperature range 700 to 1700 K, where there is experimental data. Ramunni determined the correlation factor for solute atoms using the five-frequency model. The relevant value for dilute Ni-Al alloys is 0.611 at 823 K [115], which is an upper bound of this value and therefore the diffusivities calculated using this value are also upper bounds. There isn't prior research data on correlated diffusion in either γ'(L1$_2$) or the ordered interconnected neck regions. In this study, the diffusivity is calculated using the following expression, which is at a specified temperature, for the root-mean-square diffusion distance:

$$\sqrt{4D_i^{region}t} = \sqrt{\eta_{correlation}^{region,Ni-Al} n_j \alpha^2} \; ; \qquad (18)$$

where $n_j$ is the total number of jumps made by the monovacancy for all its exchanges with an atom of type $i$ ($i$ = Ni or Al) in a time $t$ in a specified region; and $\alpha$ is the jump distance, $a_0/\sqrt{2}$

, where $a_0$ is the lattice parameter of Ni$_3$Al(L1$_2$). Time, $t$, in Eq. (18), is a rescaled monovacancy-mediated LKMC$_1$ time, which has been adjusted for the difference in the monovacancy concentrations between the LKMC$_1$ cell and a bulk material, utilizing Eq. (14) (Section 2.5). The quantity n$_j$ is multiplied by a correlation factor, $\eta_{correlation}^{region,Ni-Al}$, which reduces the value of each $D_i^{region}$ caused by correlations among diffusion fluxes. We use 0.611 from reference [115] for all regions due to the lack of the availability of diffusion data with correlation. We note that this value only applies to the disordered matrix and it should be smaller than 0.611 in both the γ'(L1$_2$)-precipitates and the interconnected ordered neck regions. The diffusivities in these two regions can be estimated from the vacancy formation energy and the migration energy using the five-frequency model with the consideration of the local atomic environment in a concentrated alloy system [116].

### 3.7.3. Four distinct aluminum diffusivities

For exchanges between Al atoms and the monovacancy the calculated diffusivity in the γ(f.c.c.)-matrix, $D_{Al}^{\gamma-matrix}$, has the largest value, (2.22 to 1.62) x 10$^{-21}$ m$^2$s$^{-1}$, of the four calculated diffusivities listed in Table 6. The calculated value of the Al diffusivity in the γ'(L1$_2$)-precipitates, $D_{Al}^{\gamma'-precipitates}$, ranges from (1.22 to 0.71) x 10$^{-21}$ m$^2$s$^{-1}$, which is 43-56% smaller than $D_{Al}^{\gamma-matrix}$. The calculated value of the Al diffusivity in the neck region, $D_{Al}^{necks}$, is (1.11 to 1.15) x 10$^{-21}$ m$^2$ s$^{-1}$, which is 46-48% smaller than $D_{Al}^{\gamma-matrix}$. The value of $D_{Al}^{necks}$ is measured only at 1 and 4 h, because too few necks between γ'(L1$_2$)-precipitates are detected at 400 h to obtain good statistics.

### 3.7.4. Four distinct nickel diffusivities

For exchanges between Ni atoms and the monovacancy, all the calculated diffusivities listed in Table 6 are smaller than those for Al exchanges with the monovacancy in the corresponding regions by an order of magnitude. This was anticipated because Al is a significantly faster diffuser in Ni than is Ni in Ni. $D_{Ni}^{\gamma-matrix}$ has the largest value, (1.11 to 0.99) x 10$^{-22}$ m$^2$ s$^{-1}$, of the four calculated diffusivities listed in Table 6. $D_{Ni}^{\gamma'-precipitates}$ ranges from (0.57 to 0.33) x 10$^{-22}$ m$^2$ s$^{-1}$, which is 55-69% smaller than $D_{Ni}^{\gamma-matrix}$. Additionally, the calculated value of $D_{Ni}^{necks}$ is (0.49 to 0.53) x 10$^{-22}$ m$^2$ s$^{-1}$, which is 51-56% smaller than $D_{Ni}^{\gamma-matrix}$. As is the Al case, $D_{Ni}^{necks}$ is measured only at 1 and 4 h because too few necks between γ'(L1$_2$)-precipitates are detected at 400 h to obtain good statistics. While $D_{Al}^{\gamma'-precipitates}$ is consistently greater than $D_{Al}^{necks}$, the values of $D_{Ni}^{\gamma'-precipitates}$ and $D_{Ni}^{necks}$ are approximately equal to one another.

### 3.7.5. The diffusivities of Ni and Al in the supercell, $D_i^{supercell}$

The values of $D_i^{supercell}$ in Table 6 are obtained for 1, 4 and 400 h by taking a weighted average of the calculated $D_i^{region}$ values, based on the fraction of the total volume of the cell that each region occupies. The volume occupied by the neck region is defined as the volume where the short-range order parameter is 80% of that of the γ'(L1$_2$)-phase. For both $D_{Ni}^{supercell}$ and $D_{Al}^{supercell}$, their weighted averages are 38-64% greater than $D_i^{\gamma'-precipitates}$, 5-36% smaller than $D_i^{\gamma-matrix}$, and 47-56% greater than $D_i^{necks}$.

## 4. Discussion

### 4.1. Temporal evolution of the Vickers microhardness values

The Vickers microhardness of Ni-12.5 Al at.% aged at 823 K (550 °C) is directly proportional to $\phi_{\gamma'}(t)$ at early aging times and inversely proportional to $\langle \lambda_{edge-edge}(t) \rangle$ at long aging times. The Vickers microhardness values increase continuously through 1024 h of aging after a small decrease at 0.25 h, Fig. 3. The peak microhardness achieved is 247 ± 11 at 1024 h, with $\phi_{\gamma'}(t)$ increasing from 2.87 ± 0.05 at 1 h to 11.96 ± 0.82% at 1024 h. The continuous increase in $\phi_{\gamma'}(t)$ between 1 and 1024 h, Fig. 5a, explains the continuous increase in the value of the Vickers microhardness during this same time frame. The slight decrease in Vickers microhardness at 2607 h to 233 ± 5 may be accounted for by the alignment of γ'(L1$_2$)-precipitates along a <100>-type direction (rafting), which begins to occur at 1024 h, Fig. 4d and g. Additionally, the value of the Vickers microhardness is inversely proportional to the value of $\langle \lambda_{edge-edge}(t) \rangle$ for the Orowan dislocation-looping strengthening mechanism, which is consistent with this strengthening mechanism at room temperature, where $\phi_{\gamma'}(t)$ is continuously increasing. The value of $\langle \lambda_{edge-edge}(t) \rangle$ is in the range 10-15 nm between 1 and 64 h of aging, Fig. 9b, before increasing to 26.80 ± 2.18 nm at 2607 h of aging.

### 4.2. Temporal evolution of the gamma-prime (L1$_2$-structure) precipitate volume fraction, $\phi_{\gamma'}(t)$, number density, $N_v(t)$, and mean radius, $\langle R(t) \rangle$

#### 4.2.1. Temporal evolution of the volume fraction of gamma-prime(L1$_2$-structure)-precipitates, $\phi_{\gamma'}(t)$

Our results for $\phi_{\gamma'}(t)$ indicate that after 4096 h of aging at 823 K (550 °C) it has achieved its equilibrium value, $\phi_{\gamma'}^{eq}$, at this temperature. Different values of $\phi_{\gamma'}^{eq}$ are calculated using the lever rule from the solvus curves, Fig. 2. The $\phi_{\gamma'}^{eq}$ values of the γ'(L1$_2$)-phase at 823 K (550 °C) are 18.33, 13.48, 15.02, and 16.75 vol.% using the Saunders, Dupin et al., Ma and Ardell, and GCMC solvus curves, respectively [22, 68-71]. At the longest aging time, $\phi_{\gamma'}(t = 4096\ h)$ is 12.59 ± 1.40%, which is 7% smaller than the Dupin et al. value of $\phi_{\gamma'}^{eq}$, and 31% smaller than the Saunders value of $\phi_{\gamma'}^{eq}$. The $\phi_{\gamma'}(t = 4096h)$ is, however, equal to the Dupin et al. value of $\phi_{\gamma'}^{eq}$ within error, implying that this alloy has most likely achieved its equilibrium volume fraction after 4096 h of aging at 823 K (550 °C). Therefore, the Dupin et al. value is in the best agreement with the 3-D APT data, Fig. 5a. The monovacancy-mediated LKMC$_1$ simulation result for 800 h has not achieved an equilibrium value of $\phi_{\gamma'}$, but an extrapolation of the monovacancy-mediated LKMC$_1$ trajectory, Fig. 5a, indicates agreement with Dupin et al.'s results as well. A summary of the $\phi_{\gamma'}^{eq}$ values is presented in Table 7.

*4.2.2. Temporal evolution of the number density of gamma-prime(L1$_2$ structure)-precipitates, $N_v(t)$*

Our results for the temporal evolution of $N_v(t)$ indicate that four regimes, from nucleation through coarsening, are observed for Ni-12.5 Al at.% aged at 823 K (550 °C) for 0.08-4096 h. We focus mainly on discussing regimes I and IV.

*4.2.2.1.   Regime I: quasi-stationary nucleation of γ'(L1$_2$)-precipitates*

The quasi-stationary γ'(L1$_2$)-precipitate nucleation regime, indicated by the initial rapid increase in $N_v(t)$, regime I in Fig. 5b, can be modeled using classical nucleation theory (CNT) [30, 104, 117-127], under the assumption that CNT obtains, which is a strong assumption. CNT states that the supersaturation of an element in a binary system depends on a Helmholtz free energy expression, which has both a chemical energy component, $\Delta F_{ch}$ (which is negative), and an elastic strain energy component, $\Delta F_{el}$ (which is always positive).

For the strain energy to depend solely on $\langle R(t) \rangle$, one must assume that there are no precipitate-precipitate elastic interactions; that is, the solid-solution is elastically isotropic and the γ'(L1$_2$)-precipitates have the same elastic moduli as the γ (f.c.c.)-matrix. The fact that rafting is observed during the later stages implies that this is not the case. Such effects are, however, negligible during the early stages of nucleation because $\langle \lambda_{edge-edge}(t) \rangle$ is large, Fig. 9b, and hence this assumption is reasonable for regime I. The Helmholtz free energy expression also contains an interfacial free energy term, $\sigma^{\gamma/\gamma'}$, which is always positive. The net reversible work required for the formation of a spherical nucleus, $W_R$, as a function of a nucleus's radius, $R$, is given by:

$$W_R = (\Delta F_{ch} + \Delta F_{el})\frac{4\pi}{3}R^3 + 4\pi R^2 \sigma^{\gamma/\gamma'} \qquad (19)$$

The critical radius, $R^*$, for nucleation is:

$$R^* = -\frac{2\sigma^{\gamma/\gamma'}}{\Delta F_{ch} + \Delta F_{el}} \qquad (20)$$

and the critical net reversible work required for the formation of a critical spherical nucleus, $W_R^*$, is:

$$W_R^* = \frac{16\pi}{3}\frac{(\sigma^{\gamma/\gamma'})^3}{(\Delta F_{ch} + \Delta F_{el})^2}. \qquad (21)$$

Alternatively, $R^*$ can be derived using capillarity theory by considering the Hessian of the Gibbs free energy of all the terms, including off-diagonal terms [128].

The chemical formation free energies are calculated based on a first-principles dilute solid-solution model, $Ni_xAl \rightarrow Ni_3Al + Ni_{x-3}$, where $x = 31$. The equation for $\Delta F_{ch}$ is given by:

$$\begin{aligned}\Delta F_{ch} &= \Delta F(Ni_3Al) + (x-3)\Delta F(Ni) - \Delta F(Ni_xAl) \\ &= [\Delta H(Ni_3Al) - \Delta H(Ni_xAl)] - T[\Delta S(Ni_3Al) - \Delta S(Ni_xAl)]\end{aligned} \qquad (22)$$

where $\Delta H$ and $\Delta S$ are the enthalpy and entropy changes of formation, respectively. We take $\Delta H$ to be equal to the internal energy change of formation because in the solid-state the pressure-volume term in $\Delta H$ is negligible compared to the internal energy change. The entropy changes of formation are based on the vibrational entropies of the ordered $Ni_3Al$ and $Ni_xAl$ cells. The calculated value of $\Delta F_{ch}$ of Ni3Al(L1$_2$), using Eq. (22), is -7.86 x 10$^7$ kJ m$^{-3}$ at 823 K (550 °C). An alternative value can be calculated using a classical thermodynamic methodology [104, 129], which yields $\Delta F_{ch}$ = -6.70 x 10$^6$ kJ m$^{-3}$, which is within the experimentally measured range of values of -4.93 x 10$^6$ to -6.97 x 10$^6$ kJ m$^{-3}$ in the temperature range 820 K (547 °C) to 920 K (647 °C) [130].

The $\Delta F_{el}$ value is small in magnitude compared to the thermodynamic driving force during nucleation, and is calculated using the approach outlined by Booth-Morrison et al. [30, 131]:

$$\Delta F_{el} = \frac{2\mu^{\gamma} B^{\gamma'} \left(V_a^{\gamma'} - V_a^{\gamma}\right)^2}{\left(3B^{\gamma'} + 4\mu^{\gamma}\right)(V_a^{\gamma'})^2} \quad . \tag{23}$$

where $\mu^{\gamma}$ is the shear modulus of the γ(f.c.c.)-matrix phase; $B^{\gamma'}$ is the bulk modulus of the γ'(L1$_2$)-phase; and $V_a^{\gamma}$ and $V_a^{\gamma'}$ are the atomic volumes of the γ(f.c.c.)-matrix and γ'(L1$_2$)-precipitates, respectively. The value of $\mu^{\gamma}$ is 100 GPa [132], and the value of $B^{\gamma'}$ is 175 GPa [133]. The values of $V_a^{\gamma}$ and $V_a^{\gamma'}$ are calculated based on the lattice parameters of the γ(f.c.c.)-matrix and γ'(L1$_2$)-precipitates, 3.58 Å and 3.59 Å, respectively [67]; this small lattice parameter mismatch, 0.01 Å, implies that the γ'(L1$_2$)-precipitates are coherent with the γ(f.c.c.)-matrix. This approach yields $\Delta F_{el}$ = 4.96 x 10$^6$ kJ m$^{-3}$ at 823 K (550 °), which is 6% of $\Delta F_{ch}$ as calculated employing Eq. (22), and 74% of $\Delta F_{ch}$ as calculated using a classical thermodynamic approach [104, 129]; the latter is dependent on the thermodynamic activities of Al and Ni. Because the thermodynamic activities of Al are unmeasured for Ni-12.5 Al at.% at 823 K (550 °), approximate estimates are used in the calculations. We, therefore, employ the value of $\Delta F_{ch}$ calculated using a first-principles approach, Eq. (22), as opposed to the classical thermodynamic approach. Booth-Morrison et al. found a similar relationship between $\Delta F_{ch}$, which was calculated using Saunders' database in Thermo-Calc [68, 69], and $\Delta F_{el}$ for a Ni-6.5 Al-9.5 Cr at.% alloy aged at 873 K; they calculated $\Delta F_{ch}$ to be -6.25 x 10$^7$ kJ m$^{-3}$ and $\Delta F_{el}$ to be 2.67 x 10$^6$ kJ m$^{-3}$, making $\Delta F_{el}$ 4% of $\Delta F_{ch}$ for this ternary Ni-Al-Cr alloy [30].

We then calculate the stationary nucleation current, $J_{CNT}^{st}$, the number of nuclei formed per unit volume per unit time (m$^{-3}$ s$^{-1}$) [22, 30], from:

$$J_{CNT}^{st} = Z\beta^* N \exp\left(-\frac{W_R(R^*)}{k_B T}\right) \tag{24}$$

where $Z$ is the Zeldovich factor, which accounts for the dissolution of supercritical clusters; $\beta^*$ is a kinetic coefficient describing the rate of condensation of single atoms on the critical nuclei; and $N$ is the total number of possible nucleation sites per unit volume. The value of $N$ is taken to be the volume density of lattice sites occupied by the precipitating solute element, Al [117, 134], *which yields an absolute upper bound to the classical homogeneous nucleation current. The quantities $Z$ and $\beta^*$ are given by:*

$$Z = \left( -\left( \frac{\partial W_{R,j}^*}{\partial n_i} \right) \left( \frac{1}{2\pi k_b T} \right) \right)^{1/2} \tag{25}$$

$$\beta^* = \frac{4\pi (R^*)^2 D_{solute}^{coarsening} C_0}{\langle a \rangle^4} \tag{26}$$

where $C_0$ is the initial concentration of the precipitating solute element. And $\langle a \rangle$ is the mean of the lattice parameters of the γ(f.c.c.)-matrix and γ'(L1$_2$)-precipitates, 0.3585 nm [67]. And $n_i$ is the average number of atoms in each precipitate. Using a value of 28.55 ± 1.61 mJ m$^{-2}$ for $\sigma^{\gamma/\gamma'}$ (calculated using the Dupin et al. thermodynamic database [70], Section 4.7), Eqs. (20) and (21) yields an $R^*$ value of 0.68 nm and a $W_R^*$ value of 7.19 x 10$^{-20}$ J. Alternatively, $R^*$ and $W_R^*$ are 0.72 nm and 8.29 x 10$^{-20}$ J, respectively, if a value of 29.94 ± 1.69 mJ m$^{-2}$ is used for $\sigma^{\gamma/\gamma'}$ (calculated using Saunders's thermodynamic database [68], Section 4.7). These values of $R^*$ are slightly smaller, by 0.07 to 0.11 and 0.12 to 0.16 nm, than the smallest values of $\langle R(t) \rangle$ measured by 3-D APT ($\langle R(t = 0.08\ h) \rangle = 0.79 \pm 0.20$ nm) and monovacancy-mediated LKMC$_1$ ($\langle R(t = 0.05\ h) \rangle = $ 0.84 nm), respectively, which are physically reasonable results. Using these values and $D_{experiment}^{coarsening}$ = 2.51 ± 0.14 x 10$^{-21}$ m$^2$ s$^{-1}$ (Section 4.8.2), $J_{CNT}^{st}$ is 1.34 x 10$^{22}$ m$^{-3}$ s$^{-1}$ (for a $\sigma^{\gamma/\gamma'}$ value of 28.55 ± 1.61 mJ m$^{-2}$). Alternatively, it is 5.20 x 10$^{21}$ m$^{-3}$ s$^{-1}$ for $\sigma^{\gamma/\gamma'}$ = 29.94 ± 1.69 mJ m$^{-2}$. This value is 66 times greater than the 3-D APT value, $J_{APT}^{st}$, of 2.03 x 10$^{20}$ m$^{-3}$ s$^{-1}$ and 57 times greater than the monovacancy-mediated LKMC$_1$ value, $J_{LKMC}^{st}$, of 2.37 x 10$^{20}$ m$^{-3}$ s$^{-1}$. In general, calculated values of $J_{CNT}^{st}$ are significantly greater than experimentally measured values of the stationary nucleation current, $J_{experiment}^{st}$, with the former being 7 to 707 times greater than the latter for Ni-Al and Ni-Al-Cr alloys, Table 8 [22, 30, 135], respectively. Xiao and Haasen obtained a value of 4.1 x 10$^{22}$ m$^{-3}$ s$^{-1}$ for $J_{CNT}^{st}$, as compared to 8.4 x 10$^{19}$ m$^{-3}$ s$^{-1}$ for $J_{experiment}^{st}$ using HREM for Ni-12 Al at.% aged at 773 K (500 °C), which makes $J_{CNT}^{st}$ 488 times greater than $J_{experiment}^{st}$ [135]. This is most likely because Xiao and Haasen calculated $\Delta F_{ch}$ and $\Delta F_{el}$ using the activity of Al from [104], which is unmeasured in the Ni-Al system at 773 K (500 °C).

The calculated value of $J_{CNT}^{st}$ is linearly proportional to $N$, Eq. (24), where we assumed initially that every Al solute atom in solid-solution serves as a possible nucleation site for a γ'(L1$_2$)-precipitate, whereas the correct effective value of $N$ is much smaller Considering pre-existing short-range ordering of solute clusters in the as-quenched samples, and therefore $J_{CNT}^{st}$ = 1.34 x $10^{22}$ m$^{-3}$ s$^{-1}$ is an absolute maximum upper bound to the true value of $J_{CNT}^{st}$. As we reported for concentrated Ni-Al-Cr alloys, solute clusters can form and diffuse faster than monomers.[136] To obtain the correct value of $J_{CNT}^{st}$, we can use the effective number of nucleation sites per unit volume, $N_{effective}$, to replace the total number of possible nucleation sites per unit volume to address the faster diffusing solute clusters. In this study, $N_{effective}$ is only about 1.52% of the value of $N$. Additionally, the calculated value of $J_{CNT}^{st}$ is strongly affected by the value of $W_R^*$, Eq. (24). $W_R^*$ is in turn proportional to $(\sigma^{\gamma/\gamma'})^3$, Eq. (21). Hence, our measured values of 28.55 ± 1.61 and 29.94 ± 1.69 mJ m$^{-2}$ for $\sigma^{\gamma/\gamma'}$ (Section 4.7) results in a range of values for $W_R^*$, and therefore a possible range of values of $J_{CNT}^{st}$ (1.44 x $10^{21}$ to 3.57 x $10^{22}$ m$^{-3}$ s$^{-1}$), which implies a factor of 25 between the largest and smallest value of $J_{CNT}^{st}$, due solely to the uncertainty in the value $\sigma^{\gamma/\gamma'}$. Additionally, the value of $\sigma^{\gamma/\gamma'}$ requires correction terms for γ'(L1$_2$)-precipitates with small values of <R(t)>, which represent contributions from precipitate edges (which are a function of $\langle R(t) \rangle$), and from vertices (a constant) [137-140].

*4.2.2.2.     Regime II: concomitant γ'(L1$_2$)-precipitate nucleation and growth*

Regime II in Fig. 5b begins at the end of regime I (the nucleation regime), t = 0.25 h, and ends when $N_v(t)$ achieves a maximum value at 1 h [$N_v(t=1h)$= 6.87 ± 0.12 x $10^{23}$ m$^{-3}$]. The maximum value of $N_v(t)$ correlates with the maximum value of $f(t)$, 39.68 ±2.81%, because there are more γ'(L1$_2$)-precipitates available to form necks between precipitates. Regime II can be differentiated from regime I because $N_v(t)$ increases in both regimes I and II, but $\phi_{\gamma'}(t)$ and $\langle R(t) \rangle$ are approximately constant in regime I, whereas they are increasing continuously in regime II, Fig. 5.

*4.2.2.3.     Regime III: Concurrent growth and coarsening via the coagulation and coalescence mechanism*

Because $N_v(t)$ is decreasing continuously throughout regime III, we may conclude that nucleation is no longer occurring. It is distinguished from regime IV in that growth and coarsening are occurring mainly via the coagulation and coalescence mechanism rather than the evaporation-condensation mechanism (the large eat the small), as assumed for the mean-field LSW and Calderon-Voorhees-Murray-Kostorz (CVMK) coarsening models, Section 4.5.

*4.2.2.4.	Regime IV: quasi-stationary coarsening*

Regime IV in Fig. 5b displays a temporal exponent of $q$ = -0.75 ± 0.03, which is smaller in magnitude than the LSW and CVMK value of -1. *We emphasize, however, that the slope of this curve, and hence the value of $q$ is increasing continuously in regime IV in accord with the Kampmann-Wagner N(umerical) model*. We anticipate that aging the alloy for a longer time at 823 K (550 °C), for example to 10,000 h, or at an aging temperature >873 K (>600 °C), would yield a value of $q$ = -1 [104]. Our earlier studies of Ni-6.5 Al-9.5 Cr at.% aged at 873 K [30] and Ni-10 Al-8.5 Cr-2 Ru at.% aged at 1073 K (800 °C) [31] both display an experimental value of -1 for $q$, implying that a higher aging temperature and therefore a larger inter-diffusivity is needed to achieve a true stationary-coarsening regime. Ternary alloys have, however, a much smaller value of $\Delta F_{el}$: Ni-5.2 Al-14.2 Cr, Ni-7.5 Al-8.5 Cr, and Ni-6.5 Al-9.5 Cr at.% have $\Delta F_{el}$ values at 873 K (600 °C) of 1.1 x $10^5$, 2.5 x $10^6$, and 2.7 x $10^6$ kJ m$^{-3}$, respectively, compared to 4.96 x $10^6$ kJ m$^{-3}$ for our current alloy, Ni-12.5 Al at.%, at 823 K (550 °C), which helps the ternary alloys to achieve a value of $q$ = -1 faster than this binary alloy due to the much smaller lattice parameter misfit and elastic strain energy in these ternary Ni-Al-Cr alloys [22, 30].

This value of $q$ = -1 differs from a mathematical equation posited by Ardell [141] and Xiao and Haasen [135], who employed the relationship $N_v(t) = c_1 t^{-1} - c_2 t^{-4/3}$, where $c_1$ and $c_2$ are given in references [135, 141]. This relationship is not generally correct, as shown by Marqusee and Ross [59, 142, 143]. In reference [142] Marqusee and Ross demonstrate rigorously that as time approaches infinity $\langle R(t) \rangle$ becomes proportional to t$^{1/3}$ and $N_v(t)$ is proportional to t$^{-1}$. If one adds mathematical corrections terms to these laws, as was posited by Ardell, and Xiao and Haasen, then it is necessary to add higher order terms to all the pertinent physical quantities, $\langle R(t) \rangle$, $N_v(t)$, $\Delta C_i^\gamma(t)$, and the PSD, which are then no longer unique.

*4.2.3. Temporal evolution of the mean radius, $\langle R(t) \rangle$ , of gamma-prime (L1$_2$ structure)-precipitates*

Fig. 5c provides very strong evidence that Ni-12.5 Al aged at 823 K (550 °C) follows a diffusion-limited coarsening process. The value of $1/p$ obtained from the data in Fig. 5c, 0.34 ± 0.02, is in excellent agreement with the LSW and CVMK coarsening models's value, $1/p$ = 1/3 [32, 33]. Because these coarsening models are based on a diffusion-limited coarsening process, we conclude that our Ni-12.5 Al at.% alloy aged at 823 K (550 °C) obeys a diffusion-limited, rather than an interface-limited (source-limited), coarsening process; the latter has a temporal exponent of 0.5 [33]. We find additional 3-D APT data and monovacancy-mediated LKMC$_1$ simulation evidence against a source-limited (interface-controlled) coarsening mechanism in the temporal evolution of $\delta(t)$, Section 4.6.

*4.3. Temporal evolution of the compositions of the gamma(f.c.c.)- and gamma-prime(L1$_2$ structure)-phases*

Our 3-D APT and monovacancy-mediated LKMC$_1$ results for Ni-12.5 Al at.% aged at 823 K (550 ºC) both indicate that the γ(f.c.c.)- and γ'(L1$_2$)-precipitate-phases are initially supersaturated in Al, and that the compositions of both phases evolve temporally with increasing aging time; this is in contrast to standard review article and text-book thermodynamic models, which state that the composition of a second-phase remains at or is very close to its equilibrium value during phase separation. And hence the compositional evolution we observe suggests that we are detecting non-standard phase separation [144, 145], which is distinctly different from standard pictures of phase separation (or phase decomposition or precipitation).

The initial 3-D APT value of $C_{Al}^{\gamma}(t)$, 12.50 ± 3.12 at.% at 0.08 h of aging, is equal to the nominal composition of the alloy, Ni-12.5 Al at.%. This was anticipated, as the corresponding initial value of $\phi_{\gamma'}(t)$ is small, 0.002 ± 0.001 %. Therefore, in a solid-solution of essentially pure γ(f.c.c.), the small number density of γ'(L1$_2$)-precipitates doesn't alter significantly the value of $C_{Al}^{\gamma}(t)$. With increasing aging time, $\phi_{\gamma'}(t)$ is increasing, and concomitantly $C_{Al}^{\gamma}(t)$ is moving toward the left along the horizontal tie-line, indicated by a bold black-arrow in Fig. 2, headed for the γ/(γ plus γ') solvus curve. At 4096 h $C_{Al}^{\gamma}(t)$ is 10.86 ± 1.21 at.% from our 3-D APT measurement, which agrees closely with the solvus curves due to Ma and Ardell [71] and Dupin et al. [70], which are identical in the temperature range 773 K (500 ºC) to 900 K (627 ºC), Fig. 2.

The γ'(L1$_2$)-precipitates nucleate far to the right of the (γ plus γ')/γ' solvus curve, and from the measured initial 3-D APT value of $C_{Al}^{\gamma'}(t)$ (33.97 ± 49 at.%) the first observed precipitates are in the [γ'(L1$_2$) plus NiAl(B2)] phase-field, Fig. 2, if this portion of the extant phase diagram is correct. The value of $C_{Al}^{\gamma'}(t)$ at the earliest aging time, 0.08 h, is determined using a proximity histogram [84, 85] for 16 effective γ'(L1$_2$)-precipitates (due to a small number density, $N_v(t = 0.08) = 0.08 \pm 0.02 \times 10^{23}$ m$^{-3}$), and thus its uncertainty is ±8.49 at.% Al, implying that $C_{Al}^{\gamma'}(t)$ could be as small as 25.48 at.% Al at 0.08 h, albeit with a very small probability. 3-D APT cannot determine a crystal structure, but powder x-ray diffraction experiments performed at the Advanced Photon Source at Argonne National Laboratory, for the earliest aging times (0.08 to 0.25 h), provide clear evidence for the presence of an L1$_2$-structure-phase, and simultaneously they failed to detect the presence of a B2 phase. The nucleation of the B2 phase requires a structural transformation from f.c.c. to b.c.c. structure, which implies a large nucleation energy barrier to overcome because of the larger initial strain associated with the B2 phase. We, therefore, conclude that the precipitates have an L1$_2$-structure at 0.08 h.

A possible reason for the γ'(L1$_2$)-precipitates nucleating in the [γ'(L1$_2$) plus NiAl(B2)] phase field, Fig. 2, can be explained by the Gibbs-Thomson-Freundlich effect [146], which gives the solute concentration at a matrix/precipitate heterophase interface as a function of its radius, $R$. The Gibbs-Thomson-Freundlich equation predicts the equilibrium concentration of Al on the γ(f.c.c)-side at the γ(f.c.c.)/γ'(L1$_2$) interface, $C_{Al}^{\gamma,eq}(R)$. The ratio of $C_{Al}^{\gamma,eq}(R)$ to the bulk equilibrium concentration of Al in the γ(f.c.c)-phase, $C_{Al}^{\gamma,eq}(\infty)$, is given by:

$$\ln\left(\frac{C_{Al}^{\gamma,eq}(R)}{C_{Al}^{\gamma,eq}(\infty)}\right) = \frac{2V_a^{\gamma'}\sigma^{\gamma/\gamma'}}{Rk_BT} \quad (27)$$

where $C_{Al}^{\gamma,eq}(R)$ is a function of the γ'(L1$_2$)-precipitate radius, $R$ [146]. This ratio depends exponentially on the inverse of $R$, implying that as $R$ increases $C_{Al}^{\gamma,eq}(R) \rightarrow C_{Al}^{\gamma,eq}(\infty)$. Employing a $V_a^{\gamma'}$ value of 1.16 x 10$^{-17}$ m$^3$ [147, 148], and taking $\sigma^{\gamma/\gamma'}$ to be 28.55 ± 1.61 mJ m$^{-2}$ (Section 4.7), implies that $C_{Al}^{\gamma,eq}(R)$ differs significantly from $C_{Al}^{\gamma,eq}(\infty)$ for γ'(L1$_2$)-precipitates for $R$ values <10 nm. For example, at 0.08 h of aging $\langle R(t=0.08\,h)\rangle = 0.79\pm0.20$, and therefore $C_{Al}^{\gamma,eq}(R)/C_{Al}^{\gamma,eq}(\infty)$ is 2.09, implying that the APT measurements should yield an Al concentration for a γ'(L1$_2$)-precipitate that is greater by this factor than a precipitate with say $R$ = 10 nm. If $\sigma^{\gamma/\gamma'}$ is taken to be 29.94 ± 1.69 mJ m$^{-2}$ (Section 4.7), then $C_{Al}^{\gamma,eq}(R)/C_{Al}^{\gamma,eq}(\infty)$ is 2.07 at 0.08 h of aging, which implies that 3-D APT measurements should yield an Al concentration for a γ'(L1$_2$)-precipitate that is still approximately twice that of a precipitate with $R$ = 10 nm.

At 4096 h, $C_{Al}^{\gamma'}$ approaches Saunders' calculated solvus curve between the [γ(f.c.c.) plus γ'(L1$_2$)] and γ'(L1$_2$) phase-fields [68]. Within the experimental uncertainty of this value [$C_{Al}^{\gamma'}(t=4096\,h)$ = 23.90 ± 2.66 at.%], the 3-D APT data is also in agreement with Ma and Ardell's experimental solvus curve and our calculated GCMC solvus curve [22, 71].

### 4.4. Temporal evolution of the γ'(L1$_2$)-precipitate size distributions (PSDs)

Fig. 8 compares experimental PSDs obtained from the 3-D APT and LKMC simulation results with four different model PSDs: (1) LSW [32, 33]; (2) modified LSW [54]; (3) BW [55]; and (4) AV [56]. Of these four, we find that both the 3D APT experimental and LKMC simulation results favor visually the BW and AV PSDs.

The experimental PSDs at short aging times (<4 h) exhibit very narrow widths (full-width at half-maximum < 1.0 $R/\langle R(t)\rangle$), indicating that the alloy is far from equilibrium because

γ'(L1$_2$)-precipitates are still nucleating in regime I, Fig. 5b. For aging times >4 h, the PSDs become wider (corresponding to a decrease in $N_v(t)$), indicating that the system is becoming quasi-stationary. The change from a left-skewed PSD to a right-skewed PSD is indicative of a system entering regime IV (quasi-stationary coarsening) [100].

Similar to the APT results, the PSD for the LKMC results, Fig. 8c, is very narrow for early aging times (<4 h) and it becomes wider after 4 h, which indicates the transformation from nucleation to growth of precipitates. The longest aging time for the LKMC simulations is 800 h and the PSDs agree better with BW and AV than with the LSW PSD, which are in excellent agreement with APT results at 4096 h, discussed below.

The LSW mean-field theory for a PSD predicts that once a stationary-regime is achieved for a two-phase system, the tail of the distribution has a cut-off at 1.5 $R/\langle R(t)\rangle$, while the most frequent γ'(L1$_2$)-precipitate in the PSD has a radius of 1.13 $\langle R(t)\rangle$ [100]. The experimental PSD for 4096 h, Fig. 8b, agrees partially with this mean-field theory, as there are no γ'(L1$_2$)-precipitates in the system larger than 1.5 $\langle R(t)\rangle$, and the peak of the distribution occurs at 1.1 $\langle R(t)\rangle$. Fig. 8b demonstrates, however, that the LSW PSD is much narrower (full-width at half-maximum = 0.4 $R/\langle R(t)\rangle$) than the experimental PSD (full-width at half-maximum = 0.8 $R/\langle R(t)\rangle$), and therefore the LSW PSD does not provide a good fit to the experimental 3-D APT data. The modified LSW PSD predicts the same full-width at half-maximum as the experimental PSD, but its shape is skewed to smaller values of $R/\langle R(t)\rangle$ than the experimental PSD. The AV and BW PSDs both capture the shape of the experimental PSDs better than either the LSW or modified LSW models. Mathematical analyses of the experimental PSDs, Appendix C, demonstrate that the LSW PSD always provides the worst fit to the experimental APT data, and that the alloy enters a quasi-stationary regime at 1024 h, which is the time at which the alignment of precipitates (rafting) is first obvious, Fig. 4.

*4.5. Temporal evolution of the fraction of γ'(L1$_2$)-precipitates interconnected by necks, $f(t)$, and the minimum edge-to-edge distances, $\langle \lambda_{edge-edge}(t)\rangle$, between neighboring γ'(L1$_2$)-precipitates*

In Section 4.2.2 we demonstrate that the temporal evolution of γ'(L1$_2$)-precipitates in Ni-12.5 Al at.% aged at 823 K is divided into four distinct regimes, Fig. 5b: (I) nucleation from 0.08 to 0.25 h; (II) nucleation and growth from 0.25 to 1 h; (III) growth and coarsening from 1 to 64 h; and (IV) quasi-stationary coarsening beyond 64 h. Our results for the temporal evolution of $f(t)$ and $\langle \lambda_{edge-edge}(t)\rangle$ prove that coagulation and coalescence is the dominant mechanism of γ'(L1$_2$)- precipitate evolution in regime III, which is in agreement with the vacancy-mediated LKMC$_1$ simulations.

*4.5.1. The evolution of the fraction of γ'(L1$_2$)-precipitates interconnected by necks, $f(t)$*

The monovacancy-mediated LKMC$_1$ results for $f(t)$ are slightly greater than the 3-D APT results for $f(t)$ in regime I, and slightly smaller than the 3-D APT results for $f(t)$ in regimes II-IV, Fig. 9a. This is most likely due to the differences in how the γ'(L1$_2$)-precipitates are defined in each case. The γ'(L1$_2$)-precipitates in the 3-D APT reconstructions are delineated using iso-concentration surfaces, Section 2.3, while γ'(L1$_2$)-precipitates in the monovacancy-mediated LKMC$_1$ simulations are delineated using iso-ordering surfaces [22, 62]. *This implies that the 3-D APT results define a neck as a region between two γ'(L1$_2$)-precipitates with approximately the same composition as the interior of the γ'(L1$_2$)-precipitates, while monovacancy-mediated LKMC$_1$ results have the added requirement that this region also exhibits L1$_2$ ordering*. Because monovacancy-mediated LKMC$_1$ simulations have a more rigid criterion for what is or is not a neck than the 3-D APT results, the monovacancy-mediated LKMC$_1$ results for $f(t)$ are smaller than the 3-D APT results for $f(t)$ in regimes II-IV. Experimental 3-D APT analyses find necks present solely when connecting two γ'(L1$_2$)-precipitates to one another. The analyses of atomic configurations generated by monovacancy-mediated LKMC$_1$ simulations find mainly two γ'(L1$_2$)-precipitates interconnected by one neck. (Occasionally the presence of three γ'(L1$_2$)-precipitates interconnected by two necks is detected.) Our prior studies have also shown definitively the presence of L1$_2$ ordered-necks utilizing 3-D APT data [18, 29].

*4.5.2. The temporal evolution of the edge-to-edge distances, $\langle \lambda_{edge-edge}(t) \rangle$, between neighboring γ'(L1$_2$)-precipitates, and the flux diffusion mechanism*

For the 3-D APT data and LKMC$_1$ simulation results, for aging times <0.25 h, the edge-to-edge distance $\langle \lambda_{edge-edge}(t) \rangle$ is greater than 200 nm, and $f(t)$ is less than 5%, Fig. 9, which implies that the evaporation-condensation coarsening mechanism ("the large precipitates eat the small precipitates") prevails because the γ'(L1$_2$)-precipitates are not sufficiently close to one another to form L1$_2$-ordered necks. For aging times greater than 256 h, the evaporation-condensation mechanism once again triumphs, as $f(t)$ is <20% and $\langle \lambda_{edge-edge}(t) \rangle$ is >18 nm; the latter is too large a distance to permit γ'(L1$_2$)-precipitates to form L1$_2$-ordered necks as a result of their overlapping concentration profiles.

To interpret our results further we utilize some basic concepts of macroscopic diffusion theory, which are reviewed briefly in Appendix D. The model Ni-Al alloy is specified by the concentrations of the atomic species, $C_{Ni}$, $C_{Al}$, and the concentration of monovacancies, $C_V$; the sum of the three concentrations is unity. Since the γ'(L1$_2$)-precipitates are coherent with the γ(f.c.c.)-matrix, lattice sites are conserved locally during phase-separation. Because the mean edge-to-edge distance, $\langle \lambda_{edge-edge}(t) \rangle$, between γ'(L1$_2$)-precipitates is small compared to the mean inter-dislocation spacing, which is of the order of <10$^3$ nm (one micron), lattice sites are conserved at all space and time scales, which are given by the mean spacing between dislocations and the time it takes for vacancies to diffuse that distance.

To understand the precipitation mechanism, a clear way to visualize diffusional couplings is to search for the eigenmodes (eigenvalues and eigenvectors) of the diffusion matrix in LKMC$_1$ simulations. We obtain two distinguishable diffusion modes from the diffusion matrix in the LKMC$_1$ simulations listed in Table 12: fast and slow modes. The fast eigenvector of the diffusion matrix is such that: $\tilde{J}^{fast} = -D^{fast} \nabla C^{fast} \Omega^{-1}$ and the eigen-diffusion vector is thus $\nabla C^{fast} = -\Omega \tilde{J}^{fast} / D^{fast}$. For the fast eigen-mode, the eigen-diffusivity coefficient is 1.05×10$^{-20}$ m$^2$ s$^{-1}$ ($D_{fast}$) with a diffusion flux coupling given by $J_{Ni}^{fast} / J_{Al}^{fast} = 0.082 \pm 0.02$. For the slow eigenmode, the eigen-diffusivity coefficient is 9.58×10$^{-22}$ m$^2$s$^{-1}$ ($D_{slow}$) with a diffusion flux coupling given by $J_{Ni}^{slow} / J_{Al}^{slow} = -12.3 \pm 0.5$. The fast eigen-diffusivity is one order of magnitude larger than the slow eigen-diffusivity. For a concentration profile to evolve at the scale of the mean edge-to-edge inter-precipitate distance, $\langle \lambda_{edge-edge}(t) \rangle$, (in the coarsening regime), a time $\tau = \frac{<\lambda_{edge-edge}(t)>^2}{4D}$, is required. Fig. 13 displays the temporal evolution of the mean edge-to-edge inter-precipitate distance, $\langle \lambda_{edge-edge}(t) \rangle$, the root-mean square diffusion distance using the 3-D APT inter-diffusion diffusivity, and the two diffusional flux-eigen-modes. Furthermore, Fig. 13 demonstrates that the fast eigen-mode contributes to growth during the first hour of aging (across the edge-to-edge distance). For LKMC$_1$, within the Ni-based γ(f.c.c.)-matrix (Ni is a slow diffuser compared to Al), the fast eigen-mode (that is, the dominant eigen-flux) is mainly Al, which drags Ni in the same direction: $J_{Ni}^{fast} / J_{Al}^{fast} = 0.082 \pm 0.02$, while the coupling between the Ni and Al flux, $J_{Ni} / J_{Al}$, required to create a γ'(L1$_2$)-precipitate with its equilibrium $J_{Ni} / J_{Al}$ composition from the supersaturated solutions with a deviation 0.121. Coagulation and coalescence of γ'(L1$_2$)-precipitates occurs mainly approximately at the aging time when the fast eigen-mode achieves the mean edge-to-edge inter-precipitate diffusion distance. The slow eigen-mode becomes significant only after 400 hours (across the edge-to-edge distance), well within the coarsening regime. Inside the γ'(L1$_2$)-precipitates, the weaker slow eigen-mode is mainly Ni, which drags small amounts of Al out of γ'(L1$_2$)-precipitates: $J_{Ni}^{slow} / J_{Al}^{slow} = -12.3 \pm 0.5$. At 400 h the γ'(L1$_2$)-precipitate morphology changes from spheroidal-to-cuboidal, which corresponds to the slow eigen-diffusion mode becoming dominant.

*4.6. Temporal evolution of the compositional interfacial widths, $\delta(t)$, between the gamma(f.c.c.)- and gamma-prime(L1$_2$ structure)-phases*

Our results for $\delta(t)$, from both 3-D APT experiments and monovacancy-mediated LKMC$_1$ simulations, demonstrate that $\delta(t)$ decreases with increasing aging time, but it never achieves a value of zero, in contrast to the trans interface diffusion coarsening (TIDC) model, which predicts the exact opposite temporal behavior for $\delta(t)$.

*4.6.1. The concentration profiles between the γ(f.c.c)- and γ'(L1$_2$)-phases*

The concentration profiles between the γ(f.c.c)- and γ'(L1$_2$)-phases, determined by 3-D APT experiments and monovacancy-mediated LKMC$_1$ simulations for our binary Ni-12.5 Al at.% alloy, Fig. 10, exhibit a smooth transition from the γ(f.c.c.)-phase to the interface between the two phases, and from this interface to the γ'(L1$_2$)-phase. This smooth transition is in strong contrast to our prior studies, which yielded concentration profiles measured by 3-D APT experiments and monovacancy-mediated LKMC$_1$ simulations for three different ternary Ni-Al-Cr alloys. In these studies, an excess of Ni and Cr atoms and a depletion of Al atoms are initially observed in the interfacial region between the two phases [12, 13, 16, 18, 22, 23, 30, 31, 61, 62, 149]. The difference between the binary and ternary concentration profiles occurs because for ternary alloys diffusion of Al into the γ'(L1$_2$)-precipitates is opposed by the diffusion of Cr and Ni into the γ(f.c.c.)-matrix. In contrast, we don't find evidence for a depletion or accumulation of Ni or Al atoms at the γ(f.c.c)/γ'(L1$_2$)-heterophase interfaces from the 3-D APT results and the monovacancy-mediated LKMC$_1$ simulations. We emphasize strongly that in both the Ni-Al and the Ni-Al-Cr alloys an additional component, monovacancies, must be taken into account to understand the diffusive fluxes between the γ(f.c.c)- and γ'(L1$_2$)-phases. Hence, binary alloys are three-component systems and ternary alloys are four-component systems because of the necessity of including the vacancy as a component.

*4.6.2. The {100}-type compositional interfacial width, $\delta(t)$*

The average value of $\delta(t)$ for {100}-type interfaces decreases with increasing aging time, Fig. 11a, which is consistent with the Cahn-Hilliard [144, 145] and Martin models [150], and with prior 3-D APT studies [27, 111, 151].

Our monovacancy-mediated LKMC$_1$ results for the Ni-12.5 Al at.% alloy agree with the 3-D APT results for $t > 1$ h, but slightly overestimate the value of $\delta(t)$ for $t<1$ h. Both our 3-D APT experimental and our monovacancy-mediated LKMC$_1$ simulation results are in disagreement with the so-called trans interface diffusion coarsening (TIDC) model for binary alloys, which posits the following ansatz: "…if $\delta(t) \propto \langle R(t) \rangle^m$ $(0 \leq m \leq 1)$ over a certain range of particle (precipitate) radii, the kinetics of particle (precipitate) growth and solute depletion will obey equations of the type $\langle R(t) \rangle^n \propto t$ and $\langle C_i^{\gamma,ff}(t) \rangle - C_i^{\gamma,eq}(\infty) \propto t^{-1/n}$, where $n = m + 2$" [42, 152]. Additionally, the authors state, "We cannot yet provide theoretical justification for the relationship $\delta(t) \propto \langle R(t) \rangle^m$, though we believe that such a relationship is credible, given the ragged nature of the γ(f.c.c.)/γ′(L1$_2$) interface" [152]. *In strong contrast, our results demonstrate that $\delta(t)$ decreases with increasing aging time and Fig. 11b establishes that $\delta(t)$ decreases continuously with increasing $\langle R(t) \rangle$*. The data in Fig. 11b fit a $\langle R(t) \rangle^m$-type relationship, where $m$ is -0.40 ± 0.03, which disagrees with the ansatz that forms the basis of the TIDC model: that is, $0 \leq m \leq 1$. The value $m = -0.40 \pm 0.03$ implies a value of 1.60 for $n$ ($n = m + 2$), which in turn implies that the TIDC model predicts $\langle R(t) \rangle \propto t^{0.63}$. *The latter relationship is clearly in disagreement with the experimental 3-D APT results, Fig. 5c, and therefore the ansatz that is the basis of the TIDC model is incorrect*. Additionally, our prior results for Ni-12.5 Al and Ni-13.4 Al at.% aged at 823 K (550 °C), and 873 K (600 °C), where we find that $\delta(t)/\langle R(t) \rangle$ varies to first order as $\langle R(t) \rangle^{-1}$ [111], and also for Ni-10.0 Al-8.5 Cr-2.0 Ta at.% aged at 1073 K (800 °C) [27], prove that $\delta(t)$ decreases with increasing aging time and increasing $\langle R(t) \rangle$.

We have previously demonstrated that the $\delta(t)$ values for three Ni-Al-Cr alloys *decrease with increasing aging time* by performing monovacancy-mediated LKMC$_1$ simulations, which include monovacancy-solute binding energies to 4$^{th}$ NN distances [45, 61] and atom-atom interactions that also extend to 4$^{th}$ NN, Section 2.5. Monovacancy-mediated LKMC$_2$ simulations of the same Ni-Al-Cr alloys produce $\delta(t)$ values that are smaller than those produced by the monovacancy-mediated LKMC$_1$ simulations, because the monovacancy-solute binding energy only extends to the 1$^{st}$ NN distance for monovacancy-mediated LKMC$_2$, Table 3, even though the atom-atom interactions extend to 4$^{th}$ NN distances. Nevertheless, these $\delta(t)$ values still decrease with increasing aging time. *Thus, the $\delta(t)$ values are determined by atomistic interactions that* *include non-zero monovacancy–solute binding energies*. The atom-atom and monovacancy-solute binding energies are not functions of aging time or $\langle R(t) \rangle$. We therefore conclude that the compositional thickness of the transition layer between the γ(f.c.c)- and γ'(L1$_2$)-phases is affected by the details of the diffusion mechanism; indeed, decreasing the range of monovacancy-solute interactions (as does the monovacancy-mediated LKMC$_2$ parameterization), modifies the coupling between the solute and solvent fluxes. The interaction distance over which atom-atom and monovacancy-solute interactions occur, $\Xi$, is a constant for a given alloy at a specified temperature. Therefore, the ratio $\Xi / \langle R(t) \rangle$ must decrease with increasing aging time, because $\Xi$ is a constant and $\langle R(t) \rangle$ is continuously increasing, which is the case for an alloy becoming quasi-stationary, Fig. 11b. Parenthetically, in the continuum limit, $\Xi$ is the origin of the so-called gradient energy coefficient in the Cahn-Hilliard equation [144, 145].

In this study, the vacancy-solute binding is also examined by employing only first NN distances up to an aging time of 400 h, which we refer to as LKMC$_2$ as discussed above. Technically, we replace $\varepsilon_{V-S}^k$ with $\varepsilon_{V-Ni}^k$ when k > 1, where k is the NN shell. We have found that the *growth and coarsening mechanism* switches from coagulation-coalescence to the classic evaporation-condensation mechanism during early aging times (<4 h). The maximum fraction of the precipitates connected by necks using LKMC$_2$ decreases to 9.6%, compared to 36% using LKMC$_1$. The compositional thickness of the transition layer using LKMC$_2$, between the γ(f.c.c)- and γ'(L1$_2$)-phases, is thinner than those utilizing LKMC$_1$; about 43 to 56 % when compared to the LKMC$_1$ value, during the *early nucleation and coarsening regimes*. The composition of the precipitates starts from 25.6 at.% Al using LKMC$_2$, compared to 30.13 at.% Al employing LKMC$_1$. This significant difference in the Al concentrations during the precipitation process demonstrates the effects of strong couplings between the solute and solvent fluxes, which have much larger values for the off-diagonal terms of the diffusion matrix when utilizing LKMC$_1$.

### 4.7. *The interfacial free energy of the gamma (f.c.c)/gamma-prime (L1$_2$) interface, $\sigma^{\gamma/\gamma'}$*

The calculated values of $\sigma^{\gamma/\gamma'}$ from 3-D APT coarsening rate constant data are in good agreement with our first-principles calculations for the {100}-type interface[153].

#### 4.7.1. *The measured interfacial free energy of the gamma (f.c.c)/gamma-prime (L1$_2$) interface determined from coarsening experiments and their comparison with first-principles calculations of $\sigma^{\gamma/\gamma'}$*

Table 9 summarizes the calculated constants that are necessary to determine $\sigma^{\gamma/\gamma'}$ utilizing Eq. (8). Each quantity listed is determined directly from our experimental 3-D APT data with the exceptions of $V_m^{\gamma'}$ [67] and $F(\phi_{\gamma'})$ [40, 59], which are calculated. Table 10 displays values of $G_m^{\gamma"}$ taken from two thermodynamic databases in Thermo-Calc [68-70] and the resulting values of $\sigma^{\gamma/\gamma'}$ obtained for each $G_m^{\gamma''}$. The values of $\sigma^{\gamma/\gamma'}$ obtained employing the Saunders and Dupin et al. $G_m^{\gamma''}$ values are 29.94 ± 1.69 and 28.55 ± 1.61 mJ m$^{-2}$, respectively, which are equal to one another within error.

We utilized first-principles calculations to calculate $\sigma^{\gamma/\gamma'}$ of the γ(Ni)/γ'(Ni$_3$Al) interface for {100}-, {110}-, and {111}-type interfaces [45]. Of these three $\sigma^{\gamma/\gamma'}$ first-principles values the one for the {100}-type interface is the correct one to compare with the experimental results, because with increasing aging time the γ'(L1$_2$)-precipitates develop a cuboidal morphology, Fig. 4e-f, and therefore the {100}-type plane is where the γ(Ni)/γ'(Ni$_3$Al)-interface lies. The first-principles calculated values of $\sigma^{\gamma/\gamma'}$ at 823 K (600 ºC) are 23.11, 26.83, and 28.42 mJ m$^{-2}$ for the {100}-, {110}-, and {111}-type interfaces, respectively [45], which are in the range of the experimental values, 26.94 to 31.16 mJ m$^{-2}$, within experimental error. Using a second-cluster expansion technique [46], Woodward et al. calculated $\sigma^{\gamma/\gamma'}$ values of 16.5, 17.7, and 15.9 mJ m$^{-2}$ for the {100}-, {110}-, and {111}-type interfaces, respectively. Because the cluster expansion technique overestimates the temperature by 335 K, these values are calculated at a scaled temperature of 488 K (215 ºC) instead of our actual aging temperature, 823 K (550 ºC). *We found that the interfacial free energy decreases with increasing temperature due to entropy effects* [153].

### 4.7.2. Equilibrium morphology of γ'(L1$_2$)-precipitates

Because the equilibrium morphology of γ'(L1$_2$)-precipitates is determined by the effects of $\sigma^{\gamma/\gamma'}$ and the elastic strain energy we calculate the ratio of the elastic strain-energy contribution to the $\sigma^{\gamma/\gamma'}$ contribution, $L$, for determining the morphology of γ'(L1$_2$)-precipitates using [154-156]:

$$L = \frac{\mu^\gamma \varepsilon^2 \langle R(t) \rangle}{\sigma^{\gamma/\gamma'}}; \qquad (28)$$

where $\varepsilon$ is the lattice parameter misfit between the two-phases. The value of $L$ is 0.02 for the shortest aging time (0.08 h), implying that the morphology of γ'(L1$_2$)-precipitates is dominated by the interfacial free energies, $\sigma^{\gamma/\gamma'}$ at 0.08 h. The values of $L$ are 0.29 at 2607 h and 0.40 at 4096 h, respectively, indicating that the elastic strain energy plays a larger role in the morphology of γ'(L1$_2$)-precipitates with increasing aging times, but it is still smaller than the role played by $\sigma^{\gamma/\gamma'}$ at 4096 h.

### 4.8. Comparison of the diffusivities calculated from atom-probe tomography experimental data and monovacancy-mediated lattice-kinetic Monte Carlo simulation results

Our results demonstrate that $D^{coarsening}_{experiment}$, derived from coarsening rate constants determined from APT experiments, is equal to *the inter-diffusion coefficient* $\tilde{D}$ for the binary Ni-12.5 Al at.% system, Eq. 29, and as determined by monovacancy-mediated LKMC$_1$ simulations.

*4.8.1. Relationship of the interdiffusivity, $\tilde{D}$, determined by monovacancy-mediated LKMC$_1$ simulations and the diffusivity determined from the coarsening experiments, $D^{coarsening}_{experiment}$*

The monovacancy-mediated LKMC$_1$ $D^{\gamma-matrix}_{Al}$ values are 2.22 ± 0.38 x 10$^{-21}$, 2.18 ± 0.36 x 10$^{-21}$, and 1.62 ± 0.26 x 10$^{-21}$ m$^2$ s$^{-1}$ for 1, 4, and 400 h of aging, respectively. And the $D^{\gamma-matrix}_{Ni}$ values are 1.11 ± 0.14 x 10$^{-22}$, 1.02 ± 0.38 x 10$^{-22}$, and 0.99 ± 0.14 x 10$^{-22}$ m$^2$ s$^{-1}$ for 1, 4, and 400 h of aging, respectively. The value of $\tilde{D}$ is related to the individual diffusivities of Ni and Al by [100]:

$$\tilde{D} = C_{Al} D_{Ni} + C_{Ni} D_{Al}; \qquad (29)$$

Using Eq. (29) and the calculated values of $D^{\gamma-matrix}_{Al}$ and $D^{\gamma-matrix}_{Ni}$ yield values of 3.75 ± 0.31 x 10$^{-22}$, 3.62 ± 0.31 x 10$^{-22}$, and 2.90 ± 0.25 x 10$^{-22}$ m$^2$ s$^{-1}$ for $\tilde{D}$, for 1, 4, and 400 h of aging, respectively, for $C_{Al}$ = 0.125 and $C_{Ni}$ = 0.875 atomic fraction. The values of $\tilde{D}$ for 1 and 4 h are equal, within experimental error, to $D^{coarsening}_{experiment}$, 2.51 ± 0.14 x 10$^{-21}$ m$^2$ s$^{-1}$ (Section 4.8.2), and the $\tilde{D}$ value at 400 h is 15% of $D^{coarsening}_{experiment}$. Thus, there is semi-quantitative agreement between $D^{coarsening}_{experiment}$ and the monovacancy-mediated LKMC$_1$ values for 1, 4, and 400 h. *In the course of phase separation, $C_{Al}$ decreases while $C_{Ni}$ increases, which improves the agreement between the APT and LKMC$_1$ results.* In conclusion, $D^{coarsening}_{experiment}$ is properly described by $\tilde{D}$, which can be measured or calculated utilizing either a Kirkendall diffusion couple or LKMC$_1$ simulations, respectively. There are not, unfortunately, Kirkendall diffusion couple measurements available for $C_{Al}$ = 0.125 and $C_{Ni}$ = 0.875 atomic fraction.

Figs. 12a, 12c, and 12e show that the monovacancy spends the majority of its time inside the two γ'(L1$_2$)-precipitates and the neck region connecting them at 1 and 4 h, Section 3.7, while the γ'(L1$_2$)-precipitates are continuously ordering. Diffusion occurring in the γ'(L1$_2$)-precipitates and the connecting necks, as demonstrated by the monovacancy-mediated LKMC$_1$ simulations, is not taken into account in any of the extant mean-field diffusion-limited models of coarsening, nor is the presence of the interconnecting neck(s) between γ'(L1$_2$)-precipitates, which is at the origin of the coagulation-coalescence mechanism of coarsening. Because of correlation effects [51, 157], the monovacancy is not able to diffuse as fast in the partially ordered γ'(L1$_2$)-precipitates nor in the partially ordered L1$_2$ neck regions connecting the γ'(L1$_2$)-precipitates, as it is in the disordered

γ(f.c.c.)-matrix. This is demonstrated quantitatively in Table 6, where $D_i^{\gamma-matrix}$ is always larger than either $D_i^{\gamma'-precipitates}$ or $D_i^{necks}$.

For both Ni and Al interchanges with the monovacancy, $D_i^{supercell}$ is closer to $D_i^{\gamma-matrix}$ than it is to $D_i^{\gamma'-precipitates}$ for the values of $\phi_{\gamma'}(t)$ for the γ'(L1$_2$)-precipitates at 1 and 4 h, 2.87 ± 0.05 and 3.64 ± 0.16%, respectively. The volume fraction of the partially ordered necks connecting the γ'(L1$_2$)-precipitates is significantly smaller than $\phi_{\gamma'}(t)$. While the value of $\phi_{\gamma'}(t)$ at 400 h is closer to its equilibrium value, 13.48%, implying that approximately 90% of the volume of the alloy is the γ(f.c.c.)-matrix, and $D_i^{supercell}$ is closer to $D_i^{\gamma-matrix}$ than it is to $D_i^{\gamma'-precipitates}$.

*4.8.2. Calculation of the diffusivity, $D_{experiment}^{coarsening}$, from atom-probe tomography coarsening data and the inter-diffusivity, $\tilde{D}$, of Al and Ni*

The calculated value of $D_{experiment}^{coarsening}$ from the 3-D APT coarsening rate data are compared with the existing archival values of the diffusivities in the Ni-Al system, Table 11, and is specifically compared to the inter-diffusion coefficient, $\tilde{D}$, measured experimentally by Swalin and Martin [158], because it is the only extant inter-diffusivity for a Ni-Al alloy.

Eq. (9) is used to calculate $D_{experiment}^{coarsening}$ for this alloy, which is 2.51 ± 0.14 x 10$^{-21}$ m$^2$ s$^{-1}$ at 823 K (550 °C). Table 11 summarizes the archival values of the pre-factor, $D_0$, and the activation energy for diffusion, $Q$, for diffusion in the Ni-Al system (solute diffusivity and inter-diffusivity), plus the applicable temperature ranges and experimental or computational methods utilized. Table 11 also lists the resulting diffusivity at 823 K (550 °C) for each reference utilizing:

$$D_{experiment}^{coarsening} = D_0 \exp\left(-\frac{Q}{k_B T}\right). \tag{30}$$

Most experimental research on the Ni-Al system was performed at higher temperatures than the current research, so calculations of $D_{experiment}^{coarsening}$ at 823 K (550 °C) are based on an extrapolation.

Our measured value of $D_{experiment}^{coarsening}$ = 2.51 ± 0.14 x 10$^{-21}$ m$^2$ s$^{-1}$ is 27% greater than the value measured by Swalin and Martin, 1.83 x 10$^{-21}$ m$^2$ s$^{-1}$ [158], for $C_{Al}$ = 7x10$^{-3}$ at.fr., utilizing a Kirkendall effect couple to measure $\tilde{D}$ for Al and Ni. We attribute this difference to the higher value of $C_{Al}$ in our Ni-Al alloy: 12.5 versus 0.7 at. % Al. Additionally, $D_{experiment}^{coarsening}$ is 72% larger than the value for the Al diffusivity calculated by C. Campbell et al. [159], 0.71 x 10$^{-21}$ m$^2$ s$^{-1}$, based on her mobility database.

## 5. Summary and Conclusions

The temporal evolution of a binary Ni-12.5 Al at.% alloy undergoing phase-separation at 823 K (550 °C) is investigated using 3-D atom-probe tomography (APT), monovacancy-mediated lattice-kinetic Monte Carlo (LKMC$_1$) simulations, microhardness measurements, and some transmission electron microscopy (TEM). The main results obtained as a function of aging time include: (1) Vickers microhardness of bulk specimens of the alloy; (2) γ'(L1$_2$)-precipitate morphology; (3) γ'(L1$_2$)-precipitate volume fraction, $\phi_{\gamma'}(t)$; (4) γ'(L1$_2$)-precipitate number density, $N_v(t)$; (5) mean radius of γ'(L1$_2$)-precipitates, $\langle R(t) \rangle$; (6) Al concentration of the γ(f.c.c.)-phase, $C_{Al}^{\gamma}(t)$; (7) Al concentration of the γ'(L1$_2$)-phase, $C_{Al}^{\gamma'}(t)$; (8) supersaturation of Al in the γ(f.c.c.)-phase, $\Delta C_{Al}^{\gamma}(t)$; (9) supersaturation of Al in the γ'(L1$_2$)-phase, $\Delta C_{Al}^{\gamma'}(t)$; (10) partitioning coefficient of Al between the γ'(L1$_2$)- and γ(f.c.c.)-phases, $K_{Al}^{\gamma'/\gamma}(t)$; (11) partitioning coefficient of Ni between the γ'(L1$_2$)- and γ(f.c.c.)-phases, $K_{Ni}^{\gamma'/\gamma}(t)$; (12) γ'(L1$_2$)-precipitate size distributions (PSDs); (13) fraction of γ'(L1$_2$)-precipitates interconnected by necks, $f(t)$; (14) minimum edge-to-edge distance between γ'(L1$_2$)-precipitates, $\langle \lambda_{edge-edge}(t) \rangle$; and (15) interfacial compositional width between the γ(f.c.c.)- and γ'(L1$_2$)-phases, $\delta(t)$, for the {100}-type interface. Experimental atom-probe tomographic measurements and monovacancy-mediated LKMC$_1$ simulation results are used to calculate kinetic and thermodynamic quantities: (i) including the quasi-stationary nucleation currents, $J_{APT}^{st}$ and $J_{LKMC}^{st}$; (ii) the diffusivity, $D_{experiment}^{coarsening}$; and (ii) the interfacial free energy of the γ(f.c.c)/γ'(L1$_2$) {100}-type interface, $\sigma^{\gamma/\gamma'}$. This is the most complete and detailed study of the temporal evolution of an alloy undergoing a first-order phase transformation made to date and it is possible because of correlative research employing 3-D APT in combination with monovacancy-mediated LKMC$_1$ simulations, which permits us to explore this four-dimensional space (positions of atoms in 3-D and chemical identity of each atom) in great detail. LKMC$_1$ includes monovacancy-solute binding energies atom-atom interaction energies out to the fourth nearest-neighbor (NN) position, whereas LKMC$_2$ is only out to first NN positions for the same energies.

- The four regimes of the γ'(L1$_2$)-phase's temporal evolution are: (I) quasi-stationary nucleation; (II) concomitant nucleation and growth; (III) concurrent growth and coarsening; and (IV) quasi-stationary coarsening, which are identified utilizing 3-D APT and monovacancy-mediated LKMC$_1$ simulations, Fig. 5.
- Monovacancy-mediated LKMC$_1$ simulations, which include monovacancy-solute binding energies out to 4$^{th}$ nearest-neighbor distances (Table 3) were performed, through 800 h, in parallel with the 3-D APT experiments. In Figs. 2, 5, 6, 7, 9, 10, and 11 we compare the LKMC$_1$ and 3-D APT results in detail.

- The γ'(L1$_2$)-precipitates undergo a spheroidal-to-cuboidal transition prior to 1024 h, Fig. 4a-f. Additionally, the TEM micrographs, Fig. 4g, at 1024 h of aging, demonstrate that the γ'(L1$_2$)-precipitates are well aligned along a <100>-type-direction indicative of rafting. The 3-D APT and TEM results collectively provide a reasonably complete picture of the morphological evolution of the γ'(L1$_2$)-precipitates.

- The Vickers microhardness values are approximately constant, while the γ'(L1$_2$)-precipitates are nucleating (<4 h of aging), and the values increase with increasing aging time, Fig. 3, because $\phi_{\gamma'}(t)$ is concomitantly increasing and the edge-to-edge distance between precipitates is decreasing. Hence, this alloy's strength is governed by the volume fraction of γ'(L1$_2$)-precipitates and the main strengthening mechanism is most likely Orowan dislocation-looping at room temperature.

- As demonstrated by 3-D APT experiments, $\phi_{\gamma'}(t=0) = 0$ in the as-quenched state, implying that the alloy consists of only the γ(f.c.c.)-phase at $t = 0$.

- The initial value of $\phi_{\gamma'}(t)$ at 0.08 h is 0.002 ± 0.001%, Fig. 5a, and it increases with increasing aging time toward its equilibrium value of 13.48% at 4096 h, where $\phi_{\gamma'}(t=4096\,h)$ is 12.59 ± 1.40 %: Fig. 2.

- The number density, $N_v(t)$, is 7.88 x 10$^{21}$ m$^{-3}$ for the earliest aging time (0.08 h), Fig. 5b. This value increases with increasing aging time as additional γ'(L1$_2$)-precipitates nucleate in the γ(f.c.c.)-matrix, and it achieves a maximum value of 6.87 x 10$^{23}$ m$^{-3}$ after 1 h of aging, representing an increase of two orders of magnitude in $N_v(t)$. With further aging, $N_v(t)$ decreases as new γ'(L1$_2$)-precipitates cease nucleating and existing γ'(L1$_2$)-precipitates grow and coarsen via a coagulation and coalescence mechanism, which we have found previously for three different Ni-Al-Cr alloys [12, 13, 16-19, 22, 23, 26, 29-31]. The temporal exponent, $q$, of $N_v(t)$ for Ni-12.5 Al at.% aged at 823 K in the range 64-4096 h is 0.75±0.03, which is slowly approaching the predicted value for regime IV of $q = -1$, Fig. 5b.

- Based on the total solute atoms as possible nucleation sites, the stationary nucleation current, $J_{CNT}^{st}$, calculated assuming classical nucleation theory (CNT), is 1.34 x 10$^{22}$ m$^{-3}$ s$^{-1}$, which is 66 times greater than the 3-D APT value, $J_{APT}^{st}$, of 2.03 x 10$^{20}$ m$^{-3}$ s$^{-1}$ and 57 times greater than the monovacancy-mediated LKMC$_1$ value, $J_{LKMC}^{st}$, of 2.37 x 10$^{20}$ m$^{-3}$ s$^{-1}$, Table 8. This value of $J_{CNT}^{st}$ is calculated using a chemical driving force for phase separation, $\Delta F_{ch}$, which is obtained from first-principles calculations, rather than utilizing the standard classical thermodynamic methodology [104, 129]. Additionally, we use $\sigma^{\gamma/\gamma'} = 28.55 \pm 1.61$ mJ m$^{-2}$, as determined from the 3-D APT coarsening experiments, Section 4.7, for calculating $J_{CNT}^{st}$, Section 4.2.2.1.

- Due to the pre-existing solute short-range order, the effective number of nucleation sites per unit volume, $N_{effective}$, is proposed to replace the total number of possible nucleation sites per unit volume to address the faster diffusing solute clusters. In this study, $N_{effective}$ is only about 1.52% of the value of $N$. This approach for calculating $J_{CNT}^{st}$ yields better agreement with $J_{APT}^{st}$ than the classical approaches [104, 129] and it depends on a value of $\sigma^{\gamma/\gamma'}$, which we measure from the coarsening experiments utilizing 3-D APT.
- γ'(L1$_2$)-precipitates are detected at the earliest aging time studied experimentally (0.08 h) with an $\langle R(t) \rangle$ value of 0.79 ± 0.20 nm, Fig. 5c. Synchrotron x-ray diffraction studies at the Advanced Photon Source at Argonne National Laboratory demonstrate that they have an L1$_2$ structure and not a B2 structure.
- After aging for 4096 h, $\langle R(t) \rangle$ increases to 14.59 ± 1.62 nm, which is 18.5 times bigger than its initial value. The temporal exponent for $\langle R(t) \rangle$ in the quasi-stationary coarsening regime is $1/p$ = 0.34 ± 0.02, which is consistent with the Lifshitz-Slyozov-Wagner (LSW) and Calderon-Voorhees-Murray-Kostorz (CVMK) mean-field diffusion-limited coarsening models' value of 1/3, thereby indicating diffusion-limited coarsening behavior. The rate constant, $K$, for $\langle R(t) \rangle$ is 2.09 ± 0.10 x 10$^{-31}$ m$^3$ s$^{-1}$.
- γ'(L1$_2$)-precipitates nucleate with an initial high Al-supersaturation (excess of Al), and then proceed to become less supersaturated in Al with increasing aging time, Fig. 6. Both supersaturations, $\Delta C_{Al}^{\gamma}(t)$ and $\Delta C_{Al}^{\gamma'}(t)$, decrease continuously with increasing aging time. The longest aging time studied (4096 h) is insufficient for the alloy to achieve equilibrium at 823 K (550 °C). Nevertheless, the temporal exponent for the decrease in supersaturations in both phases is $1/r$ = 0.33±0.03. And this value is also in agreement with the LSW and CVMK mean-field diffusion-limited coarsening models, which is consistent with *diffusion-limited coarsening, but it is inconsistent with the trans-interface diffusion-controlled (TIDC) coarsening model*.
- The calculated rate constants for $\Delta C_{Al}^{\gamma}(t)$ and $\Delta C_{Al}^{\gamma'}(t)$, $\kappa^{\gamma}$ and $\kappa^{\gamma'}$, are 0.25 ± 0.01 and 0.68 ± 0.03 s$^{1/3}$, respectively. Because of the difficulties involved with measuring $C_{Al}^{\gamma}(t)$ and $C_{Al}^{\gamma'}(t)$ with methods other than 3-D APT, it is rare to find reliable measured values for $\kappa^{\gamma}$ and $\kappa^{\gamma'}$ in the archival literature.
- Utilizing the LSW and CVMK equations for supersaturation, the equilibrium concentrations of the γ(f.c.c.)- and γ'(L1$_2$)-phases are 11.14 ± 0.32 and 23.14 ± 0.47 at.%, respectively, extrapolated to infinite aging time, Fig. 6a-b.
- The temporal evolution of $C_{Al}^{\gamma}(t)$ approaches the solvus curves due to Ma and Ardell [71] and Dupin et al. [70], and $C_{Al}^{\gamma'}(t)$ approaches the solvus curve due to Saunders [68] for the partial Ni-Al phase diagram, as indicated by the heavy black horizontal arrows, Fig. 2.

- In contrast to the conventional wisdom that the second-phase γ'(L1$_2$)-precipitates should nucleate with a composition that is close to its equilibrium composition [104], we find that they nucleate far from their equilibrium composition and then evolve temporally toward it, Fig. 2, where the Al concentration trajectory is denoted by the heavy black horizontal arrow. *This indicates the nucleation is occurring by a different mechanism than the one commonly found in review articles and text books on phase transformations, as opposed to what we find experimentally using atom-probe tomography and vacancy-mediated lattice kinetic Monte Carlo simulations*.
- The PSDs evolve temporally with increasing aging time from a narrow PSD in regime I to a wider distribution in regimes II and III, and finally to a narrow PSD in regime IV, Fig 8. This last PSD is similar to the the model PSDs generated from the Akaiwa-Voorhees (AV) simulations and Brailsford-Wynblatt (BW) calculations based on our semi-quantitative observations of PSDs.
- The maximum value of $N_v(t)$, Fig. 5b, corresponds to the maximum value of the fraction of γ'(L1$_2$)-precipitates interconnected by necks, $f(t)$, 39.68±2.81%, and the minimum value of $\langle \lambda_{edge-edge}(t) \rangle$, Fig. 9. These results are consistent with a coagulation and coalescence mechanism of coarsening, over a range of aging times about the maximum value of $f(t)$, as opposed to the classic evaporation-condensation ("the large precipitates eat the small precipitates") mechanism, which is assumed implicitly in the LSW and CVMK mean-field models.
- Monovacancy-mediated LKMC$_1$ simulations are in reasonably good agreement with all the experimental 3-D APT data for all measured physical quantities, Figs. 2, 5, 6, 7, 9, 10, and 11.
- The monovacancy-mediated LKMC$_1$ results are consistent with experimental results for $C^{\gamma'}_{Al}(t)$, Fig. 6b, but underestimate slightly $C^{\gamma}_{Al}(t)$ for $t>0.25$ h, Fig. 6a. Additionally, the monovacancy-mediated LKMC$_1$ results are in better agreement with the experimental 3-D APT data for aging times longer than 4 h.
- An important assumption of the so-called trans-interface diffusion-coarsening (TIDC) model concerns the dependence of the {100}-type interfacial composition width, $\delta(t)$, on aging time and $\langle R(t) \rangle$, which are in disagreement with our 3-D APT experimental and monovacancy-mediated LKMC$_1$ simulation results for this Ni-Al alloy. The value of $\delta(t)$ decreases with increasing $t$ and $\langle R(t) \rangle$, Fig. 11, as $t^{-0.08\pm0.01}$ and $\langle R(t) \rangle^{-0.47\pm0.03}$, respectively, but is never equal to zero.
- The value of $\sigma^{\gamma/\gamma'}$ is calculated from the coarsening data using the rate constants determined from the experimental 3-D APT data for $\langle R(t) \rangle$ and $\Delta C^{\gamma}_{Al}(t)$, and two different values of the curvature of the molar Gibbs free energy of mixing with respect to concentration, yielding values of 29.94 ± 1.69 and 28.55 ± 1.61 mJ m$^{-2}$, respectively, which are consistent with our prior first-principles calculations performed on the γ(Ni)/γ'(Ni$_3$Al) interface for the {100}-, {110}- and {111}-type interfaces at 823 K (550 ºC ) [45].

- It is suggested strongly that all experimental values of $\sigma^{\gamma/\gamma'}$ determined that do not use Ardell's approach (1995) not be taken seriously because they all utilize a value of a diffusivity from the archival literature, rather than determining both values independently of one another.

- The diffusivity, $D_{experiment}^{coarsening}$, at 823 K (550 ºC) is calculated using the rate constants obtained from the 3-D APT experiments for $\langle R(t) \rangle$ and $\Delta C_{Al}^{\gamma}(t)$, which is equal to 2.51 ± 1.41 x 10$^{-21}$ m$^2$ s$^{-1}$.

- The calculated diffusivities for the entire monovacancy-mediated LKMC$_1$ supercell, $D_{Al}^{supercell}$ and $D_{Ni}^{supercell}$, are determined by counting the number of jumps of Al or Ni atoms for exchanges with the monovacancy. For both Al and Ni, $D_i^{supercell}$ is closer to the calculated $D_i^{\gamma-matrix}$ value than it is to the calculated $D_i^{\gamma'-precipitates}$ because of the small values of $\phi_{\gamma'}(t)$ of the γ'(L1$_2$)-precipitates at 1 and 4 h, 2.14 ± 0.26 and 2.12 ± 0.26%, respectively, Table 4. The volume fraction of the partially ordered necks connecting the γ'(L1$_2$)-precipitates is smaller than $\phi_{\gamma'}(t)$. The value of $\phi_{\gamma'}(t)$ at 400 h is closer to its equilibrium value, 13.48%, where approximately 90% of the volume is the γ(f.c.c.)-matrix, and thus $D_i^{supercell}$ is closer to $D_i^{\gamma-matrix}$ than it is to $D_i^{\gamma'-precipitates}$.

- $D_{experiment}^{coarsening}$ is equal to the value of the inter-diffusion coefficient, $\tilde{D}$, derived from the monovacancy-mediated LKMC$_1$ calculations for 1 and 4 h, 3.75 ± 0.31 x 10$^{-22}$ and 3.62 ± 0.31 x 10$^{-22}$ m$^2$ s$^{-1}$, respectively, while $\tilde{D}$ at 400 h, 2.90 ± 0.25 x 10$^{-22}$ m$^2$ s$^{-1}$, is 15% of $D_{experiment}^{coarsening}$. From this semi-quantitative agreement between simulations and experiment we conclude that our experimental results permit us to use the experimentally obtained coarsening rate constants to calculate the inter-diffusion coefficient of Ni and Al in our binary Ni-12.5 Al at.% alloy. This is an important result, as there are many examples in the archival literature where the proper diffusivity for coarsening is stated to be $D$, without any indication as to whether it is an inter-diffusivity or the diffusivity of the solute species.

## 6. Acknowledgements

This research was supported by the National Science Foundation (NSF), Division of Materials Research (DMR) grant number DMR-1610367 001; Profs. Diana Farkas and Gary Shiflet, grant monitors. Atom-probe tomography was performed at the Northwestern University Center for Atom-Probe Tomography (NUCAPT). The LEAP tomograph at NUCAPT was purchased and upgraded with grants from the NSF-MRI (DMR-0420532) and ONR-DURIP (N00014-0400798, N00014-0610539, N00014-0910781, N00014-1712870) programs. NUCAPT received support

from the MRSEC program (NSF DMR-1720139) at the Materials Research Center, the SHyNE Resource (NSF ECCS-1542205), and the Initiative for Sustainability and Energy (ISEN) at Northwestern University. A portion of this research was performed at the DuPont-Northwestern-Dow Collaborative Access Team (DND-CAT) located at Sector 5 of the Advanced Photon Source (APS). DND-CAT is supported by E.I. DuPont de Nemours & Co., The Dow Chemical Company, and Northwestern University. Use of the APS, an Office of Science User Facility operated for the U.S. Department of Energy (DOE) Office of Science by Argonne National Laboratory, was supported by the U.S. DOE under Contract No. DE-AC02-06CH11357. Ms. Elizaveta Y. Plotnikov was initially supported by a W. P. Murphy Fellowship and then this NSF grant. Drs. Zugang Mao and Sung-Il Baik were partially supported by this NSF grant. Prof. Yongsheng Li was supported by the China Scholarship Council. Dr. Mehmet Yildirim was supported by the Scientific HR Development Program of the Middle East Technical University. The authors thank Dr. Nathalie Dupin for generous access to her Thermo-Calc data for the partial Ni-Al phase diagram. Additionally, Prof. Peter Voorhees is thanked for stimulating and enlightening discussions and Prof. Pascal Bellon is thanked for important discussions concerning correlated diffusion. Mr. Pavithran Maris (visiting undergraduate scholar) is thanked for helping with atom-probe tomography experiments during the summer of 2013; Dr. John Thompson for calculating the AV PSD and $F(\phi_\gamma)$; and Ms. Yanyan (Ashley) Huang for assisting in performing Vickers microhardness measurements.

**Appendix A: Methodologies for Calculating Temporal Exponents from Experimental Data**

*A1.    Determination of the temporal exponent of the mean radius, $\langle R(t) \rangle$*

First, we emphasize that while Fig. 5 is plotted on a log-log scale, *we do not use log-log plots to determine the temporal exponent, $p$, from Eq. (1)*. While the data in this article is presented in a log-log format for clarity, $p$ is always calculated using a nonlinear multivariate regression analysis [58] of the LSW relationship for $\langle R(t) \rangle$, Eq.(1) *with no assumptions being made about the values of $p$ or the rate constant, $K$*. The Solver package in Microsoft Excel is utilized to vary these parameters to minimize the residual squared error when the data is fit to Eq. 1. Thus, $p$ can be calculated with the same accuracy independent of how the data is finally plotted. Our approach is more accurate than making $\langle R(t) \rangle^p$ versus time plots for different values of $p$ and stating that the correct temporal exponent is the value of $p$ value that yields the largest coefficient of determination, $\xi^2$ [152]. To illustrate this point, Fig. A1 displays $\langle R(t) \rangle^p$ versus time plots for Ni-12.5 Al at.% aged at 823 K (550 ºC) for the $p$ values 2, 2.4, 3, and 4, as well as the associated linear fits. This is not, however, statistically the best way to find which exponent yields the best fit to the data. While a $\xi^2$ value of 0.99 is certainly better than a $\xi^2$ value of say 0.70, all this quantity tells us is how much better a linear fit to the data is than simply taking the mean of all the data points. The quantity $\xi^2$ is calculated using two different sums of squares, the total sum of squares, $SS_{tot}$, and the sum of square residuals, $SS_{res}$:

$$SS_{tot} = \sum_i \left(y_i - \langle y \rangle\right)^2 \tag{A1}$$

$$SS_{res} = \sum_i \left(f_i - y_i\right)^2 \tag{A2}$$

$$\xi^2 = 1 - \frac{SS_{res}}{SS_{tot}} \tag{A3}$$

where $y_i$ is a measured quantity (in this case $\langle R(t) \rangle^p$ at a given aging time); $\langle y \rangle$ is the mean of all $y_i$ values; and $f_i$ is the associated modeled value (in this case from a linear fit) for each data point [160, 161]. Thus, $\xi^2$ is by definition a number between 0 and 1 that measures how much better a given model is than a control (i.e., a horizontal line), but it is not particularly useful for determining whether one model is better than another. Fig. A1a-c demonstrates that fitting either $\langle R(t) \rangle^2$, $\langle R(t) \rangle^{2.4}$, or $\langle R(t) \rangle^3$, respectively, versus time for a linear model produces approximately the same value of $\xi^2$, which implies that each value of $p$ is equally probable. Any fitting technique where plotting the cube of some measured quantity produces approximately the same result as plotting the square of the same quantity cannot be used to judge the true temporal behavior of the model. Furthermore, the best fit linear equation for Fig. A1c is $\langle R(t) \rangle^3 = 0.6827t - 2.7054$, implying that a linear fit predicts a negative value of $\langle R(t) \rangle$ at $t = 0$, which is physically impossible. When $\langle R(t) \rangle^4$ is plotted versus time, Fig. A1b, $\xi^2$ is still relatively high, 0.910. In each case, the coefficient of determination is high simply because $\langle R(t) \rangle$ is increasing rather than remaining constant with time, but this isn't necessarily indicative of a good fit because $\langle R(t) \rangle$ is not increasing *linearly* with time. Therefore, an $\langle R(t) \rangle^p$ versus time plot is not only a misleading approach for analyzing data, but it is also a poor predictor of how the data actually behave. Ideally, one should use a Box-Cox transformation to find the value of $p$ that yields a regression error of constant variance [162, 163], but our data for $\langle R(t) \rangle$ contains too few data points to determine whether or not the regression error variance is a constant.

To summarize, determining $p$ using a nonlinear multivariate regression analysis is the most appropriate and accurate way of analyzing the data as opposed to picking a value of $p$ and then plotting $\langle R(t) \rangle^p$ versus time and calculating the coefficient of determination.

## A2. Determination of the temporal exponent of the supersaturations, $\Delta C_{Al}^{\gamma}(t)$ and $\Delta C_{Al}^{\gamma'}(t)$

The same approach used above also applies for determining the temporal exponents for the values of $\Delta C_{Al}^{\gamma}(t)$ and $\Delta C_{Al}^{\gamma'}(t)$. This is sometimes performed in the archival literature by plotting the Al concentration versus $t^{-1/2}$ or $t^{-1/3}$ and using the plot that yields the highest value of $\xi^2$ as the temporal exponent [152], as opposed to the method we employ, which involves plotting $\Delta C_{Al}^{\gamma}(t)$ or $\Delta C_{Al}^{\gamma'}(t)$ versus time and determining the temporal exponent, $r$, from a nonlinear multivariate regression analysis [58] of the experimental 3-D APT data, Figs. 6c and 6d. To demonstrate this point, Fig. A2 displays a plot of $\Delta C_{Al}^{\gamma}(t)$ versus $t^{-1/2}$, $t^{-1/2.4}$, $t^{-1/3}$, and $t^{-1/4}$ with the corresponding values of $\xi^2$ indicated on the graphs. While one may conclude that $t^{-1/3}$ is the best fit to the data than the other temporal exponents, $t^{-1/2}$, $t^{-1/2.4}$, and $t^{-1/4}$ ($\xi^2=0.952$ versus $\xi^2=0.911, 0.937$, and $0.944$, respectively), it is clear that the data plotted this way is linear, and thus it cannot be used to identify accurately the correct value of $r$. Therefore, in conclusion, the most appropriate and accurate way of plotting the data is to plot $\Delta C_{Al}^{\gamma}(t)$ or $\Delta C_{Al}^{\gamma'}(t)$ versus time and determine the temporal exponent using a nonlinear multivariate regression analysis [58].

**Appendix B: Normalization of Composition Measurements**

The preferential evaporation of Ni in certain cases can affect the measured composition of a data set obtained using 3-D APT. For electropolished specimens whose shank angle may not always be the same, this preferential evaporation can lead to a difference in the overall measured specimen composition for the same alloy. To compare data sets from different aging times, the overall composition of each dataset is normalized to the composition obtained by the ICP-AES analytical measurements. Schmuck et al. performed such a normalization of the overall composition of each dataset [9]. This method leads, however, to fluctuations in the supersaturation measured for each dataset, which are inconsistent with the LSW and CVMK prediction of a continuous decrease.

Therefore, a normalization procedure based on the concentrations obtained for each phase using the proximity histogram method [84, 85] is employed. For a binary alloy, the concentration of the solute, Al, can be calculated using the Al concentrations in each of the two phases from:

$$C_{Al} = \phi_{\gamma'} C_{Al}^{\gamma'} + (1-\phi_{\gamma'}) C_{Al}^{\gamma} \qquad (B1)$$

Where $C_{Al}$ is calculated for each dataset using the values of $C_{Al}^{\gamma}$ and $C_{Al}^{\gamma'}$ taken from the plateaus of the proximity histogram in each phase. This value is then normalized to the ICP-AES analytical value that is measured for this alloy.

**Appendix C: Empirical analyses of experimental precipitate size distributions (PSDs)**

The experimental PSDs, Fig. 8, are examined mathematically using the Bhattacharyya coefficient (BC) [164] and the Kolmogorov-Smirnov (KS) test [165] to determine how well they fit the four theoretical PSDs: (1) LSW [32, 33]; (2) modified LSW [54]; (3) BW [55]; and (4) AV [56]. The BC describes how close two probability distributions are to one another; the distributions may be discrete or continuous. If they are discrete, the two distributions must have the same number of bins and same bin size. The BC is a number between 0 and 1, where 0 indicates no overlap at all between the two distributions, and 1 implies that the two distributions are identical. When comparing the experimental PSD in Fig. 8 with each theoretical distribution the BC is relatively high for each aging time (>0.80), but it is consistently higher for the modified LSW, BW, and AV PSDs than for the LSW PSD. The BCs for these latter three distributions are approximately equal to one another at each aging time, which is a result of their closeness to one another. For our analyses, we care greatly about the exact shapes of the distributions, about which the BC is agnostic. Additionally, the BC method yields more accurate information if there are more bins. Our experimental data is divided into 15 bins from 0 to 3 (bin size = 0.2). If we had a much larger data set (for example, many, many thousands of precipitates), we would be able to utilize more bins, and the BC would become a useful number for determining how the different distributions fit our PSDs. At long aging times we can only obtain datasets containing about 100 precipitates, which limits the accuracy of the PSDs. Additionally, the BC does not change with aging time, even though our experimental PSDs are temporally evolving. If we use the BC to determine how close the system is to equilibrium, we would conclude that it is the same at 0.25 h as at 4096 h, given that they are the same. From all of our analyses we know, however, that this is untrue. Therefore, while the BC indicates strongly that the LSW PSD is not a good fit to our PSDs, Fig. 8, the PSDs contain an insufficient number of data points to prove conclusively that one of the other three theoretical PSDs is the best one.

The KS test is a method for assessing if a certain distribution has the potential to have come from a larger theoretical distribution within some confidence interval. We find that our PSDs do not conform to any of the four theoretical distributions until 1024 h of aging, at which point the three remaining theoretical PSDs are all candidates for the experimental PSDs with a 0.01 level of significance, indicating that the PSDs enter a quasi-stationary coarsening regime at approximately 1024 h. Given, however, the small number of precipitates in each bin in our PSDs, we need significantly more APT results to rank conclusively one of the theoretical PSDs above another in terms of the maximum deviation alone. The small number density of precipitates, $N_v(t)$, in the long aging time datasets, in regime IV, decreases as $t^{-1}$, and $\langle R(t) \rangle$ increases as $t^{1/3}$, implying that each dataset (whose size is a constant) contains fewer and fewer precipitates that are becoming larger and larger. Qualitatively we find that the BW and AV PSDs fit the experimental PSDs best at the longest aging time.

**Appendix D: The complex flux diffusion theory**

A model Ni-Al alloy is specified by the concentrations of the atomic species, $C_{Ni}$, $C_{Al}$, and the concentration of monovacancies, $C_V$; these three concentrations, in atomic fraction, sum to unity. Since the γ'(L1$_2$)-precipitates are coherent with the γ(f.c.c.)-matrix, lattice sites are conserved locally during phase-separation. Because the mean edge-to-edge distance between γ'(L1$_2$)-precipitates is small compared to the inter-dislocation spacing, lattice sites are conserved at all space and time scales, which are given by the mean spacing between dislocations, $10^3$ nm (one micron), and the time it takes for vacancies to diffuse that distance. Hence, a microstructure, specifically, the 3-D spatial distribution of γ'(L1$_2$)-precipitates, their compositions, the mean composition of the γ(f.c.c.)-matrix, and the Ni and Al concentration profiles immediately adjacent to γ'(L1$_2$)-precipitates) are defined by two independent composition fields. There are two diffusion potential fields, $(\mu_{Ni} - \mu_V)$ and $(\mu_{Al} - \mu_V)$, which drive three independent diffusional-fluxes, whose magnitudes imply three Onsager coefficients, $L_{NiNi}$, $L_{AlAl}$, $L_{NiAl}$. The flux of matter, in the lattice frame of reference, is given by:

$$\tilde{J} = -\bar{\bar{L}} \nabla \tilde{\mu} (\Omega kT)^{-1} = -\bar{\bar{D}} \nabla \tilde{C} \Omega^{-1}; \tag{D1}$$

where $\tilde{J}$ is a column vector with elements $J_{Ni}$ and $J_{Al}$, similarly for the diffusion potential, $\tilde{\mu}$, the $i^{th}$ component of which is $(\mu_i - \mu_V)$, and the concentrations of atomic species, $\tilde{C}$; $\Omega$ is the atomic volume of an atom in the γ(f.c.c)-matrix phase and kT has its standard meaning. Since lattice sites are conserved far from the dislocations, that have a mean-spacing, $10^3$ nm, which is significantly greater than the mean edge-to-edge inter-precipitate distance, the vacancy flux, $J_v$, in the lattice frame of reference is the negative of the sum of the two elemental solute fluxes: that is, $-J_v = J_{Ni} + J_{Al}$. The diffusion matrix, $\bar{\bar{D}}$, is given by the product of the kinetic factor, described by the Onsager matrix, $\bar{\bar{L}}$, and the thermodynamic factor, imbedded in the susceptibility matrix, $\bar{\bar{\chi}}$:

$$\bar{\bar{D}} = \bar{\bar{L}} \frac{\bar{\bar{\chi}}}{kT} \text{, where} \quad \chi_{ij} = \overline{\left(\frac{\partial(\mu_i - \mu_v)}{\partial C_j}\right)} \text{ with } i,j = \text{Ni, Al.} \tag{D2}$$

The diffusion matrix is in general *non-diagonal*, that is, according to Eq. (D1), the flux of any species *i* is a linear combination of all the concentration gradients. Two distinct physical processes contribute to the off-diagonal terms of the diffusion matrix:

(a) Firstly, the chemical potential of species *i* depends on the concentrations of all other species, as embedded in the off-diagonal terms of the susceptibility matrix, $\chi_{i \neq j}$; and

(b) Secondly, even in the absence of the above effect (that is, $\chi_{i \neq j} = 0$), the diffusion mechanism (monovacancy jumps) introduces kinetic couplings among the fluxes; these induced fluxes are described by the off-diagonal terms of the Onsager matrix. Specifically, $\chi_{i \neq j} = 0$, but $L_{i \neq j} \neq 0$ yields $D_{i \neq j} \neq 0$.

The $\bar{\bar{L}}$ and $\bar{\bar{\chi}}$ matrices are both symmetrical. They can be calculated by Monte Carlo techniques for LKMC$_1$. We have used the semi-grand Canonical Monte Carlo technique (SGCMC) [166] to estimate the equilibrium solid-solution composition, which is chosen as the terminal solid-solution composition expected after the completion of phase separation at long aging times. The $\bar{\bar{L}}$ matrix is calculated, following Einstein's definition, along an LKMC trajectory in the equilibrium solid-solution [167] The diffusion potential potentials, $(\mu_i - \mu_v)$ are computed by the SGCMC technique applied to a ternary solution consisting of the two chemical components and vacancies. The $\bar{\bar{\chi}}$ matrix is obtained by first computing the two diffusion potentials for five distinct compositions in the vicinity of the equilibrium composition; we then perform a linear multivariate regression vs. composition changes. The same technique was used to determine the values of the susceptibilities in the $\gamma'(L1_2)$-phase, $G'_{ij}$.

# Tables

**Table 1:** Summary of prior studies conducted to determine the interfacial free energy, $\sigma^{\gamma/\gamma'}$, between the $\gamma$(f.c.c.) and $\gamma'$(L1$_2$)-phases. All the values are displayed graphically in Fig. 1.

| Alloy Composition (at.%) | Aging Temperature (K) | Interfacial Free Energy (mJ m$^{-2}$) | Method | Reference |
|---|---|---|---|---|
| Ni-13.5 Al | 898 | 31.2 | Electron microscopy | [34] |
| Ni-13.5 Al | 1023 | 27.2 | Electron microscopy | [34] |
| Ni-13.5 Al | 1048 | 32.3 | Electron microscopy | [34] |
| Ni-13.1 Al | 898 | 14.4 | Ferromagnetic Curie temperature measurements | [36] |
| Ni-12.8 Al | 988 | 14.2 | Ferromagnetic Curie temperature measurements | [36] |
| Ni-14.1 Al | 1073 | 6.2 | Magnetic analysis and transmission electron microscopy | [37] |
| Ni-15.9 Al | 1073 | 8.9 | Magnetic analysis and transmission electron microscopy | [37] |
| Ni-17.7 Al | 1073 | 11.9 | Magnetic analysis and transmission electron microscopy | [37] |
| Ni-19.3 Al | 1073 | 8.3 | Magnetic analysis and transmission electron microscopy | [37] |
| Ni-12.3 Al | 943 | 17.4 | Electron microscopy | [38] |
| Ni-12.3 Al | 953 | 16.6 | Electron microscopy | [38] |
| Ni-12.3 Al | 963 | 19.8 | Electron microscopy | [38] |
| Ni-12.3 Al | 968 | 24.3 | Electron microscopy | [38] |
| Ni-14 Al | 823 | 19 | Atom-probe field microscopy | [168] |
| Ni-12.5 Al | 823 | 16.9 | X-ray diffraction | [39] |
| Ni-12.5 Al | 873 | 21.7 | X-ray diffraction | [39] |
| Ni-12.5 Al | 923 | 16.6 | X-ray diffraction | [39] |
| Ni-12.5 Al | 973 | 10.3 | X-ray diffraction | [39] |
| Ni-12 Al | 773 | 14 | High resolution electron microscopy | [135] |
| Ni-12.8 Al | 848 | 42±2 | Small angle neutron scattering and transmission electron microscopy | [40] |
| Ni-12.8 Al | 863 | 68±6, 80±8 | Small angle neutron scattering and transmission electron microscopy | [40] |
| Ni-13.1 Al | 898 | 8.2 | Reanalysis of [36] | [41] |
| Ni-12.8 Al | 988 | 8.0 | Reanalysis of [36] | [41] |
| Ni-12.5 Al | 823 | 8.7 | Reanalysis of [39] | [41] |
| Ni-12.5 Al | 873 | 8.6 | Reanalysis of [39] | [41] |
| Ni-12.5 Al | 923 | 3.0 | Reanalysis of [39] | [41] |
| Ni-12.5 Al | 973 | 0.9 | Reanalysis of [39] | [41] |
| Ni-12.86 Al | 898 | 4.29±0.35 | Reanalysis of [169] | [152] |
| Ni-12.86 Al | 988 | 3.71±0.74 | Reanalysis of [169] | [152] |
| Ni-13.1 Al | 898 | 22.33±1.31 | Reanalysis of [36] | [42] |
| Ni-12.8 Al | 988 | 19.52±0.90 | Reanalysis of [36] | [42] |

**Table 2**: Summary of computationally determined values of $\sigma^{\gamma/\gamma'}$ (mJ m$^{-2}$) for the {100}-, {110}-, and {111}-type planes for the γ(Ni)/γ'(Ni$_3$Al) interface in Ni-Al alloys.

| Method | Temperature (K) | {100} | {110} | {111} | Reference |
| --- | --- | --- | --- | --- | --- |
| Density functional theory | 0 | 63 | --- | --- | [43] |
| Embedded atom method | 700 | 46 | 28 | 12 | [44] |
| First-principles calculations | 823[a] | 23.11 | 26.83 | 28.42 | [45] |
| First-principles calculations | 823 | 16.5 | 17.7 | 15.9 | [46] |
| Capillary fluctuation method | 800 | 14 | --- | --- | [47] |

[a]In reference [45] $\sigma^{\gamma/\gamma'}$ is determined as a function of {hkl}, for temperatures between 0 and 1100 K, plus the effects of coherency strain, phonon vibrational entropy, and ferromagnetism are included, which makes this the most complete study to date.

**Table 3:** Thermodynamic and kinetic parameters used for monovacancy-mediated LKMC$_1$ simulations: (1) pair-wise interactions between atoms out to the 4$^{th}$ NN, $\varepsilon_{i-j}^k$ ; (2) saddle-point energies for Ni and Al, $E_{sp-p,q}^i$ ; (3) attempt frequencies for Ni and Al, $v^i$. The $\varepsilon_{i-j}^k$ values are calculated from first-principles [61, 62], while the $E_{sp-p,q}^i$ and $v^i$ values are from [11]; and (4) monovacancy-Al (V-Al) binding energies out to 4$^{th}$ NN distances, LKMC$_1$ and the V-Al binding energy for 1$^{st}$ NN, LKMC$_2$. A positive value of $\varepsilon_{i-j}^k$ or $E_{sp-p,q}^i$ indicates a repulsive force, while a negative value designates an attractive force, and similarly for V-Al binding energies.

**Thermodynamic parameters**

| $\varepsilon_{i-j}^k$ (eV) | Ni-Ni | Al-Al | Ni-Al | V-Al (LKMC$_1$) | V-Al (LKMC$_2$) |
|---|---|---|---|---|---|
| 1$^{st}$ NN | -0.7485 | -0.6845 | -0.7495 | -0.055 | -0.055 |
| 2$^{nd}$ NN | -0.0135 | -0.0265 | -0.0349 | -0.048 | 0 |
| 3$^{rd}$ NN | 0.0142 | 0.0084 | -0.0285 | 0.042 | 0 |
| 4$^{th}$ NN | -0.0066 | -0.0121 | 0.0125 | -0.019 | 0 |

**Kinetic parameters**

| | Ni | Al |
|---|---|---|
| $E_{sp-p,q}^i$ (eV) | -9.750 | -9.412 |
| $v^i$ (s$^{-1}$) | 1.10 x 10$^{15}$ | 1.10 x 10$^{15}$ |

**Table 4:** The temporal evolution of the following quantities for the γ'(L1$_2$)-precipitates: volume fraction, $\phi_{\gamma'}(t)$; number density, $N_v(t)$; mean radius, $<R(t)>$; fraction of γ'(L1$_2$)-precipitates interconnected by necks, $f(t)$; and the minimum edge-to-edge distance between γ'(L1$_2$)-precipitates, $<\lambda_{edge\text{-}to\text{-}edge}(t)>$, and their standard errors, 2σ, for Ni-12.5 Al at.% aged at 823 K.

| Aging time (h) | $N_{ppt}$ [a] | $\langle R(t) \rangle \pm 2\sigma$ (nm) | $N_v(t) \pm 2\sigma$ (10$^{23}$ m$^{-3}$) | $\phi_{\gamma'}(t) \pm 2\sigma$ (%) | $f(t) \pm 2\sigma$ (%) | $\langle \lambda_{edge-edge}(t) \rangle \pm 2\sigma$ (nm) |
|---|---|---|---|---|---|---|
| 0.08 | 16 | 0.79 ± 0.20 | 0.08 ± 0.02 | 0.002 ± 0.001 | ND[b] | 272.71 ± 15.19 |
| 0.17 | 24 | 0.78 ± 0.16 | 0.14 ± 0.03 | 0.003 ± 0.001 | 4.88 ± 0.43 | 218.32 ± 10.11 |
| 0.25 | 1808 | 1.34 ± 0.03 | 2.72 ± 0.06 | 1.19 ± 0.03 | 20.63 ± 1.46 | 18.47 ± 0.39 |
| 0.5 | 2918 | 1.36 ± 0.03 | 5.74 ± 0.11 | 1.89 ± 0.04 | 34.29 ± 2.42 | 20.81 ± 0.23 |
| 1 | 3333.5 | 1.47 ± 0.03 | 6.87 ± 0.12 | 2.87 ± 0.05 | 39.68 ± 2.81 | 13.14 ± 0.19 |
| 2 | 2119 | 2.06 ± 0.04 | 6.09 ± 0.13 | 2.33 ± 0.08 | 39.53 ± 2.80 | 10.37 ± 0.23 |
| 4 | 2389.5 | 2.29 ± 0.10 | 5.78 ± 0.26 | 3.64 ± 0.16 | 33.42 ± 0.68 | 14.94 ± 0.20 |
| 16 | 1289.5 | 3.11 ± 0.16 | 3.82 ± 0.22 | 8.62 ± 0.24 | 34.16 ± 2.42 | 12.97 ± 0.32 |
| 64 | 1725 | 3.46 ± 0.10 | 2.67 ± 0.21 | 9.23 ± 0.24 | 25.22 ± 1.78 | 13.99 ± 0.30 |
| 256 | 657.5 | 5.65 ± 0.22 | 0.85 ± 0.03 | 10.00 ± 0.39 | 18.46 ± 2.40 | 18.66 ± 0.66 |
| 1024 | 211 | 9.43 ± 0.65 | 0.24 ± 0.02 | 11.96 ± 0.82 | 15.00 ± 2.47 | 26.39 ± 1.67 |
| 2607 | 153 | 10.72 ± 0.87 | 0.17 ± 0.01 | 13.37 ± 1.08 | 5.13 ± 0.83 | 26.80 ± 2.18 |
| 4096 | 127 | 14.02 ± 2.14 | 0.08 ± 0.01 | 12.80 ± 1.40 | ND[b] | 37.38 ± 4.15 |

[a]The number of γ'(L1$_2$)-precipitates analyzed, $N_{ppt}$, is smaller than the total number of γ'(L1$_2$)-precipitates detected by atom-probe tomography (APT). γ'(L1$_2$)-precipitates that intersect the sample volume's surface contribute 0.5 to the quantity $N_{ppt}$, and are included in estimates of $N_v(t)$ and $\phi_{\gamma'}(t) \pm 2\sigma$, plus the phase compositions, but not in the measurement of $\langle R(t) \rangle$. 3-D APT requires a small (<50 nm) nanotip radius [81] and the 3-D reconstructions have to first-order a cylindrical volume. As a result, once $\langle R(t) \rangle$ intersects the cylinder's surface, a fraction of γ'(L1$_2$)-precipitates are not fully enclosed within the 3-D reconstruction. As a result, while 78% of γ'(L1$_2$)-precipitates in the 0.08 h data set are fully enclosed within the 3-D reconstruction, only 26% of the γ'(L1$_2$)-precipitates in the 4096 h data set are fully enclosed within the 3-D reconstruction. This accounts for the difference between the number of γ'(L1$_2$)-precipitates listed for each PSD in Fig. 8 and the total number of γ'(L1$_2$)-precipitates listed for each aging time.
[b]ND means not detected

**Table 5:** Temporal evolution of the γ(f.c.c.)- and γ'(L1$_2$)-phase compositions as a function of aging time for Ni-12.5 Al at.% aged at 823 K (550 °C), as measured by 3-D APT. Because this is a binary alloy, the concentrations of Ni and Al at each aging time sum to 100 at.%, and therefore only the Al concentrations are listed.

| Aging time (h) | $C_{Al}^{\gamma}$ (at.%) | $C_{Al}^{\gamma'}$ (at.%) |
| --- | --- | --- |
| 0.08 | 12.5 ± 3.12 | 33.97 ± 8.49 |
| 0.17 | 12.5 ± 2.55 | 30.87 ± 6.3 |
| 0.25 | 12.74 ± 0.3 | 28.02 ± 0.66 |
| 0.5 | 12.68 ± 0.23 | 26.13 ± 0.48 |
| 1 | 12.56 ± 0.22 | 25.93 ± 0.45 |
| 2 | 12.18 ± 0.26 | 24.6 ± 0.53 |
| 4 | 12.19 ± 0.25 | 26.06 ± 0.53 |
| 16 | 11.74 ± 0.33 | 24.54 ± 0.68 |
| 64 | 11.6 ± 0.28 | 24.26 ± 0.58 |
| 256 | 11.43 ± 0.45 | 23.95 ± 0.93 |
| 1024 | 11.25 ± 0.77 | 23.77 ± 1.64 |
| 2607 | 11.37 ± 0.10 | 23.83 ± 0.07 |
| 4096 | 11.25 ±0.09 | 23.73± 0.07 |

**Table 6:** Relative rescaled monovacancy-mediated lattice-kinetic Monte Carlo (LKMC$_1$) simulations times spent by the monovacancy in the γ(f.c.c.)-matrix, the γ'(L1$_2$)-precipitates, the neck between two γ'(L1$_2$)-precipitates, and in the supercell for the aging times 1, 4 and 400 h, Fig. 12, are normalized to the volume of the supercell. The different diffusivities for Al and Ni for each region, $D_i^{region}$, calculated utilizing monovacancy-mediated LKMC$_1$ simulations, Eq. (12), for 1, 4, and 400 h, are listed in the last three columns. Too few interconnected γ'(L1$_2$)-precipitates are detected for 400 h to yield satisfactory statistics, and therefore there aren't $D_i^{necks}$ values listed. All values include a correlation factor for a monovacancy diffusion mechanism in a f.c.c. lattice, 0.611.

| Region | Ratio of relative travel time, $t/t_{matrix}$ (normalized to volume of cell) | | $D_{Al}^{region}$ (x 10$^{-21}$ m$^2$ s$^{-1}$) | | | $D_{Ni}^{region}$ (x 10$^{-22}$ m$^2$ s$^{-1}$) | | |
|---|---|---|---|---|---|---|---|---|
| | 1 h | 4 h | 1 h | 4 h | 400 h | 1 h | 4 h | 400 h |
| 1. γ(f.c.c.)-matrix | 1 | 1 | 2.22±0.38 | 2.18±0.36 | 1.62±0.26 | 1.11±0.14 | 1.02±0.38 | 0.99±0.14 |
| 2. γ'(L1$_2$)-precipitates | 3.12 | 3.25 | 1.22±0.17 | 1.16±0.15 | 0.71±0.09 | 0.57±0.06 | 0.45±0.17 | 0.33±0.03 |
| 3. Neck region | 3.85 | 3.58 | 1.11±0.19 | 1.15±0.19 | --- | 0.49±0.07 | 0.53±0.19 | --- |
| 4. supercell | --- | --- | 2.19±0.26 | 2.04±0.26 | 1.51±0.14 | 1.07±0.11 | 0.98±0.26 | 0.93±0.09 |

**Table 7:** Equilibrium volume fraction of the γ'(L1$_2$)-phase, $\phi_{\gamma'}^{eq}$, as determined by: (a) Thermo-Calc using two different thermodynamic databases (Saunders and Dupin et al.); (b) the Ni-Al phase diagram (Ma and Ardell); (c) experimental 3-D atom-probe tomographic data from this study at 4096 h; and (d) our grand canonical Monte Carlo (GCMC) simulation for Ni-12.5 Al at.% aged at 823 K.

| Source | $\phi_{\gamma'}^{eq}$ |
| --- | --- |
| Saunders [68] | 18.33 % |
| Dupin et al. [70] | 13.48 % |
| Ma and Ardell [71] | 15.02 % |
| 3-D APT data for 4096 h | 12.59 ± 1.40 % |
| GCMC [22] | 16.56 % |

**Table 8**: Comparison of calculated and atom-probe tomographic measured values of the stationary nucleation current, number of nuclei per unit volume per unit time, from the archival literature and current research. The ratio of $J_{CNT}^{st}$, calculated assuming classical nucleation theory (CNT) to $J_{experiment}^{st}$ measured experimentally is given for each case. For the Ni-12 Al at.% alloy $J_{experiment}^{st}$ was measured using high-resolution transmission electron microscopy [135], whereas for the Ni-12.5 Al at.% and three Ni-Al-Cr alloys atom-probe tomography was utilized.

| Alloy composition (at.%) | Aging Temperature (K) | $J_{CNT}^{st}$ from CNT (m$^{-3}$ s$^{-1}$) | $J_{experiment}^{st}$ from experimental data (m$^{-3}$ s$^{-1}$) | Ratio of $J_{CNT}^{st}$ to $J_{experiment}^{st}$ | $N_{effective}/N$ (%) | Reference |
|---|---|---|---|---|---|---|
| Ni-12 Al | 773 | 4.1 x 10$^{22}$ | 8.4 x 10$^{19}$ | 488 | 0.21 | [135] |
| Ni-7.5 Al-8.5 Cr | 873 | 4.0 x 10$^{22}$ | 5.4 x 10$^{21}$ | 7 | 14.29 | [22] |
| Ni-5.2 Al-14.2 Cr | 873 | 3.2 x 10$^{23}$ | 5.9 x 10$^{21}$ | 54 | 1.85 | [22] |
| Ni-6.5 Al-9.5 Cr | 873 | 1.06 x 10$^{23}$ | 1.5 x 10$^{20}$ | 707 | 0.14 | [30] |
| Ni-12.5 Al | 823 | 1.34 x 10$^{22}$ | 2.03 x 10$^{20}$ | 66 | 1.52 | Present study |

**Table 9:** Relevant physical constants measured or calculated from experimental data (3-D atom-probe tomography) utilized to determine the interfacial free energy, $\sigma^{\gamma/\gamma'}$, at 823 K (550 ºC) for Ni-12.5 Al at.%, and the diffusivity, $D_{experiment}^{coarsening}$.

| Measured Constant | Value |
|---|---|
| $V_m^{\gamma'}$ | 6.98 x 10$^{-6}$ m$^3$ mol$^{-1}$ |
| $C_{Al}^{\gamma,eq}(\infty)$ | 11.14 ± 0.32 at.% |
| $C_{Al}^{\gamma',eq}(\infty)$ | 23.14 ± 0.47 at.% |
| $K$ | 2.09 ± 0.10 x 10$^{-31}$ m$^3$ s$^{-1}$ |
| $\kappa^{\gamma}$ | 0.25 ± 0.01 s$^{1/3}$ |
| $\kappa^{\gamma'}$ | 0.68 ± 0.03 s$^{1/3}$ |
| $F(\phi_{\gamma'})$ | 1.8308 |
| $l^{\gamma}$ | 1.23 ± 0.07 x 10$^{-11}$ m |
| $D_{experiment}^{coarsening}$ | 2.51 ± 0.14 x 10$^{-21}$ m$^2$ s$^{-1}$ |

**Table 10:** Curvature of the molar Gibbs free energy of mixing with respect to concentration in the γ(f.c.c.)- phase, $G_m^\gamma$, and interfacial free energy, $\sigma^{\gamma/\gamma'}$, calculated using two different thermodynamic databases and the rate constants listed in Table 9 for a Ni-12.5 Al at.% alloy aged at 823 K. The two databases result in approximately equal values.

| Thermo-Calc Database | Saunders [68] | Dupin et al. [70] |
|---|---|---|
| $G_m^\gamma$ | 283,056 J mol$^{-1}$ | 269,968 J mol$^{-1}$ |
| $\sigma^{\gamma/\gamma'}$ | 29.94 ± 1.69 mJ m$^{-2}$ | 28.55 ± 1.61 mJ m$^{-2}$ |

**Table 11:** Solute diffusion coefficient pre-factors, $D_0$, and activation energies, $Q$, for Al solute diffusion in Ni and Ni$_3$Al and Ni solute diffusion in Ni$_3$Al, taken from different experimental and computational archival sources. Also listed are the resulting values of $D_{Al}$ for Al in Ni and Ni$_3$Al at 823 K and $D_{Ni}$ in Ni$_3$Al at 823 K.

| $D_0$ ($10^{-4}$ m$^2$ s$^{-1}$) | $Q$ (kJ mol$^{-1}$) | Temperature Range (K) | $D$ at 823 K (x $10^{-21}$ m$^2$ s$^{-1}$) | Reference | Techniques |
|---|---|---|---|---|---|
| Inter-diffusivities in Ni-Al alloys | | | | | |
| 1.87 | 267.8 | 1372-1553 | 1.83 | [158] | Kirkendall diffusion couples |
| Al diffusivity in Ni | | | | | |
| 1.10 | 249.1 | 1073-1243 | 17.01 | [170] | Electron diffraction |
| 10.0 | 272.0 | 1273-1573 | 5.44 | [171] | Electron probe microanalysis |
| 1.0 | 260 | 914-1212 | 3.14 | [172] | Vapor deposition; SIMS |
| 7.1 | 276.6 | 898-973 | 1.97 | [41] | Analysis of coarsening data |
| 29 | 290 | 1273-1623 | 1.14 | [173] | Laser Induced Breakdown Spectrometry |
| 7.52 | 284 | 1173-1673 | 0.71 | [159] | Mobility database |
| 9.03 | 282 | 600-900 | 1.14 | [174] | First-principles calculations |
| Al diffusivity in Ni$_3$Al | | | | | |
| 3.7 | 223.85 | 1418-1585 | 2292.17 | [175] | MD simulations |
| 0.00505 | 243 | 1173-1473 | 0.19 | [176] | Electron probe microanalysis |
| Ni diffusivity in Ni$_3$Al | | | | | |
| 3.23 | 302 | 1195-1523 | 0.02 | [177] | Electron probe microanalysis |
| 0.00917 | 116 | 1400-1550 | 3.98 x 10$^7$ | [178] | MD simulations |
| 0.00583 | 108 | 1300-1550 | 8.14 x 10$^7$ | [179] | MD simulations |

**Table 12:** The diffusion matrix and its related eigen-modes for the Ni-12.5 Al at.% alloy aged at 823 K (550 °C) as calculated employing LKMC$_1$. The calculations are performed in the γ(f.c.c.)-phase solid-solution with equilibrium composition, Ni-10.76 Al at.%. The last column displays the coupling between the Ni and Al flux, $J_{Ni}/J_{Al}$, required to build a γ'(L1$_2$)-precipitate with its equilibrium $J_{Ni}/J_{Al}$ composition from the supersaturated solutions.

| Diffusion matrix (m²s⁻¹) | Eigen diffusion coefficients (m²s⁻¹) | Diffusion flux coupling $J_{Ni}/J_{Al}$ | $J_{Ni}/J_{Al}$ Equilibrium |
|---|---|---|---|
| $\begin{pmatrix} 1.14 \times 10^{-22} & 2.95 \times 10^{-21} \\ -2.95 \times 10^{-21} & 1.13 \times 10^{-20} \end{pmatrix}$ | $D_{fast} = 1.05 \times 10^{-20}$ <br> $D_{slow} = 9.58 \times 10^{-22}$ | $J_{Ni}^{fast}/J_{Al}^{fast} = 0.082 \pm 0.02$ <br> $J_{Ni}^{slow}/J_{Al}^{slow} = -12.3 \pm 0.5$ | 0.121 |

**Figure captions**

**Fig. 1.** Summary of prior values determined for the interfacial free energy, $\sigma^{\gamma/\gamma'}$, between the γ(f.c.c.) and γ'(L1$_2$)-phases as a function of Al concentration and aging temperature. The values are also displayed in tabular form in Table 1.

**Fig. 2.** The overall composition of the alloy, 12.5 at.% Al, is indicated by a vertical dashed-line in the pertinent portion of the Ni-Al phase diagram. On the left-hand side, the solvus curves are displayed for calculations performed using Saunders's [68] (dot-dashed green-curve), quasi-grand canonical Monte Carlo (GCMC) simulation (dashed black-curve) [22], Dupin et al.'s thermodynamic databases (purple doted-curve) [70], employing Thermo-Calc [69], and Ma and Ardell's (solid blue-curve) experimental curve. Dupin et al.'s γ(f.c.c.)/(γ(f.c.c.) plus γ'(L1$_2$)) solvus curve (purple doted-curve) overlaps with Ma's and Ardell's γ/(γ plus γ') solvus curve (solid blue-curve) [71]. On the right-hand side, the solvus curves are as follows: solid-blue curve, Ma and Ardell [71]; the dashed black-curve, GCMC simulation [22]; dot-dashed green-curve, Saunders [68]; purple dotted-curve [70]; and dashed red-curve on the extreme right-hand side, Ardell and Nicholson [34]. The (γ plus γ') /γ' solvus curves due to Ma and Ardell, GCMC simulation, and Saunders overlap approximately, while Dupin's solvus curve is significantly to the right of those three solvus curves. The γ'/(γ'(L1$_2$) plus NiAl(B2 structure) phase field lies to the right of the dashed red-curve. Note that both the APT and LKMC simulations indicate that that nucleation of a second-phase occurs in this phase field. While synchrotron x-ray diffraction studies demonstrate that the crystal structure of the second phase is the L1$_2$ structure. Also note the large supersaturation of Al in the γ'(L1$_2$)-precipitate-phase. The horizontal heavy black arrows, pointing to the left, indicate the direction in which the compositions of the γ(f.c.c.)- and γ'(L1$_2$)-phases are temporally evolving. The 3-D APT and LKMC$_1$ results are offset for clarity, but they represent data for the same aging temperature, 823 K (550 °C).

**Fig. 3.** The temporal evolution of the Vickers microhardness of the Ni-12.5 Al at.% alloy aged at 823 K (550 °C). The increase in Vickers microhardness is consistent with the increase in volume fraction of the γ'(L1$_2$) -phase, Fig. 5a.

**Fig. 4.** A display of 3-D-APT reconstructions of a Ni-12.5 Al at. % alloy aged for: (a) 0.25 h; (b) 1; (c) 256; (d) 1024; (e) 2607; (f) 4096 h at 823 K (550 °C), and a dark-field TEM micrograph of a specimen aged for 1024 h at 823 K (550 °C); and (g) The γ'(L1$_2$)-precipitates are indicated by red Al iso-concentration surfaces. The value of the Al iso-concentration surface was determined for each dataset using the inflection point method [13]. Aluminum atoms are represented by red dots, while the Ni atoms are omitted for clarity. The γ'(L1$_2$)-precipitate number density is increasing in the nucleation regime, (a) and (b). $\langle R(t) \rangle$ increases and $N_v(t)$ decreases beyond 1 h due to growth and coarsening, (c)-(f). The γ'(L1$_2$)-precipitates nucleate and grow as spheroids,

(a), (b), and (c). In (d) they commence becoming facetted cuboids on the {100}-type planes as a result of a spheroid-to-cuboid transformation. Finally, in (e) and (f) they complete the transformation to cuboids, which are aligned along a <100>-type direction due to elastic interactions, which is commonly called rafting.

**Fig. 5.** The temporal evolution of: (a) the γ'(L1$_2$)-precipitate volume fraction, $\phi_{\gamma'}(t)$; (b) number density, $N_v(t)$; and (c) mean radius, $\langle R(t) \rangle$, aged at 823 K (550 °C). The quantity $\langle R(t) \rangle$ is proportional to $t^{-0.34 \pm 0.02}$, during quasi-stationary coarsening for aging times of 16 h and longer, as predicted by the LSW and Calderon, Voorhees et al. (CVMK) diffusion-limited mean–field coarsening models. Once $\phi_{\gamma'}(t)$ is within 25% of the equilibrium volume fraction (>64 h of aging), the temporal dependence of the quantity $N_v(t)$ commences to approach $t^{-1}$, as predicted by the LSW and CVMK diffusion-limited mean-field coarsening models.

**Fig. 6.** Concentrations of Al in: (a) γ(f.c.c.)-matrix and (b) γ'(L1$_2$)-precipitates (numerical values in Table 5), and supersaturations of Al in (c) γ(f.c.c.)-matrix and (d) γ'(L1$_2$)-precipitates from 3-D atom-probe tomography data and lattice kinetic Monte Carlo (LKMC$_1$) simulations. Both phases are initially supersaturated in Al and with increasing aging time the Al concentrations of the γ(f.c.c.)- and γ'(L1$_2$)- phases approach quasi-stationary values of 11.30 and 23.65 at.%, respectively. At 4096 h of aging, the supersaturation of Al in the γ(f.c.c.)-matrix is essentially zero and hence it is not plotted on the log$_{10}$ scale.

**Fig. 7.** The partitioning coefficients, $K_i^{\gamma'/\gamma}(t)$, of Al and Ni demonstrate that the Ni-12.5 Al at.% alloy aged at 823 K (550 °C) exhibits partitioning of Al to the γ'(L1$_2$)-precipitates and Ni to the γ(f.c.c.)-matrix.

**Fig. 8.** The γ'(L1$_2$)-precipitate size distributions (PSDs) aged at 823 K for: (a) 0.08 to 2607 h and (b) 4096 h. The total number of γ'(L1$_2$)-precipitates for each aging time is smaller than the effective number of γ'(L1$_2$)-precipitates, $N_{ppt}$, listed in Table 4 because only γ'(L1$_2$)-precipitates that are fully enclosed within the 3-D APT reconstruction volume (it varies from 4.2 x 10$^5$ to 9.8 x 10$^6$ nm$^3$ per data set) are used to generate the PSDs. The monovacancy-mediated LKMC$_1$ data are not used to generate PSDs because the computational volume analyzed is too small to yield satisfactory statistics.

**Fig. 9.** (a) Fraction of γ'(L1$_2$)-precipitates interconnected by necks as compared to (b) the mean minimum edge-to-edge distance between neighboring γ'(L1$_2$)-precipitates. The black solid-circles correspond to the APT results and the red outlined triangles represent the monovacancy-mediated LKMC$_1$ data.

**Fig. 10.** Concentration profiles for Ni and Al on either side of the γ(f.c.c.)/γ'(L1$_2$) interface for: (a) 0.25, 1, 256, and 4096 h extracted from 3-D APT data; and (b) 0.25, 4, 16, and 256 h from LKMC$_1$ simulations, Section 2.5. Positive distances are defined as into the γ'(L1$_2$)-precipitates, while negative distances are into the γ(f.c.c.)-matrix. The values of $\langle R(t) \rangle$ for these aging times are 1.34 ± 0.03, 1.47 ± 0.03, 5.65 ± 0.22, and 14.59 ± 1.87 nm, respectively.

**Fig. 11.** (a) Effect of aging time on the interfacial compositional width between the γ(f.c.c.)- and γ'(L1$_2$)-phases for a {100}-type plane. The interfacial compositional width is defined using the 10 and 90% points when fitting the proximity histogram for each dataset to a spline curve: (a) the value of $\delta(t)$ varies as $t^{-0.08 \pm 0.01}$; (b) Combination of the $\langle R(t) \rangle$ data from Fig. 5c with the $\delta(t)$ data from Fig. 11(a) to display the relationship between $\delta(t)$ and $\langle R(t) \rangle$ as the alloy is aged. *The quantity $\delta(t)$ decreases continuously with increasing $\langle R(t) \rangle$, varying as $\langle R(t) \rangle^{-0.47 \pm 0.03}$, which is in very strong contrast to the relationship obtained based on the so-called trans-interface-diffusion-controlled (TIDC) ansatz.*

**Fig. 12.** LKMC$_1$ snapshots of two γ'(L1$_2$)-precipitates in a γ(f.c.c.)-matrix (yellow background) at (a) 1 and (c) 4 h of aging time at 823 K (550 °C) for a 128x128x128 cell; and (e) an LKMC$_1$ snapshot of a single γ'(L1$_2$)-precipitate in a γ(f.c.c.)-matrix at 400 h of aging time at 823 K (550 °C) for a 256x256x256 cell. (a), (c), and (e) display the positions of a single monovacancy's trajectory (red circles) using monovacancy-solute binding energies through 4$^{th}$ NN distance. Note that the monovacancy spends the majority of its time inside the γ'(L1$_2$)-precipitates and the partially ordered necks between them, (a), (c), and (e). Additionally, the corresponding positions of the Ni (green) and Al (red) atoms are displayed for the γ'(L1$_2$)-precipitates for (b) 1, (d) 4, and (f) 400 h of aging. The partial ordering of the atoms in an L1$_2$ structure is evident in (b), (d), and (f). At 400 h the γ'(L1$_2$)-precipitate has {100}-type facets and appears to be essentially fully ordered; its interface is qualitatively diffuse. The interfacial region between the γ(f.c.c.)-matrix and γ'(L1$_2$)-precipitates, (b) 1 and (d) 4 h, is also qualitatively diffuse. The compositional interfacial width is quantified in figure 11.

**Fig. 13.** The temporal evolution of the edge-to-edge inter-precipitate spacing $d$ (open circles): 3-D APT; bold symbols LKMC$_1$) in this alloy compared to the root-mean-square (RMS) diffusion distance ($2\sqrt{Dt}$) for the fast and the slow modes of the diffusion-limited model, respectively. Fast mode: 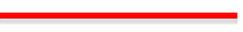; slow mode: 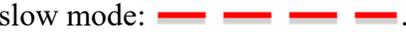.

**Fig. A1.** $\langle R(t) \rangle^p$ versus time plots at 823 K (550 °C) for $p$ values of: (a) $p = 2$; (b) $p = 2.4$; (c) $p = 3$; and (d) $p = 4$, as well as their associated linear fits. The coefficient of determination, $\xi^2$, is given for each linear fit to the data. It is emphasized strongly that this is not the best method

for determining a temporal exponent given that $\langle R(t) \rangle^2$, $\langle R(t) \rangle^{2.4}$, and $\langle R(t) \rangle^3$ versus aging time produce approximately the same value of $\xi^2$.

**Fig. A2**. The supersaturation $\Delta C_i^\gamma(t)$ in the γ(f.c.c.)-phase plotted versus aging time: (a) $t^{-1/2}$; (b) $t^{-1/2.4}$; (c) $t^{-1/3}$; and (d) $t^{-1/4}$, plus their associated linear fits to the data. The coefficient of determination, $\xi^2$, is given for each linear fit. It is emphasized strongly that this is not the best method for determining a temporal exponent.

*Figures*

*Figure 1*

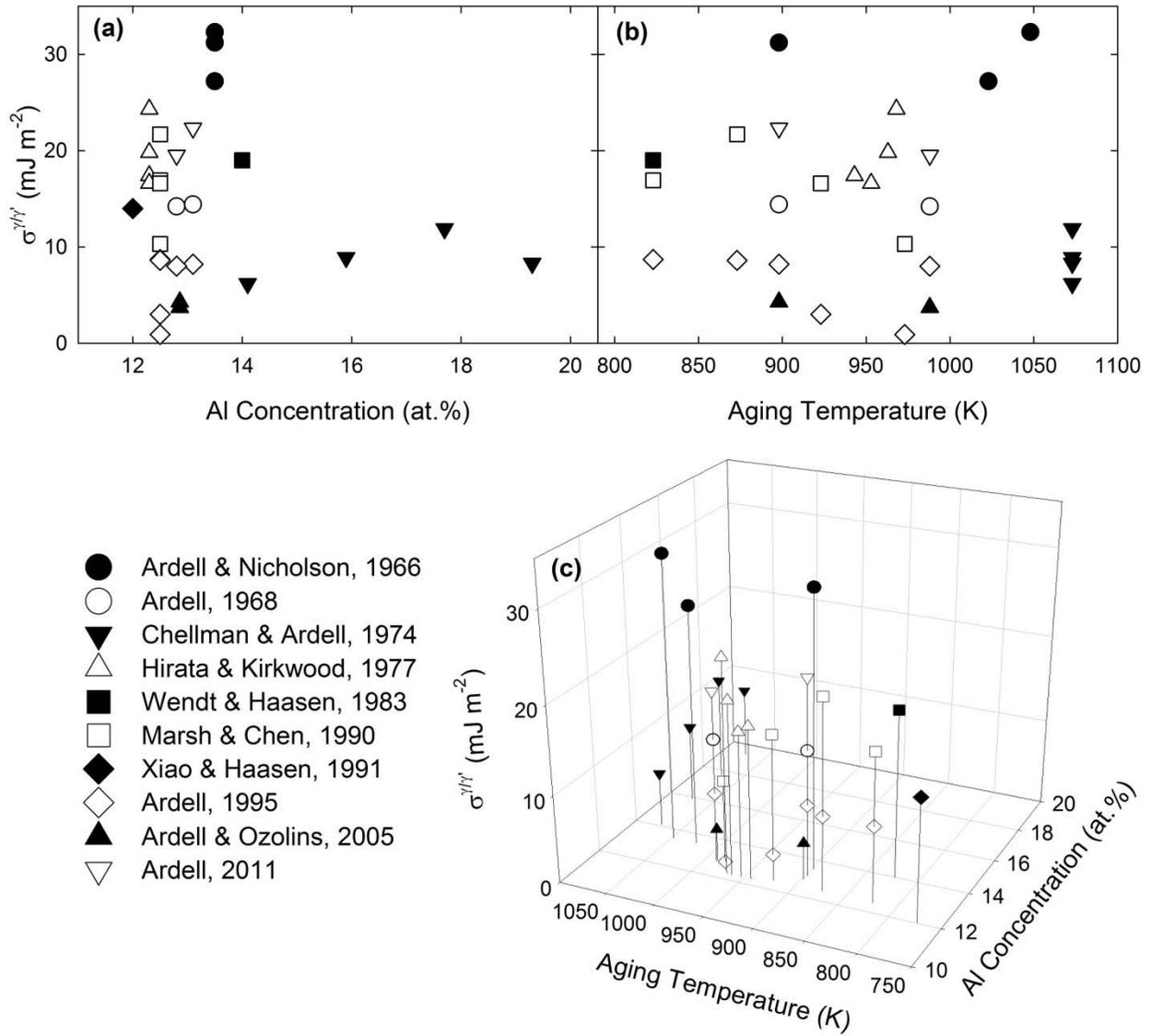

**Fig. 1.** Summary of prior values determined for the interfacial free energy, $\sigma^{\gamma/\gamma'}$, between the γ(f.c.c.) and γ'(L1$_2$)-phases as a function of Al concentration and aging temperature. The values are also displayed in tabular form in Table 1.

*Figure 2*

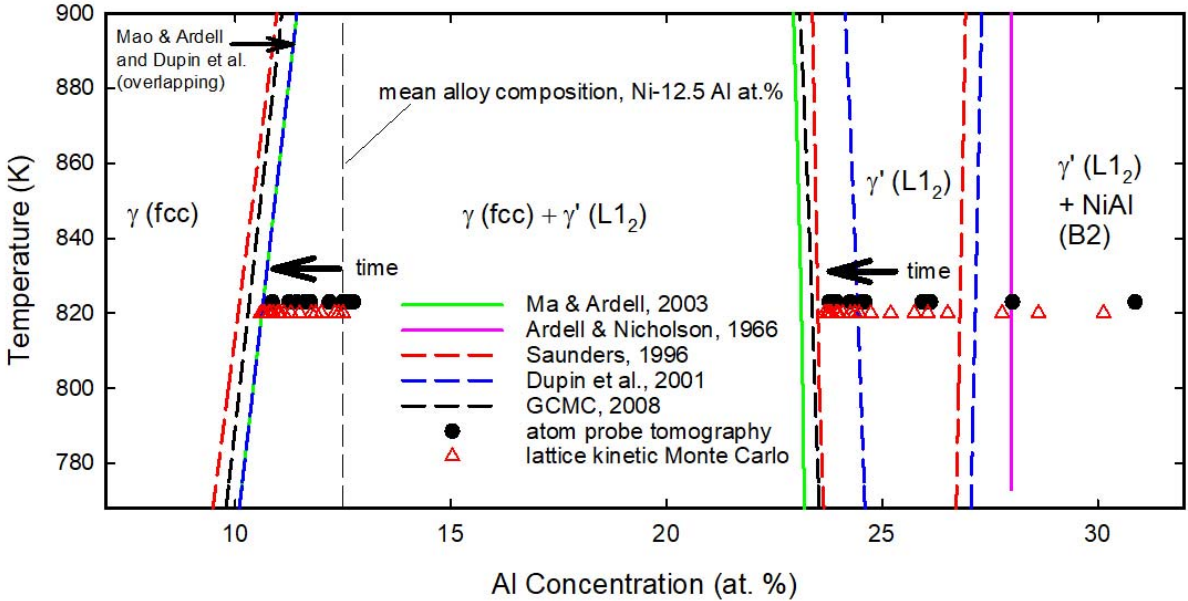

**Fig. 2.** The overall composition of the alloy, 12.5 at.% Al, is indicated by a vertical dashed-line in the pertinent portion of the Ni-Al phase diagram. On the left-hand side, the solvus curves are displayed for calculations performed using Saunders's [68] (dot-dashed green-curve), quasi-grand canonical Monte Carlo (GCMC) simulation (dashed black-curve) [22], Dupin et al.'s thermodynamic databases (purple doted-curve) [70], employing Thermo-Calc [69], and Ma and Ardell's (solid blue-curve) experimental curve. Dupin et al.'s γ(f.c.c.)/(γ(f.c.c.) plus γ'(L1$_2$)) solvus curve (purple doted-curve) overlaps with Ma's and Ardell's γ/(γ plus γ') solvus curve (solid blue-curve) [71]. On the right-hand side, the solvus curves are as follows: solid-blue curve, Ma and Ardell [71]; the dashed black-curve, GCMC simulation [22]; dot-dashed green-curve, Saunders [68]; purple dotted-curve [70]; and dashed red-curve on the extreme right-hand side, Ardell and Nicholson [34]. The (γ plus γ')/γ' solvus curves due to Ma and Ardell, GCMC simulation, and Saunders overlap approximately, while Dupin's solvus curve is significantly to the right of those three solvus curves. The γ'/(γ'(L1$_2$) plus NiAl(B2 structure) phase field lies to the right of the dashed red-curve. Note that both the APT and LKMC simulations indicate that that nucleation of a second-phase occurs in this phase field. While synchrotron x-ray diffraction studies demonstrate that the crystal structure of the second phase is the L1$_2$ structure. Also note the large supersaturation of Al in the γ'(L1$_2$)-precipitate-phase. The horizontal heavy black arrows, pointing to the left, indicate the direction in which the compositions of the γ(f.c.c.)- and γ'(L1$_2$)-phases are temporally evolving. The 3-D APT and LKMC$_1$ results are offset for clarity, but they represent data for the same aging temperature, 823 K (550 °C).

*Figure 3*

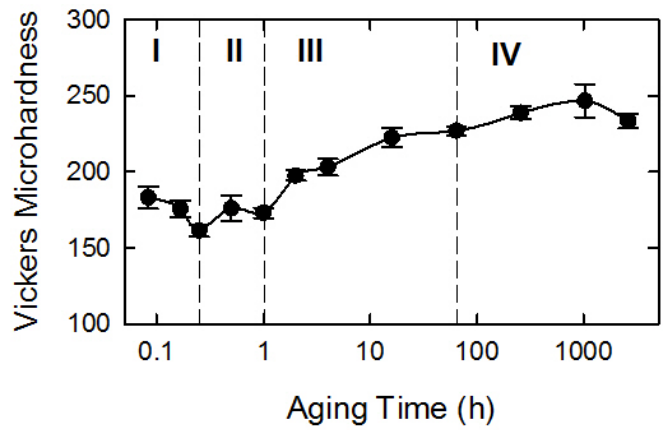

**Fig. 3.** The temporal evolution of the Vickers microhardness of the Ni-12.5 Al at.% alloy aged at 823 K (550 °C). The increase in Vickers microhardness is consistent with the increase in volume fraction of the γ'(L1$_2$) -phase, Fig. 5a.

*Figure 4*

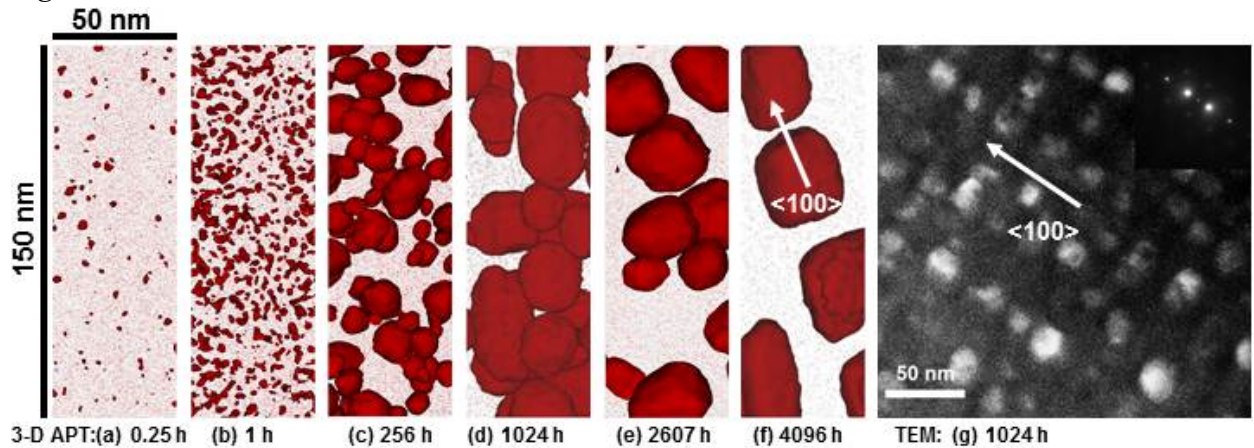

**Fig. 4.** A display of 3-D-APT reconstructions of a Ni-12.5 Al at. % alloy aged for: (a) 0.25 h; (b) 1; (c) 256; (d) 1024; (e) 2607; (f) 4096 h at 823 K (550 °C), and a dark-field TEM micrograph of a specimen aged for 1024 h at 823 K (550 °C); and (g) The γ'(L1$_2$)-precipitates are indicated by red Al iso-concentration surfaces. The value of the Al iso-concentration surface was determined for each dataset using the inflection point method [13]. Aluminum atoms are represented by red dots, while the Ni atoms are omitted for clarity. The γ'(L1$_2$)-precipitate number density is increasing in the nucleation regime, (a) and (b). $\langle R(t) \rangle$ increases and $N_v(t)$ decreases beyond 1 h due to growth and coarsening, (c)-(f). The γ'(L1$_2$)-precipitates nucleate and grow as spheroids, (a), (b), and (c). In (d) they commence becoming facetted cuboids on the {100}-type planes as a result of a spheroid-to-cuboid transformation. Finally, in (e) and (f) they complete the transformation to cuboids, which are aligned along a <100>-type direction due to elastic interactions, which is commonly called rafting.

*Figure 5*

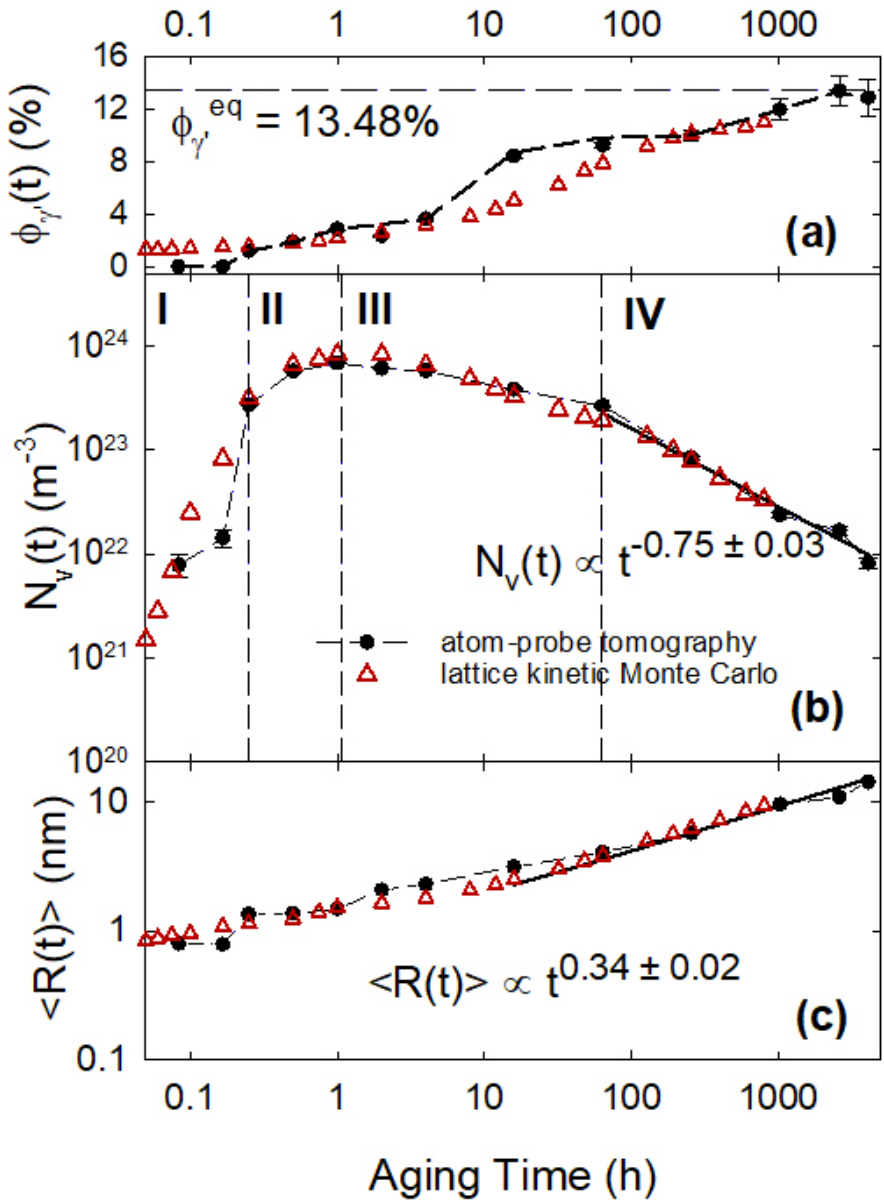

**Figure 5.** The temporal evolution of: (a) the γ'(L1₂)-precipitate volume fraction, $\phi_{\gamma'}(t)$; (b) number density, $N_v(t)$; and (c) mean radius, $\langle R(t) \rangle$, aged at 823 K (550 °C). The quantity $\langle R(t) \rangle$ is proportional to $t^{-0.34 \pm 0.02}$, during quasi-stationary coarsening for aging times of 16 h and longer, as predicted by the LSW and Calderon, Voorhees et al. (CVMK) diffusion-limited mean–field coarsening models. Once $\phi_{\gamma'}(t)$ is within 25% of the equilibrium volume fraction (>64 h of aging), the temporal dependence of the quantity $N_v(t)$ commences to approach $t^{-1}$, as predicted by the LSW and CVMK diffusion-limited mean-field coarsening models.

*Figure 6*

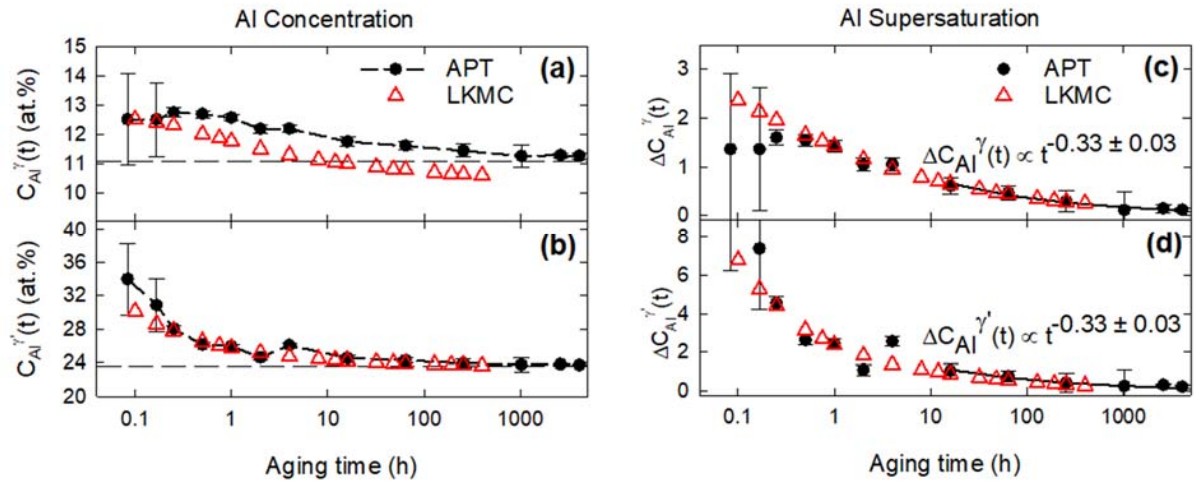

**Fig. 6.** Concentrations of Al in: (a) γ(f.c.c.)-matrix and (b) γ'(L1$_2$)-precipitates (numerical values in Table 5), and supersaturations of Al in (c) γ(f.c.c.)-matrix and (d) γ'(L1$_2$)-precipitates from 3-D atom-probe tomography data and lattice kinetic Monte Carlo (LKMC$_1$) simulations. Both phases are initially supersaturated in Al and with increasing aging time the Al concentrations of the γ(f.c.c.)- and γ'(L1$_2$)- phases approach quasi-stationary values of 11.30 and 23.65 at.%, respectively. At 4096 h of aging, the supersaturation of Al in the γ(f.c.c.)-matrix is essentially zero and hence it is not plotted on the log$_{10}$ scale.

*Figure 7*

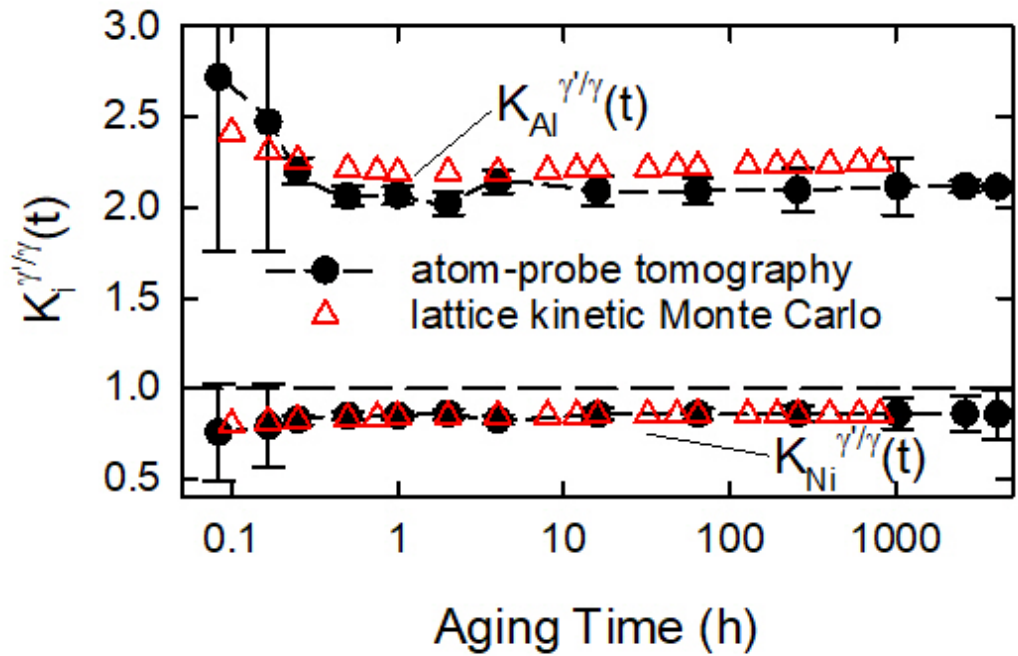

**Fig.7.** The partitioning coefficients, $K_i^{\gamma'/\gamma}(t)$, of Al and Ni demonstrate that the Ni-12.5 Al at.% alloy aged at 823 K (550 °C) exhibits partitioning of Al to the γ'(L1$_2$)-precipitates and Ni to the γ(f.c.c.)-matrix.

*Figure 8a*

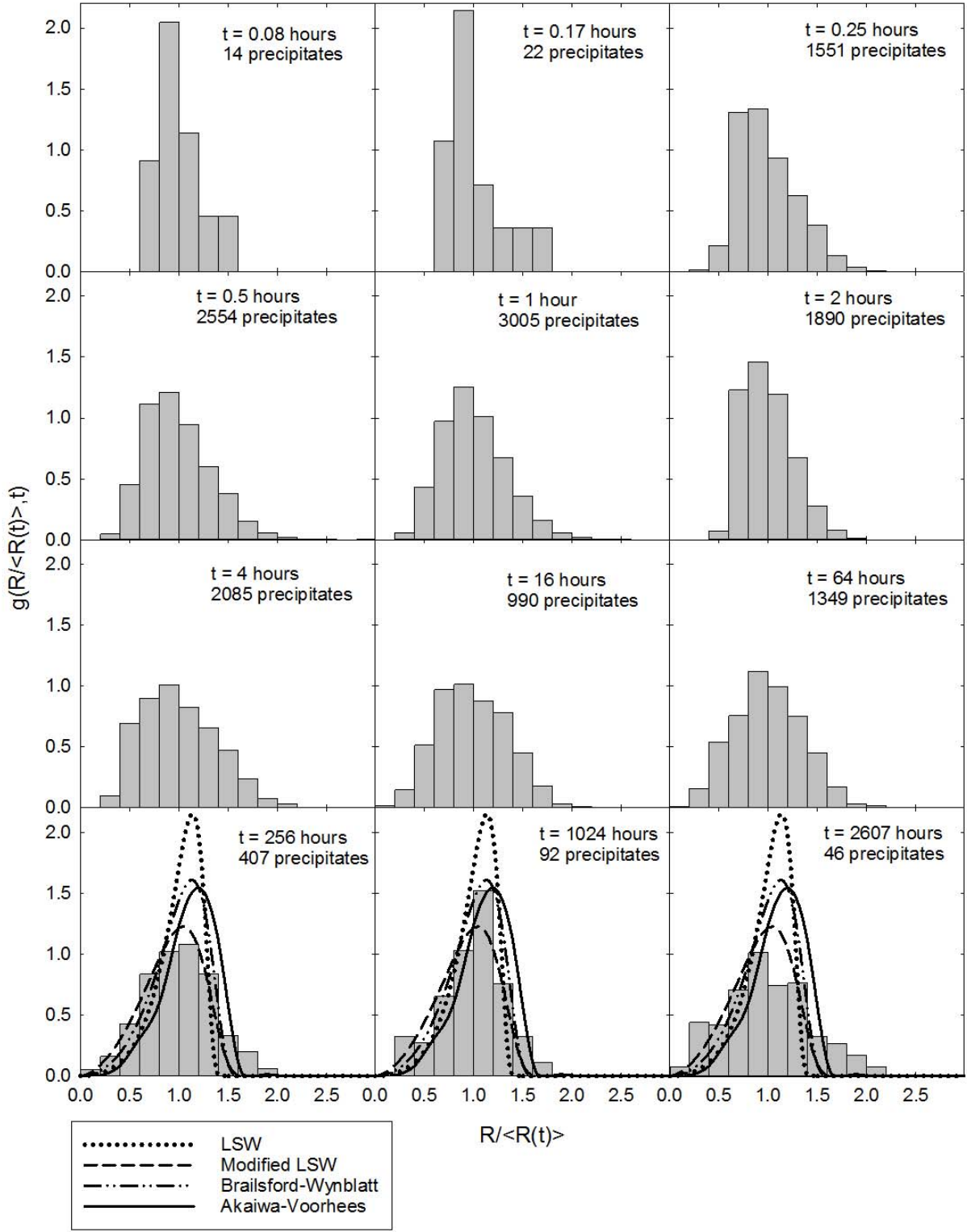

*Figure 8b*

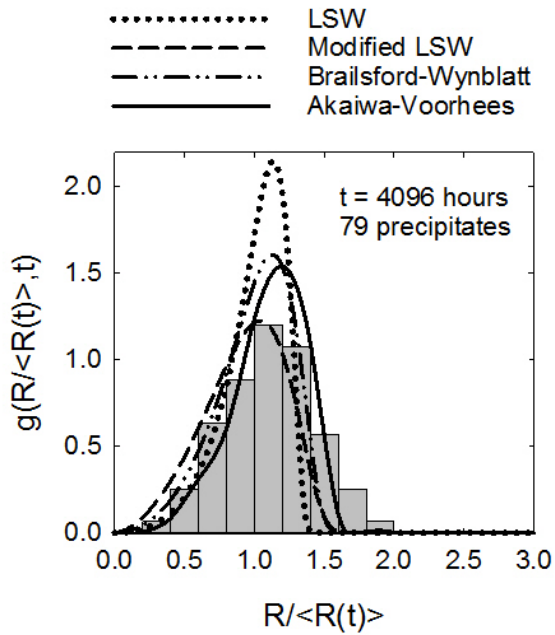

**Fig. 8.** The γ'(L1$_2$)-precipitate size distributions (PSDs) aged at 823 K for: (a) 0.08 to 2607 h and (b) 4096 h. The total number of γ'(L1$_2$)-precipitates for each aging time is smaller than the effective number of γ'(L1$_2$)-precipitates, $N_{ppt}$, listed in Table 4 because only γ'(L1$_2$)-precipitates that are fully enclosed within the 3-D APT reconstruction volume (it varies from 4.2 x 10$^5$ to 9.8 x 10$^6$ nm$^3$ per data set) are used to generate the PSDs. The monovacancy-mediated LKMC$_1$ data are not used to generate PSDs because the computational volume analyzed is too small to yield satisfactory statistics.

*Figure 8c*

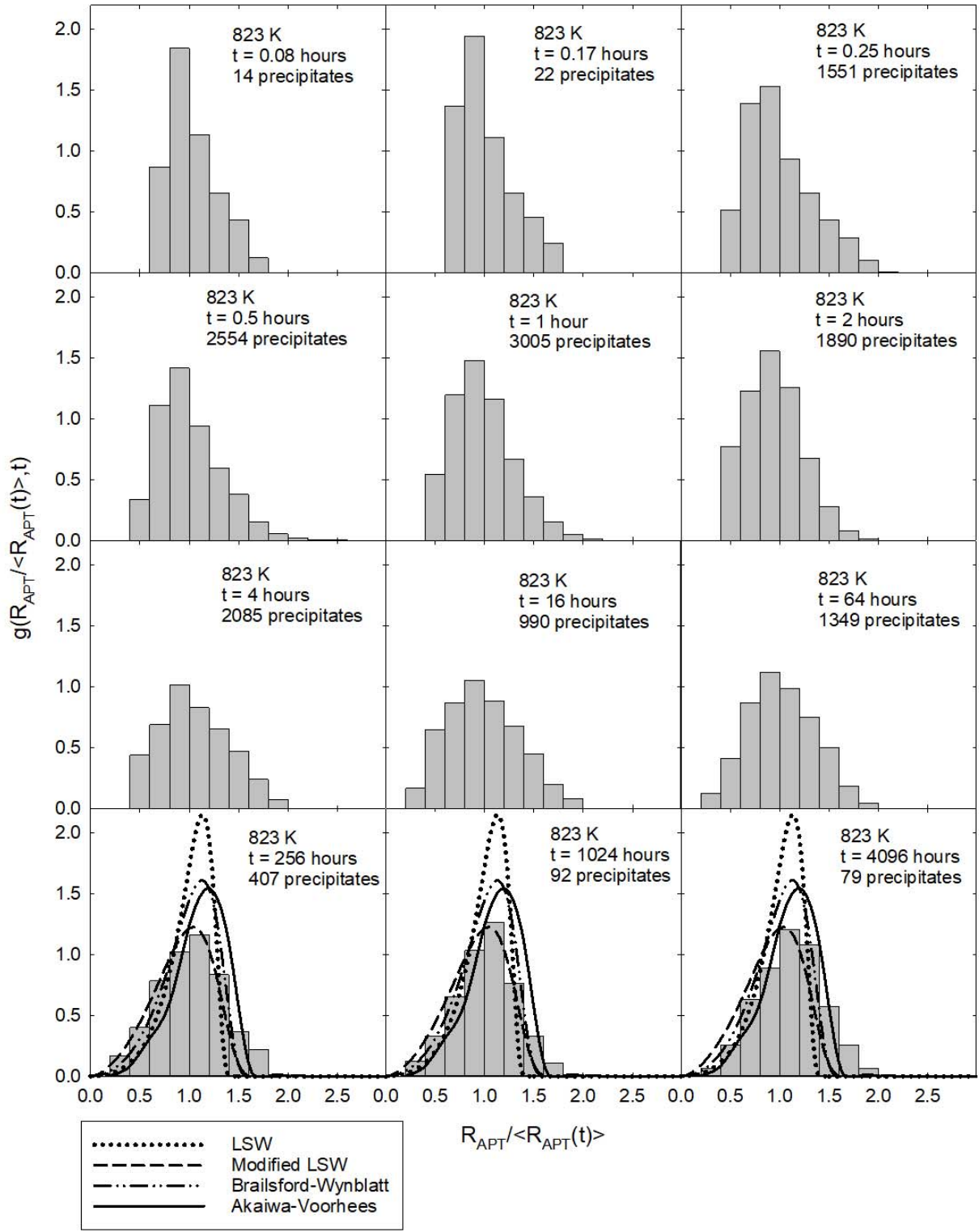

*Figure 9*

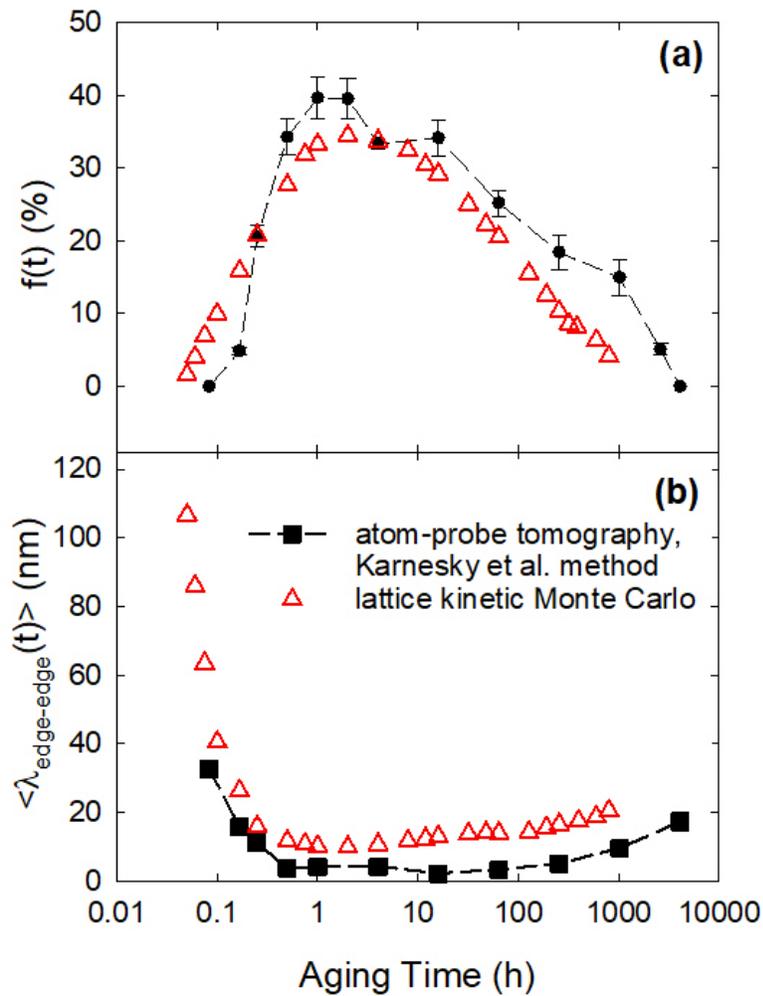

**Fig. 9.** (a) Fraction of γ'(L1$_2$)-precipitates interconnected by necks as compared to (b) the mean minimum edge-to-edge distance between neighboring γ'(L1$_2$)-precipitates. The black solid-circles correspond to the APT results and the red outlined triangles represent the monovacancy-mediated LKMC$_1$ data.

*Figure 10*

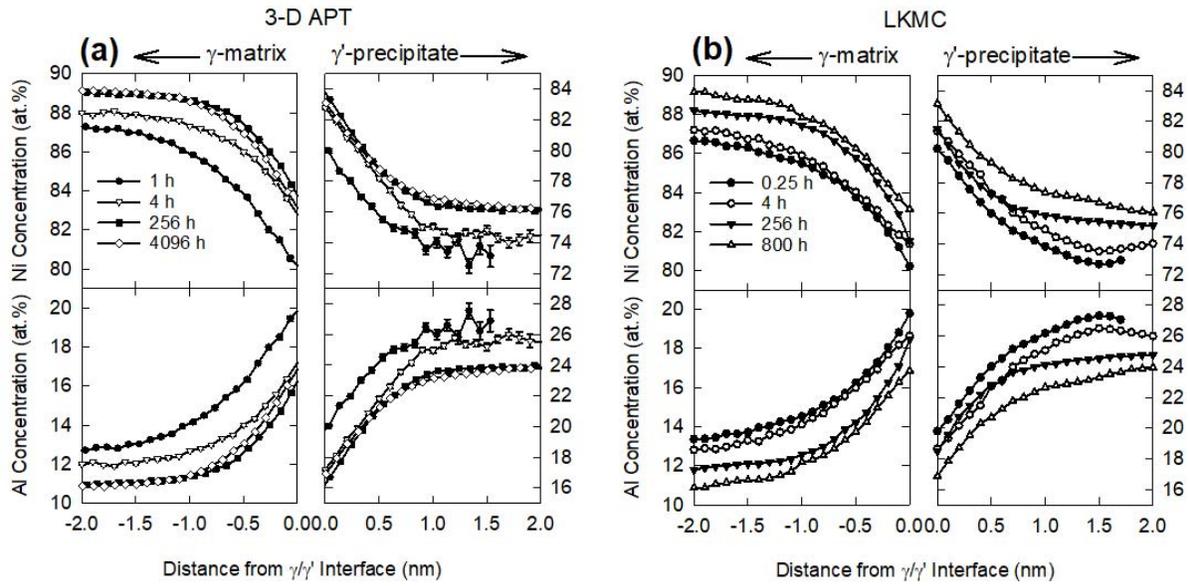

**Fig. 10.** Concentration profiles for Ni and Al on either side of the γ(f.c.c.)/γ'(L1$_2$) interface for: (a) 0.25, 1, 256, and 4096 h extracted from 3-D APT data; and (b) 0.25, 4, 16, and 256 h from LKMC$_1$ simulations, Section 2.5. Positive distances are defined as into the γ'(L1$_2$)-precipitates, while negative distances are into the γ(f.c.c.)-matrix. The values of $\langle R(t) \rangle$ for these aging times are 1.34 ± 0.03, 1.47 ± 0.03, 5.65 ± 0.22, and 14.59 ± 1.87 nm, respectively.

*Figure 11*

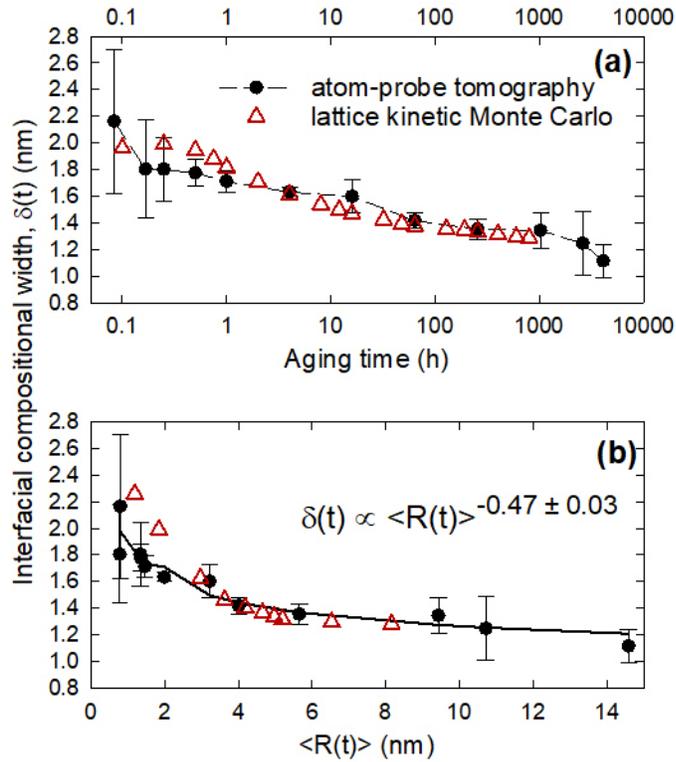

**Fig. 11.** (a) Effect of aging time on the interfacial compositional width between the γ(f.c.c.)- and γ'(L1$_2$)-phases for a {100}-type plane. The interfacial compositional width is defined using the 10 and 90% points when fitting the proximity histogram for each dataset to a spline curve: (a) the value of $\delta(t)$ varies as $t^{-0.08\pm0.01}$; (b) Combination of the $\langle R(t) \rangle$ data from Fig. 5c with the $\delta(t)$ data from Fig. 11(a) to display the relationship between $\delta(t)$ and $\langle R(t) \rangle$ as the alloy is aged. *The quantity $\delta(t)$ decreases continuously with increasing $\langle R(t) \rangle$, varying as $\langle R(t) \rangle^{-0.47\pm0.03}$, which is in very strong contrast to the relationship obtained based on the so-called trans-interface-diffusion-controlled (TIDC) ansatz.*

*Figure 12*

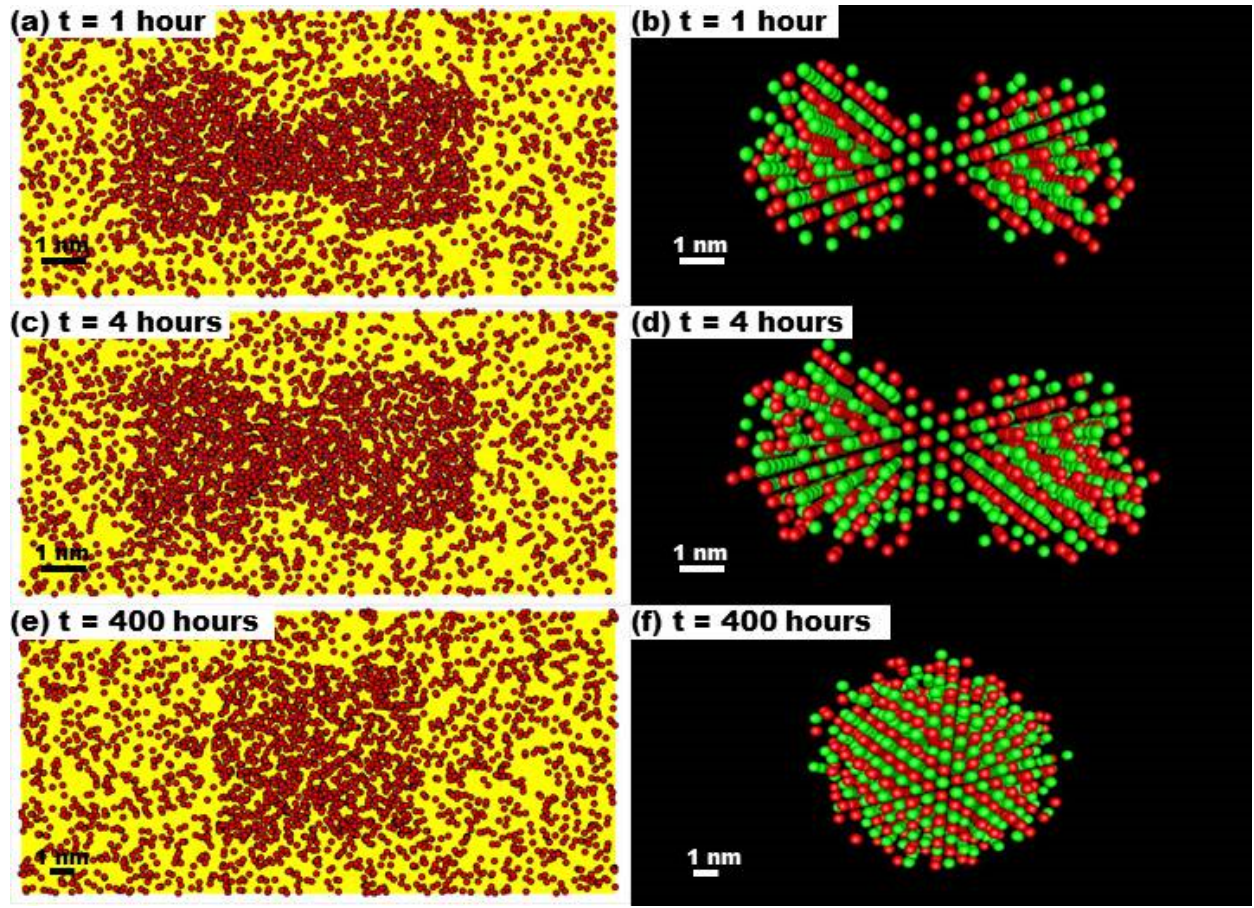

**Fig. 12.** LKMC$_1$ snapshots of two γ'(L1$_2$)-precipitates in a γ(f.c.c.)-matrix (yellow background) at (a) 1 and (c) 4 h of aging time at 823 K (550 °C) for a 128x128x128 cell; and (e) an LKMC$_1$ snapshot of a single γ'(L1$_2$)-precipitate in a γ(f.c.c.)-matrix at 400 h of aging time at 823 K (550 °C) for a 256x256x256 cell. (a), (c), and (e) display the positions of a single monovacancy's trajectory (red circles) using monovacancy-solute binding energies through 4$^{th}$ NN distance. Note that the monovacancy spends the majority of its time inside the γ'(L1$_2$)-precipitates and the partially ordered necks between them, (a), (c), and (e). Additionally, the corresponding positions of the Ni (green) and Al (red) atoms are displayed for the γ'(L1$_2$)-precipitates for (b) 1, (d) 4, and (f) 400 h of aging. The partial ordering of the atoms in an L1$_2$ structure is evident in (b), (d), and (f). At 400 h the γ'(L1$_2$)-precipitate has {100}-type facets and appears to be essentially fully ordered; its interface is qualitatively diffuse. The interfacial region between the γ(f.c.c.)-matrix and γ'(L1$_2$)-precipitates, (b) 1 and (d) 4 h, is also qualitatively diffuse. The compositional interfacial width is quantified in figure 11.

Figure 13

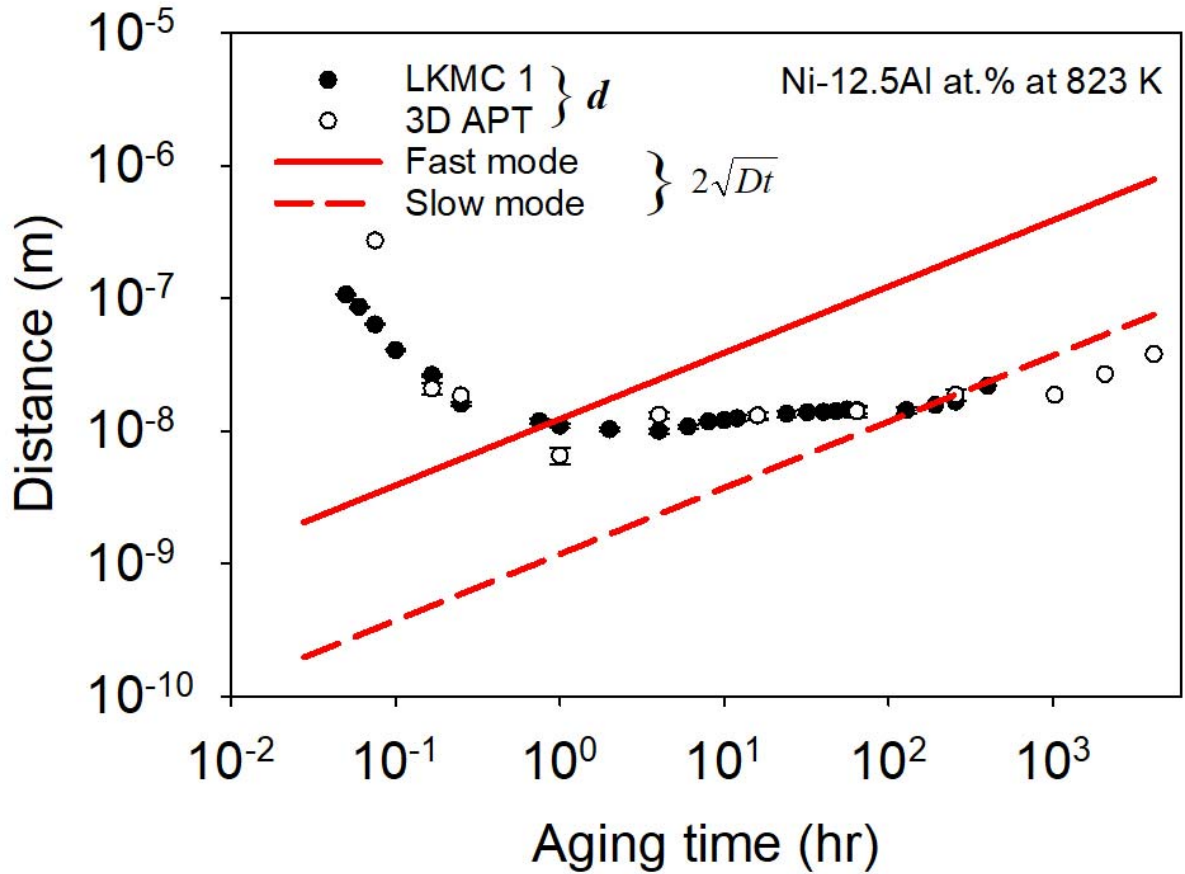

**Fig. 13**. The temporal evolution of the edge-to-edge inter-precipitate spacing $d$ (open circles): 3-D APT; bold symbols LKMC$_1$) in this alloy compared to the root-mean-square (RMS) diffusion distance ($2\sqrt{Dt}$) for the fast and the slow modes of the diffusion-limited model, respectively. Fast mode: ─────────; slow mode: ─ ─ ─ ─ ─.

*Figure A1*

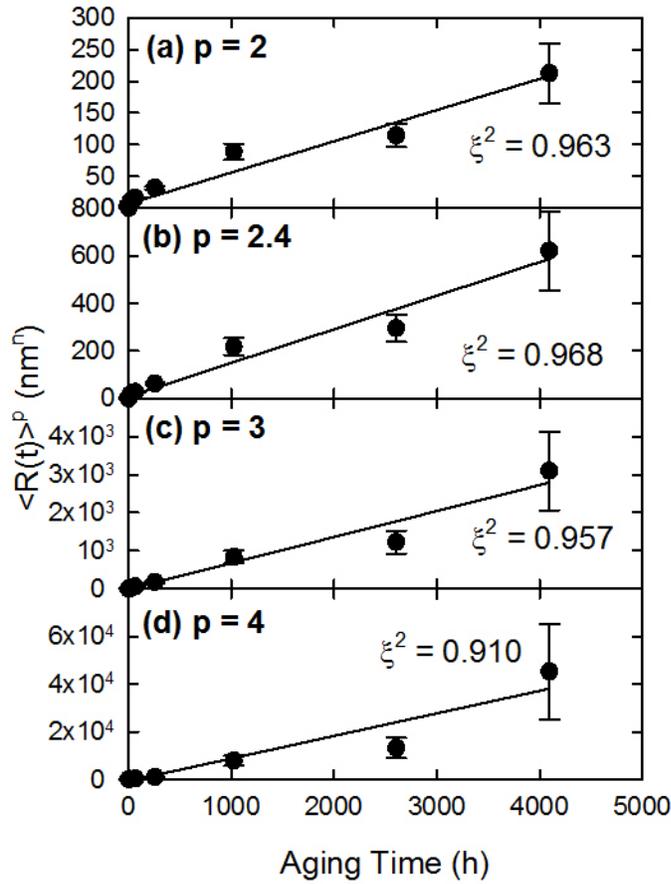

**Fig. A1.** $\langle R(t) \rangle^p$ versus time plots at 823 K (550 °C) for $p$ values of: (a) $p = 2$; (b) $p = 2.4$; (c) $p = 3$; and (d) $p = 4$, as well as their associated linear fits. The coefficient of determination, $\xi^2$, is given for each linear fit to the data. It is emphasized strongly that this is not the best method for determining a temporal exponent given that $\langle R(t) \rangle^2$, $\langle R(t) \rangle^{2.4}$, and $\langle R(t) \rangle^3$ versus aging time produce approximately the same value of $\xi^2$.

*Figure A2*

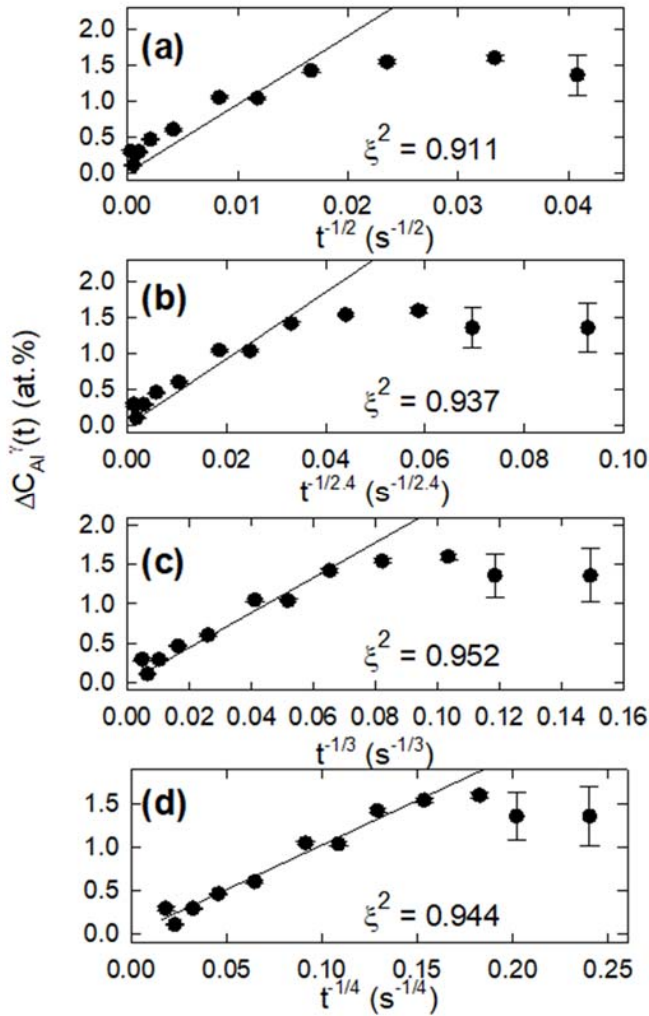

**Fig. A2**. The supersaturation $\Delta C_i^\gamma(t)$ in the γ(f.c.c.)-phase plotted versus aging time: (a) $t^{-1/2}$; (b) $t^{-1/2.4}$; (c) $t^{-1/3}$; and (d) $t^{-1/4}$, plus their associated linear fits to the data. The coefficient of determination, $\xi^2$, is given for each linear fit. It is emphasized strongly that this is not the best method for determining a temporal exponent.